\documentclass[prd, aps, nofootinbib, preprint, letterpaper, superscriptaddress]{revtex4-1}

\usepackage[colorlinks=true]{hyperref}
\usepackage{xcolor}
\hypersetup{colorlinks=true, citecolor=blue, urlcolor=blue, linkcolor=blue}
\usepackage{mathtools}
\usepackage{color}
\usepackage{graphicx}
\usepackage{tabularx}
\definecolor{darkred}{RGB}{196,0,0}

\usepackage{amsmath, amssymb, amsfonts}

\usepackage{ulem}

\begin{document}

\title {Study of the molecular Properties of the $P_c$ and $P_{cs}$ States}

\author{Jing-Zhi Cao}
\affiliation{Department of Physics, Guangxi Normal University, Guilin 541004, China}

\author{Huan-Yu Wei}
\affiliation{Department of Physics, Guangxi Normal University, Guilin 541004, China}

\author{Jiao-Xue Yang}
\affiliation{Department of Physics, Guangxi Normal University, Guilin 541004, China}

\author{Jian Sun}
\affiliation{School of Physics, Hunan Key Laboratory of Nanophotonics and Devices, Central South University, Changsha 410083, China}

\author{Chu-Wen Xiao}
\email{xiaochw@gxnu.edu.cn}
\affiliation{Department of Physics, Guangxi Normal University, Guilin 541004, China}
\affiliation{Guangxi Key Laboratory of Nuclear Physics and Technology, Guangxi Normal University, Guilin 541004, China}
\affiliation{School of Physics, Hunan Key Laboratory of Nanophotonics and Devices, Central South University, Changsha 410083, China}

\begin{abstract}

In the present work, we investigate the molecular properties of the hidden charm pentaquark states $P_c$ and $P_{cs}$ with a coupled channel framework that combines heavy quark spin symmetry and the local hidden gauge formalism. By solving the Bethe-Salpeter equation with the cutoff method, we obtain the pole trajectories, wave functions, and root-mean-square radii. For the hidden charm system, the full coupled channel interactions respecting the heavy quark spin symmetry are essential to generate the $P_c$ states, which significantly affect the poles' widths. The dominant bound channels are $\bar{D} \Sigma_c$ and $\bar{D}^* \Sigma_c$, which couple strongly to the lower decay channels. In contrast, for the hidden charm strange system, the full heavy quark spin symmetry treatment is not necessary, where the splitting PB and VB sectors yield similar results. The main bound channels $\bar{D} \Xi_c$ and $\bar{D}^* \Xi_c$ couple strongly to $\bar{D}_s \Lambda_c$ and $\bar{D}_s^* \Lambda_c$, respectively, but weakly to the lower decay channels, different from the hidden charm case. The trajectories of the pole widths for the loosely bound channels $\bar{D} \Xi'_c$, $\bar{D}^* \Xi'_c$, and $\bar{D}^* \Xi_c^*$ exhibit distinct behaviors. Notably, all the primary bound channels have similar binding energies in the single channel interactions due to equally attractive potentials. Furthermore, we also calculate the wave functions and root-mean-square radii of the corresponding poles. The wave functions are localized within $0\sim 6$ fm and vanish fast beyond $4$ fm. The root-mean-square radii, evaluated by two consistent methods, typically lie between $0.5$ and $2$ fm, comparable to the characteristic scale of molecular states. The root-mean-square radii depend on the pole trajectories and differ among the full coupled channel case, the split PB and VB sectors, and the single channel interactions.

\end{abstract}

\maketitle 

\section{Introduction}

The study of the internal structure of hadrons represents a cornerstone of modern nuclear and particle physics. For decades, the quark model~\cite{Gell-Mann:1964ewy,Zweig:1964ruk} has provided a remarkably successful framework for classifying hadrons, postulating that the mesons are bound states of a quark and an antiquark ($q\bar{q}$), and the baryons are composed of three quarks ($qqq$). This simple and elegant scheme accounts for the vast majority of the observed hadronic spectrum, leading to the widespread belief that these are the only possible configurations. However, the fundamental theory of the strong interaction, Quantum Chromodynamics (QCD), does not inherently forbid the existence of more complex, or ``exotic" hadronic states, such as glueballs (bound states of gluons), hybrids (quarks and gluons), and multiquark states like tetraquarks ($qq\bar{q}\bar{q}$) and pentaquarks ($qqqq\bar{q}$). And thus, the search for these exotic states becomes a critical test of QCD in its non-perturbative regime, which catches much attention both in the experiments and theories~\cite{Klempt:2009pi,Richard:2016eis,Esposito:2016noz,Guo:2017jvc,Karliner:2017qhf}.

The hunt for the pentaquark states has a long and controversial history. It was not until 2015 that the field witnessed a definitive breakthrough. 
In 2015, the LHCb Collaboration reported the unambiguous observation of two hidden-charm pentaquark candidates, named as $P_c(4380)^+$ and $P_c(4450)^+$, in the $\Lambda_b^0 \to J/\psi K^- p$ decay~\cite{LHCb:2015yax}. Subsequently, in 2019, with the high-statistics data, the LHCb Collaboration discovered that the state $P_c(4450)^+$ was actually composed of two states with similar masses but different narrow widths, $P_c(4440)^+$ and $P_c(4457)^+$, while also a lower-mass new state $P_c(4312)^+$ was found~\cite{LHCb:2019kea}. Although there was some evidence supporting the $P_c(4380)^+$ signal with the significance less than $5\sigma$. 
Extending these studies to the strange sector, in 2020, the LHCb Collaboration reported a new hidden-charm pentaquark state $P_{cs}(4459)^0$ containing a strange quark in the $\Xi_b^- \to J/\psi \Lambda K^-$ decay~\cite{LHCb:2020pdt}. In 2022, a narrow state $P_{cs}(4338)^0$ was observed in the $B^- \to J/\psi \Lambda \bar{p}$ decay~\cite{LHCb:2022ogu}. 
These discoveries have transformed the pentaquark states from a speculative concept into a vibrant experimental field, see more discussions in the forthcoming reviews~\cite{Richard:2016eis,Esposito:2016noz,Guo:2017jvc,Karliner:2017qhf,Chen:2016spr,Lebed:2016hpi,Ali:2017jda,Olsen:2017bmm,Yuan:2018inv,Liu:2019zoy,Brambilla:2019esw}.

Theoretically, the interpretation of these states has inspired widely debate. The proximity of these pentaquark states' masses to the thresholds of conventional charmed meson-baryon systems, such as the $\bar{D}^{(*)}\Sigma_c^{(*)}$ or $\bar{D}^{(*)}\Xi_c^{(*)}$ channels, had led to a natural conclusion that these states were not compact pentaquark bound states in the traditional sense, but rather ``molecular" states, analogous to the deuteron, implying that these pentaquark states might be bound states or resonances formed by a charmed baryon and a charmed anti-meson through the long-distant strong interactions in the hadronic level~\cite{Chen:2019asm,Liu:2019tjn,Xiao:2019mvs,Du:2019pij,Guo:2019kdc,Xiao:2019aya,Burns:2019iih,Zhu:2019iwm,Azizi:2016dhy,Azizi:2018bdv,Azizi:2020ogm,Chen:2020uif,Wang:2020eep,Chen:2020kco,Liu:2020hcv,Zhu:2021lhd,Xiao:2021rgp,Azizi:2021utt,Wang:2022neq,Meng:2022wgl,Zhu:2022wpi,Feijoo:2022rxf}. This interpretation ties the study of pentaquark states intimately to the dynamics of the strong interaction near threshold and the nature of exotic hadronic molecules~\cite{Guo:2017jvc,Liu:2019zoy,Dong:2020hxe,Dong:2021juy,Dong:2021bvy,Azizi:2017bgs,Azizi:2018dva}. 
Even though this significant threshold proximity provides strong support for the hadronic molecular picture, numerous fundamental questions remain open, such as assigning them as the compact pentaquark states or hadrocharmonium states~\cite{Giannuzzi:2019esi,Ali:2019npk,Eides:2019tgv,Shi:2021wyt,Giron:2021fnl,Mohan:2026blk}. The nature of the binding mechanism is still under debate that whether it is predominantly driven by the long-range boson exchange, short-range QCD dynamics, or a combination thereof~\cite{Liu:2019zoy,Burns:2019iih,Yamaguchi:2019seo,Chen:2022asf,Zou:2021sha,Meng:2022ozq,Hanhart:2025bun}. And thus, the molecular picture and compact multiquark states are also key unresolved issues~\cite{Gross:2022hyw,Chen:2022asf}. 
Furthermore, the existence of some of these pentaquark states were also questionable with the kinematic effects of the triangle singularity~\cite{Guo:2015umn,Liu:2015fea,Mikhasenko:2015vca,Bayar:2016ftu,Guo:2019twa,Shen:2020gpw,Nakamura:2021qvy,Duan:2023dky} or the cusp effects~\cite{Kuang:2020bnk,Nakamura:2021dix}, which was not supported by a deep learning framework~\cite{Co:2024bfl}, see more discussions on the triangle singularity in Refs.~\cite{Zhang:2024qkg,Sakthivasan:2024uwd} .

In the present work, to shed light on the internal structure of the $P_c$ and $P_{cs}$ states and provide a deeper understanding of their molecular properties, we systematically investigate the strong interactions of the systems $\bar{D}^{(*)}\Sigma_c^{(*)}$, $\bar{D}^{(*)}\Xi_c^{(*)}$ with their coupled channels and the molecular nature of these $P_c$, $P_{cs}$ states in detail within a coupled channel interaction approach, where the pole structures, wave functions, and the radii of these resonances are evaluated. 
In the next section, we make a brief introduction of our formalism of the coupled channel interaction approach. Following, our study results for different cases are shown in detail. Finally, a short conclusion is made at the end.

\section{Theoretical Framework}\label{model}

Within the framework of the coupled channel approach, the scattering amplitude $T$ can be obtained by solving the on-shell Bethe-Salpeter equation in algebraic matrix form~\cite{Oset:1997it},
\begin{equation}
T = [1 - VG]^{-1}V , \label{eq:BS}
\end{equation}
where $G$ is the diagonal matrix made of the  loop functions with the propagator of the intermediate meson-baryon system, and $V$ is the potential matrix for the coupled-channel interactions. 
In the isospin $I = 1/2$ and spin-parity $J^P = 1/2^-$ sector of the hidden charm system, where the $P_c$ states appear, there are seven coupled channels, $\eta_c N$, $J/\psi N$, $ \bar{D}\Lambda_c$, $\bar{D}\Sigma_c$, $\bar{D}^*\Lambda_c$, $\bar{D}^*\Sigma_c$, and $\bar{D}^*\Sigma_c^*$. Following Ref.~\cite{Xiao:2013yca}, the potential matrix elements $V_{ij}$ for these seven coupled channels are shown in Table~\ref{tab:I1J1c7}, which had been taken into account the constraint of the heavy quark spin symmetry (HQSS)~\cite{Isgur:1989vq,Neubert:1993mb}, see more detail in Ref.~\cite{Xiao:2013yca}. 
Besides, for the isospin $I = 1/2$ and spin-parity $J^P = 3/2^-$ sector, there are five coupled channels, $J/\psi N$, $\bar{D}^*\Lambda_c$, $\bar{D}^*\Sigma_c$, $\bar{D}\Sigma_c^*$, and $\bar{D}^*\Sigma_c^*$, where corresponding coupled channel potentials are given in Eq.~(31) of Ref.~\cite{Xiao:2013yca}, not listed here.

\begin{table}[ht]
\centering
\caption{Interaction potentials $V_{ij}$ for the seven coupled channels in the $I = 1/2, J^P = 1/2^-$ sector.}\label{tab:I1J1c7}
\footnotesize
\renewcommand{\arraystretch}{1.2} 
\setlength{\tabcolsep}{2pt}     
\begin{tabular}{ccccccc}
\hline\hline
$\eta_c N$ & $J/\psi N$ & $\bar{D}\Lambda_c$ & $\bar{D}\Sigma_c$ & $\bar{D}^*\Lambda_c$ & $\bar{D}^*\Sigma_c$ & $\bar{D}^*\Sigma_c^*$ \\ \hline
$\mu_1$ & 0 & $\dfrac{\mu_{12}}{2}$ & $\dfrac{\mu_{13}}{2}$ & $\dfrac{\sqrt{3}\mu_{12}}{2}$ & $-\dfrac{\mu_{13}}{2\sqrt{3}}$ & $\sqrt{\dfrac{2}{3}}\mu_{13}$ \\
0 & $\mu_1$ & $\dfrac{\sqrt{3}\mu_{12}}{2}$ & $-\dfrac{\mu_{13}}{2\sqrt{3}}$ & $-\dfrac{\mu_{12}}{2}$ & $\dfrac{5\mu_{13}}{6}$ & $\dfrac{\sqrt{2}\mu_{13}}{3}$ \\
$\dfrac{\mu_{12}}{2}$ & $\dfrac{\sqrt{3}\mu_{12}}{2}$ & $\mu_2$ & 0 & 0 & $\dfrac{\mu_{23}}{\sqrt{3}}$ & $\sqrt{\dfrac{2}{3}}\mu_{23}$ \\
$\dfrac{\mu_{13}}{2}$ & $-\dfrac{\mu_{13}}{2\sqrt{3}}$ & 0 & $\dfrac{1}{3}(2\lambda_2 + \mu_3)$ & $\dfrac{\mu_{23}}{\sqrt{3}}$ & $\dfrac{2(\lambda_2 - \mu_3)}{3\sqrt{3}}$ & $\dfrac{1}{3}\sqrt{\dfrac{2}{3}}(\mu_3 - \lambda_2)$ \\
$\dfrac{\sqrt{3}\mu_{12}}{2}$ & $-\dfrac{\mu_{12}}{2}$ & 0 & $\dfrac{\mu_{23}}{\sqrt{3}}$ & $\mu_2$ & $-\dfrac{2\mu_{23}}{3}$ & $\dfrac{\sqrt{2}\mu_{23}}{3}$ \\
$-\dfrac{\mu_{13}}{2\sqrt{3}}$ & $\dfrac{5\mu_{13}}{6}$ & $\dfrac{\mu_{23}}{\sqrt{3}}$ & $\dfrac{2(\lambda_2 - \mu_3)}{3\sqrt{3}}$ & $-\dfrac{2\mu_{23}}{3}$ & $\dfrac{1}{9}(2\lambda_2 + 7\mu_3)$ & $\dfrac{1}{9}\sqrt{2}(\mu_3 - \lambda_2)$ \\
$\sqrt{\dfrac{2}{3}}\mu_{13}$ & $\dfrac{\sqrt{2}\mu_{13}}{3}$ & $\sqrt{\dfrac{2}{3}}\mu_{23}$ & $\dfrac{1}{3}\sqrt{\dfrac{2}{3}}(\mu_3 - \lambda_2)$ & $\dfrac{\sqrt{2}\mu_{23}}{3}$ & $\dfrac{1}{9}\sqrt{2}(\mu_3 - \lambda_2)$ & $\dfrac{1}{9}(\lambda_2 + 8\mu_3)$ \\ \hline\hline
\end{tabular}
\end{table}

Note that, in Table~\ref{tab:I1J1c7} the low energy constants $\mu_i$ (or $\mu_{ij}$), $\lambda_i$ are not specified by the HQSS, and thus, they should be determined by the other models, such as the local hidden gauge (LHG) formalism~\cite{Bando:1984ej,Bando:1987br,Meissner:1987ge,Nagahiro:2008cv}, as done in Ref.~\cite{Xiao:2013yca}. Within the LHG framework and using the vector meson exchange mechanism, these low energy constants after the $S$-wave projection are given by~\cite{Xiao:2013yca},
\begin{equation}
\begin{aligned}
&\mu_1 = 0, \quad \mu_{23} = 0, \quad \lambda_2 = \mu_3, \quad \mu_{13} = -\mu_{12}, \\
&\mu_2 = \frac{1}{4f^2}(k^0 + k'^0), \quad \mu_3 = -\frac{1}{4f^2}(k^0 + k'^0), \\
&\mu_{12} = -\sqrt{6}\frac{m_\rho^2}{p_{D^*}^2 - m_{D^*}^2} \frac{1}{4f^2}(k^0 + k'^0) , 
\end{aligned}
\label{eq:lechc}
\end{equation}
where $f_{\pi} = 93 \, \text{MeV}, m_{\rho} = 775 \, \text{MeV}$ are taken, $k^0$ and $k'^0$ represent the center-of-mass energies of the incoming and outgoing mesons in the transition process $MB \to M'B'$, respectively, given by 
$k^0 = \frac{s+m^2-M^2}{2\sqrt{s}}$,
with $m$ and $M$ the masses of the meson and baryon in the corresponding channel, and $s$ the total energy square of the system. 
Additionally, the transfer momentum squared $p_{D^*}^2$ is kept for the non-diagonal elements, which is taken as $p_{D^*}^2\approx m^2 + m'^2 - 2k^0k'^0$.

As discussed in the last section, to systematically investigate the molecular nature of the $P_c$ and $P_{cs}$ states, we also study the strong interactions of the $\bar{D}^{(*)}\Xi_c^{(*)}$ channels in the hidden charm strange system compared with the ones of the $\bar{D}^{(*)}\Sigma_c^{(*)}$ channels. 
Note that, Ref.~\cite{Xiao:2019gjd} extended the framework of Ref.~\cite{Xiao:2013yca} to the charmed and strange sector with the results of Ref.~\cite{Xiao:2019aya}. 
Thus, known from Ref.~\cite{Xiao:2019gjd}, there are nine coupled channels, $\eta_c\Lambda$, $J/\psi\Lambda$, $\bar{D}\Xi_c$, $\bar{D}_s\Lambda_c$, $\bar{D}\Xi'_c$, $\bar{D}^*\Xi_c$, $\bar{D}^*_s\Lambda_c$, $\bar{D}^*\Xi'_c$, and $\bar{D}^*\Xi^*_c$, in the isospin $I = 0$ and spin-parity $J^P = 1/2^-$ sector, of which the interaction potential matrix ($V_{ij}$) is shown in Table~\ref{tab:I0J1cs9} with the constraint of the HQSS. 
Furthermore, for the isospin $I = 0$ and spin-parity $J^P = 3/2^-$ sector, the system consists of six coupled channels, $J/\psi \Lambda$, $\bar{D}^*\Xi_c$, $\bar{D}_s^* \Lambda_c$, $ \bar{D}^*\Xi'_c$, $\bar{D}\Xi_c^*$, and $\bar{D}^*\Xi_c^*$, of which the corresponding potentials are given in Eq.~(6) of Ref.~\cite{Xiao:2019gjd}.

\begin{table}[ht]
\centering
\caption{Interaction potentials $V_{ij}$ for the nine coupled channels in the $I = 0, J^P = 1/2^-$ sector.}\label{tab:I0J1cs9}
\footnotesize
\renewcommand{\arraystretch}{1.2}
\setlength{\tabcolsep}{2pt} 
\begin{tabular}{ccccccccc}
\hline\hline
$\eta_c \Lambda$ & $J/\psi \Lambda$ & $\bar{D}\Xi_c$ & $\bar{D}_s \Lambda_c$ & $\bar{D}\Xi'_c$ & $\bar{D}^*\Xi_c$ & $\bar{D}^*_s \Lambda_c$ & $\bar{D}^*\Xi'_c$ & $\bar{D}^*\Xi^*_c$ \\ \hline
$\mu_1$ & 0 & $-\dfrac{1}{2}\mu_{12}$ & $-\dfrac{1}{2}\mu_{13}$ & $\dfrac{1}{2}\mu_{14}$ & $\dfrac{\sqrt{3}}{2}\mu_{12}$ & $\dfrac{\sqrt{3}}{2}\mu_{13}$ & $\dfrac{1}{2\sqrt{3}}\mu_{14}$ & $\sqrt{\dfrac{2}{3}}\mu_{14}$ \\
0 & $\mu_1$ & $\dfrac{\sqrt{3}}{2}\mu_{12}$ & $\dfrac{\sqrt{3}}{2}\mu_{13}$ & $\dfrac{1}{2\sqrt{3}}\mu_{14}$ & $\dfrac{1}{2}\mu_{12}$ & $\dfrac{1}{2}\mu_{13}$ & $\dfrac{5}{6}\mu_{14}$ & $-\dfrac{\sqrt{2}}{3}\mu_{14}$ \\
$-\dfrac{1}{2}\mu_{12}$ & $\dfrac{\sqrt{3}}{2}\mu_{12}$ & $\mu_2$ & $\mu_{23}$ & 0 & 0 & 0 & $\dfrac{1}{\sqrt{3}}\mu_{24}$ & $-\sqrt{\dfrac{2}{3}}\mu_{24}$ \\
$-\dfrac{1}{2}\mu_{13}$ & $\dfrac{\sqrt{3}}{2}\mu_{13}$ & $\mu_{23}$ & $\mu_3$ & 0 & 0 & 0 & $\dfrac{1}{\sqrt{3}}\mu_{34}$ & $-\sqrt{\dfrac{2}{3}}\mu_{34}$ \\
$\dfrac{1}{2}\mu_{14}$ & $\dfrac{1}{2\sqrt{3}}\mu_{14}$ & 0 & 0 & $\dfrac{1}{3}(2\lambda + \mu_4)$ & $\dfrac{1}{\sqrt{3}}\mu_{24}$ & $\dfrac{1}{\sqrt{3}}\mu_{34}$ & $-\dfrac{2}{3\sqrt{3}}(\lambda - \mu_4)$ & $\dfrac{1}{3}\sqrt{\dfrac{2}{3}}(\mu_4 - \lambda)$ \\
$\dfrac{\sqrt{3}}{2}\mu_{12}$ & $\dfrac{1}{2}\mu_{12}$ & 0 & 0 & $\dfrac{1}{\sqrt{3}}\mu_{24}$ & $\mu_2$ & $\mu_{23}$ & $\dfrac{2}{3}\mu_{24}$ & $\dfrac{\sqrt{2}}{3}\mu_{24}$ \\
$\dfrac{\sqrt{3}}{2}\mu_{13}$ & $\dfrac{1}{2}\mu_{13}$ & 0 & 0 & $\dfrac{1}{\sqrt{3}}\mu_{34}$ & $\mu_{23}$ & $\mu_3$ & $\dfrac{2}{3}\mu_{34}$ & $\dfrac{\sqrt{2}}{3}\mu_{34}$ \\
$\dfrac{1}{2\sqrt{3}}\mu_{14}$ & $\dfrac{5}{6}\mu_{14}$ & $\dfrac{1}{\sqrt{3}}\mu_{24}$ & $\dfrac{1}{\sqrt{3}}\mu_{34}$ & $-\dfrac{2}{3\sqrt{3}}(\lambda - \mu_4)$ & $\dfrac{2}{3}\mu_{24}$ & $\dfrac{2}{3}\mu_{34}$ & $\dfrac{1}{9}(2\lambda + 7\mu_4)$ & $\dfrac{\sqrt{2}}{9}(\lambda - \mu_4)$ \\
$\sqrt{\dfrac{2}{3}}\mu_{14}$ & $-\dfrac{\sqrt{2}}{3}\mu_{14}$ & $-\sqrt{\dfrac{2}{3}}\mu_{24}$ & $-\sqrt{\dfrac{2}{3}}\mu_{34}$ & $\dfrac{1}{3}\sqrt{\dfrac{2}{3}}(\mu_4 - \lambda)$ & $\dfrac{\sqrt{2}}{3}\mu_{24}$ & $\dfrac{\sqrt{2}}{3}\mu_{34}$ & $\dfrac{\sqrt{2}}{9}(\lambda - \mu_4)$ & $\dfrac{1}{9}(\lambda + 8\mu_4)$ \\ \hline\hline
\end{tabular}
\end{table}

Analogously, using the LHG formalism, the derived low energy constants are determined as~\cite{Xiao:2019gjd},
\begin{equation}
\begin{aligned}
&\mu_1 = \mu_3 = \mu_{24} = \mu_{34} = 0, \\
&\mu_2 = \frac{\mu_{23}}{\sqrt{2}} = \mu_4 = \lambda = - \frac{1}{4f^2}(k^0 + k'^0), \\
&\mu_{12} = -\frac{\mu_{13}}{\sqrt{2}} = \frac{\mu_{14}}{\sqrt{3}} = -\sqrt{\frac{2}{3}}\frac{m_V^2}{m_{D^*}^2} \frac{1}{4f^2}(k^0 + k'^0),
\end{aligned}
\label{eq:lechcs}
\end{equation}
where we take $f_\pi = 93 \, MeV$ and $m_V = 800 \, MeV$, with $k^0$ and $k'^0$ as the ones above. 
It should be mentioned that in the non-diagonal transition matrix elements involving $D^*$ meson exchange, we introduce a reduction factor $\frac{m_V^2}{m_{D^*}^2}$  to approximately account for the exchange effect. Additionally, since the contribution of single-pion exchange to the potential is relatively small in the $S$-wave interactions, the pion exchange contribution is neglected in our formalism, and thus, $\mu_{24} = \mu_{34} = 0$.

Furthermore, in Eq.~\eqref{eq:BS}, the diagonal matrix $G$ are constructed by the meson-baryon loop functions. 
Note that, in the prediction works~\cite{Xiao:2013yca,Xiao:2019gjd}, the loop functions were taken the form of the dimensional regularization scheme~\cite{Oller:2000fj,Oller:1998zr}, see more discussions in Ref.~\cite{Xiao:2013yca}. 
In the present work, in order to better understand the behaviours between different bound systems, we explore the three-momentum cutoff method to the loop functions~\cite{Oller:1997ti}, where the analytical expression for the loop function $G_{ll}(s)$ of the $l$-th channel is given by~\cite{Guo:2005wp},
\begin{equation}
\begin{aligned}
G_{ll}(s) &= \frac{2M_l}{16\pi^2 s} \left\{ \sigma \left( \arctan \frac{s + \Delta}{\sigma \lambda_1} + \arctan \frac{s - \Delta}{\sigma \lambda_2} \right) \right. \\
&\left. - \left[ (s + \Delta)\ln \left( \frac{q_{maxl}}{M_l}(1 + \lambda_1) \right) + (s - \Delta)\ln \left( \frac{q_{maxl}}{m_l}(1 + \lambda_2) \right) \right] \right\} ,
\end{aligned}
\end{equation}
with the definitions 
\begin{equation}
\begin{aligned}
\sigma &= \sqrt{-\lambda(s, M_l^2, m_l^2)} = \sqrt{-[s - (M_l + m_l)^2][s - (M_l - m_l)^2]} , \\
\Delta &= M_l^2 - m_l^2, \quad \lambda_1 = \sqrt{1 + \frac{M_l^2}{q_{maxl}^2}}, \quad \lambda_2 = \sqrt{1 + \frac{m_l^2}{q_{maxl}^2}} ,
\end{aligned}
\end{equation}
and the cutoff parameter $q_{max}$ as free parameter, see the discussions later. Besides, $\lambda(a, b, c)$ is the usual K\"allen triangle function $\lambda(a, b, c)=a^{2}+b^{2}+c^{2}-2(a b+a c+b c)$. Using the cutoff scheme, it is also to be consistent with the evaluation of the wave functions of the resonances as discussed below. 

To search for the poles of the $T$-matrix corresponding to the resonances by looking for the zeros of the determinant $det\,[I - V \cdot G] = 0$ on the complex energy plane, the loop functions $G_{ll}(s)$ need to be analytically extrapolated from the first Riemann sheet to the second Riemann sheet~\cite{Oller:1997ti,Oset:1997it}. The pole of the resonance is obtained as $\sqrt{s_{\text{pole}}} = M_{\text{pole}} - i\Gamma_{\text{pole}}/2$, implying that the real part of the pole $M_{\text{pole}}$ corresponds to the mass of the resonance, and the imaginary part is one half of its decay width $\Gamma_{\text{pole}}$. 
With the analytical continuity condition, it was easy to obtain the relation~\cite{Oller:1997ti},
\begin{equation}
\begin{aligned}
G_{ll}^{(II)}(\sqrt{s} + i\epsilon) &= G_{ll}^{(I)}(\sqrt{s} + i\epsilon) - 2i \text{Im} G_{ll}^{(I)}(\sqrt{s} + i\epsilon) \\
&= G_{ll}^{(I)}(\sqrt{s} + i\epsilon) + 2M_l \frac{i}{4\pi} \frac{p_{cml}}{\sqrt{s}} ,
\end{aligned}
\end{equation}
with the three momentum in the center-of-mass frame
\begin{equation}
p_{cml} = \frac{\lambda^{1/2}(s, M_l^2, m_l^2)}{2\sqrt{s}} = \frac{\sqrt{[s-(M_l+m_l)^2][s-(M_l-m_l)^2]}}{2\sqrt{s}}. 
\end{equation}

To quantitatively characterize the coupling strengths between the poles and other channels, the scattering amplitude can be rewritten by a Laurent series expansion near the pole $s_{pole}$ on the complex energy plane~\cite{Guo:2006fu,Oller:2004xm},
\begin{equation}
T_{ij} = \frac{g_i g_j}{\sqrt{s} - \sqrt{s_{pole}}} + \gamma_0 + \gamma_1(s - s_{pole}) + \dots ,
\end{equation}
where $g_i$ and $g_j$ represent the effective coupling constants of the $i$-th and $j$-th channels, respectively, defined as
\begin{equation}
g_i g_j = \lim_{s \to s_{pole}} (\sqrt{s}-  \sqrt{s_{pole}}) T_{ij} .
\end{equation}
Using the Cauchy residue theorem, the square of the coupling constant $g_i^2$ can also be obtained by calculating the residues of the $T_{ij}$ around the pole $s = s_{pole}$ on the complex energy plane~\cite{Oller:1998zr,Ozpineci:2013zas}, given by
\begin{equation}
g_i^2 = \frac{1}{2\pi i} \oint T_{ii} ds .
\end{equation}

To investigate more properties of the resonances, we further study the wave function of the resonance at small distance to learn more about the sources of the resonance. As done in Ref.~\cite{Yamagata-Sekihara:2010kpd}, the wave function $\phi(\vec{r})$ is defined via a Fourier transform,
\begin{equation}
\phi(\vec{r}) = \int_{q_{max}} \frac{d^3\vec{p}}{(2\pi)^{3/2}} e^{i\vec{p}\cdot\vec{r}} \langle \vec{p} | \psi \rangle .
\end{equation}
After performing the angular integration over the momentum, the specific expression of the wave function is given by~\cite{Ozpineci:2013zas}
\begin{equation}
\phi(\vec{r}) = \frac{1}{(2\pi)^{3/2}} \frac{4\pi}{r} \frac{1}{C} \int_{q_{max}} p dp \sin(pr) \times \frac{\Theta(q_{max} - |\vec{p}|)}{E - \omega_1(\vec{p}) - \omega_2(\vec{p})} \frac{m_V^2}{\vec{p}^2 + m_V^2} , \label{eq:wavf}
\end{equation}
where $\omega_i = (\vec{q}^2 + m_i^2)^{1/2}$, $C$ is the normalization constant, and $E \equiv \sqrt{s_{pole}}$. 
Note that, in Eq.~\eqref{eq:wavf} we introduce an additional form factor $f(\vec{q}) = \frac{m_V^2}{\vec{p}^2 + m_V^2}$ to regulate the dynamical behaviour at short distances. One can take $m_V =  m_\rho$ to account for the light vector meson exchanges in the main bound systems, where the $P_c$ and $P_{cs}$ states appear. Even if this additional form factor is removed, the line shapes of the wave functions will not substantially change. 
As one can see the results later, the wave functions will go to zero after a few fm, which is in fact the confined size for a molecular state, coincided with the results of the radius defined below. 
With the wave functions obtained, one can evaluate the form factor $F(\vec{q}^2)$ of corresponding resonance. By the definition, the form factor can be calculated from the wave function~\cite{Yamagata-Sekihara:2010kpd}, 
\begin{equation}
\begin{aligned}
F(\vec{q}) &= \int d^3\vec{r} \phi(\vec{r})\phi^*(\vec{r})e^{-i\vec{q}\cdot\vec{r}} \\
&= \int d^3\vec{p} \times \frac{\theta(\Lambda - p)\theta(\Lambda - |\vec{p} - \vec{q}|) f(\vec{q}) f(\vec{p}-\vec{q})}{[E - \omega_1(p) - \omega_2(p)][E - \omega_1(\vec{p}-\vec{q}) - \omega_2(\vec{p}-\vec{q})]} , \label{eq:ffq}
\end{aligned}
\end{equation}
where $\Lambda$ is a cutoff parameter, taken $\Lambda = q_{\text{qmax}}$ as the one in the loop functions for the consistency. One should keep in mind that a normalization factor is introduced to keep $F(q=0) \equiv 1$ in Eq.~\eqref{eq:ffq}. 
In the limit of low momentum transfer ($|\vec{q}| \to 0$), the form factor can be expanded as a Taylor expansion,
\begin{equation}
F(\vec{q}^2) \approx F(0) - \frac{1}{6} \langle r^2 \rangle \vec{q}^2 + \dots ,
\end{equation}
with the normalization condition $F(0) \equiv 1$. 
Accordingly, the mean-square radius of the state can be extracted from the derivative of the form factor with respect to $\vec{q}^2$~\cite{Ahmed:2020kmp},
\begin{equation}
\langle r^2 \rangle = -6 \left[ \frac{dF(q)}{dq^2} \right]_{q^2=0} , \label{eq:msr1}
\end{equation}
where a soft step function should be chosen for the $\theta(\Lambda - p)$ and $\theta(\Lambda - |\vec{p} - \vec{q}|)$ functions to meet the form factor converge when $q^2 \to 0$. 
On the other hand, for the bound states, the mean-square radius can also be estimated using the derivative of the loop function $G$ with respect to energy and the binding energy $B_{E,i}$~\cite{Sekihara:2013wlq}, 
\begin{equation}
\langle r^2 \rangle_i = \frac{-g_i^2 \left[ \frac{dG_i(s)}{d\sqrt{s}} \right]_{\sqrt{s}=\sqrt{s_{pole}}}}{4\mu_i B_{E,i}} , \label{eq:msr2}
\end{equation}
where the binding energy is obtained with $B_{E,i} = m_i + M_i - M_B$, and the reduced mass $\mu_i = \frac{m_i M_i}{m_i + M_i}$, $g_i$ the coupling constant defined above. $\langle r^2 \rangle_i$ is the mean square distance of the bound state in the $i$-th channel. In this case, the mean-square radius also depends on the binding energy for a certain pole respected to the threshold. When the pole is very close to the threshold, Eq.~\eqref{eq:msr2} may lead to numerical instability due to the binding energy in the denominator becoming zero. In such cases, the result of Eq.~\eqref{eq:msr1} are more stable, see our results later. 

\section{Results}

As discussed in last section, to better investigate the properties of the bound systems, we take the three momentum cutoff method to regularize the loop functions, and then obtain the poles of the $P_c$ and $P_{cs}$ states in coupled-channel interactions. 
Subsequently, we further investigate the wave functions and mean-square radii of these states to discuss their internal properties in more detail. 
To obtain more dynamical information regarding with these poles appeared in the coupled channel interactions, we split the full coupled-channel systems into the pseudoscalar meson-baryon (PB) and vector meson-baryon (VB) subsystems without the HQSS constraint, and also compare the results with those of the single-channel interactions. 
Note that in the present work, the only free parameter is the cutoff $q_{max}$. In order to check the properties of different bound systems, we vary the values of $q_{max}$ , which also indicate the uncertainties of the results obtained. 

\subsection{Results of coupled-channel interactions in hidden charm sector}

\begin{figure}[htbp]
\centering
\includegraphics[width=0.8\textwidth]{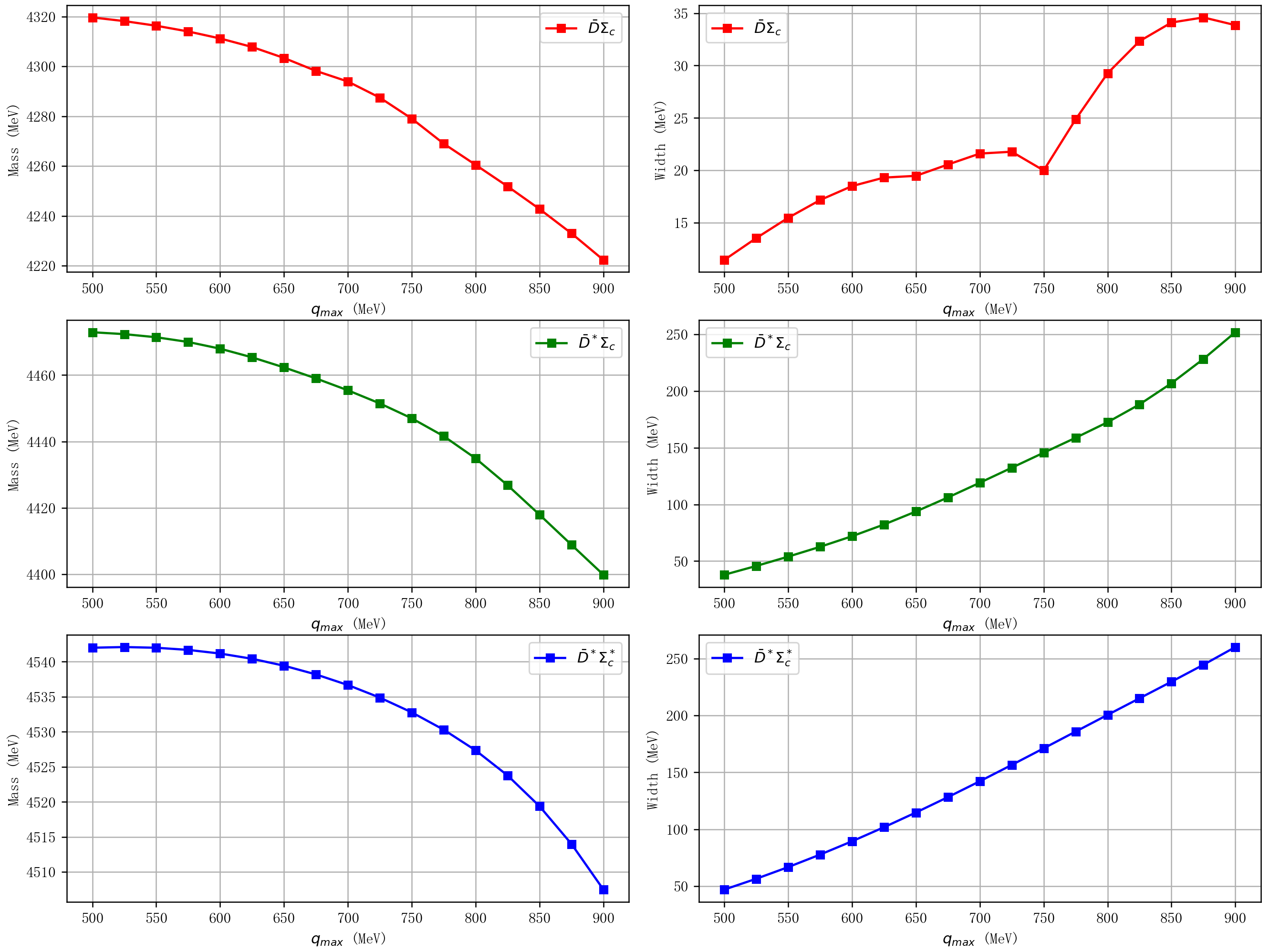}
\caption{Mass (left) and width (right) trajectories of the poles in the second Riemann sheets for the $I = 1/2,\, J^P = 1/2^-$ sector as a function of the cutoff $q_{max}$ in the seven coupled-channel case.}
\label{fig:I1J1c7}
\end{figure}

We first investigate the hidden charm system binding the $P_c$ states. For the isospin $I = 1/2$ and spin-parity $J^P = 1/2^-$ sector, see Table~\ref{tab:I1J1c7} for all seven coupled channels under the HQSS constraint, we calculated the pole positions on the second Riemann sheets by varying the cutoff $q_{max}$ from $500$ to $900$ MeV. 
The obtained masses and widths trajectories of the poles are shown in Fig.~\ref{fig:I1J1c7}, and some parts of the results with $q_{max} = 600, \, 700, \, 800\, \text{MeV}$ are listed in Table~\ref{tab:resI1J1c7}. 
These poles primarily couple to the channels $\bar{D}\Sigma_c$, $\bar{D}^*\Sigma_c$, and $\bar{D}^*\Sigma_c^*$, respectively. 

The results of Fig.~\ref{fig:I1J1c7} indicate that the masses of the three main poles show a monotonic downward trend as the cutoff $q_{max}$ increases, while their corresponding widths go up nearly linearly when the $q_{max}$ increases. 
But, for the $q_{max}$ varying range from $500$ to $900$ MeV, the changing of their masses and widths are quite different, where the mass of the first pole can be bound largely and the widths of last two poles increase strongly. 
For the pole of the $\bar{D}\Sigma_c$ channel, the mass drops down from 4320 MeV to 4220 MeV nearly having 100 MeV differences, while the width increases from about 11 MeV to 34 MeV with 23 MeV differences, where there is a fluctuation near $q_{max} = 750 \, \text{MeV}$ due to the pole crossed the open channel threshold of the $D^* \Lambda_c$ to affect the decay properties. 
The mass of the second pole, mainly bound by the $\bar{D}^*\Sigma_c$ channel, decreases from about 4472 MeV to 4400 MeV with 72 MeV dropping, whereas, the width changes about 210 MeV from 40 MeV to 250 MeV.
For the third one, mainly bound by the $\bar{D}^*\Sigma_c^*$ channel, the mass reduces about 34 MeV from 4542 MeV to 4508 MeV and the width enhances about 200 MeV from 50 MeV to 250 MeV, which is similar to the one of the second pole. 
These results show that the interaction of the $\bar{D}\Sigma_c$ channel is strong and can be bound strongly compared to the other two channels, $\bar{D}^*\Sigma_c$ and $\bar{D}^*\Sigma_c^*$, of which the poles have moved to above their thresholds for lower values of the cutoff $q_{max}$. 
Therefore, from these results of Fig.~\ref{fig:I1J1c7}, one also can easily find that it is difficult to get a ``good" value for the cutoff $q_{max}$ to obtain three main poles matching the masses and widths of the three $P_c$ states well as the results obtained in Ref.~\cite{Xiao:2019aya} with one value of $a_\mu$ under the dimensional regularization scheme, as indicated in the results of the hidden charm and strange sector~\cite{Feijoo:2022rxf}. Indeed, as shown in Table~\ref{tab:resI1J1c7}, they match to the experimental findings of the $P_c$ states should be taken different value of the cutoff $q_{max}$ with different uncertainties. 

\begin{table}[ht]
\centering
\caption{Pole positions $(M, \Gamma)$ in the second Riemann sheets for the $I = 1/2,\, J^P = 1/2^-$ sector with seven coupled channels.} 
\label{tab:resI1J1c7}
\renewcommand{\arraystretch}{0.7}
\begin{tabular}{cccccc}
\hline\hline
$q_{max}$ & Mass & Width & Main & & Experimental \\
{[MeV]} & {[MeV]} & {[MeV]} & channel & $J^P$ & states \\ \hline
600 & 4311.22 & 18.50 & $\bar{D}\Sigma_c$ & $1/2^-$ & $P_c(4312)$ \\
700 & 4293.93 & 21.60 & $\bar{D}\Sigma_c$ & $1/2^-$ &  \\
800 & 4260.47 & 29.26 & $\bar{D}\Sigma_c$ & $1/2^-$ &  \\ \hline
600 & 4467.93 & 71.95 & $\bar{D}^*\Sigma_c$ & $1/2^-$ &  \\
700 & 4455.39 & 119.08 & $\bar{D}^*\Sigma_c$ & $1/2^-$ & $P_c(4440)$  \\
800 & 4434.89 & 172.42 & $\bar{D}^*\Sigma_c$ & $1/2^-$ &  \\ \hline
600 & 4541.17 & 89.41 & $\bar{D}^*\Sigma_c^*$ & $1/2^-$ & -- \\
700 & 4536.68 & 142.14 & $\bar{D}^*\Sigma_c^*$ & $1/2^-$ & -- \\
800 & 4527.35 & 200.45 & $\bar{D}^*\Sigma_c^*$ & $1/2^-$ & -- \\ \hline\hline
\end{tabular}
\end{table}
\begin{figure}[htbp]
\centering
\includegraphics[width=0.8\textwidth]{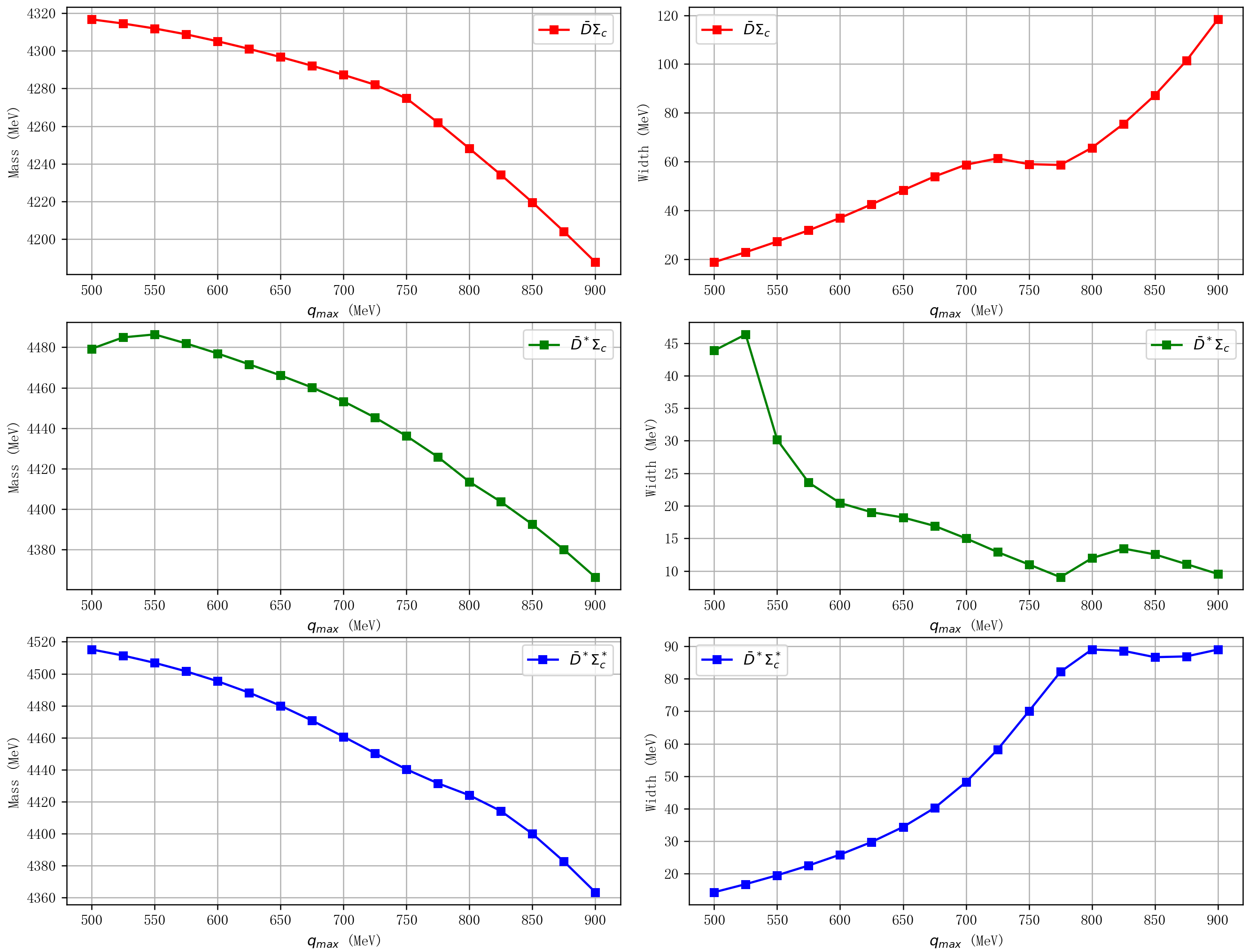}
\caption{Mass (left) and width (right) trajectories of the poles in the second Riemann sheets for the $I = 1/2,\, J^P = 1/2^-$ sector as a function of the cutoff $q_{max}$ in the splitting PB and VB sectors.}
\label{fig:I1J1c7PV}
\end{figure}
\begin{table}[ht]
\centering
\caption{Pole positions $(M, \Gamma)$ in the second Riemann sheets for the $I = 1/2,\, J^P = 1/2^-$ sector with the splitting PB and VB sectors.}
\label{tab:resI1J1c7PV}
\renewcommand{\arraystretch}{0.7}
\begin{tabular}{ccccccc}
\hline\hline
$q_{max}$ & Mass & Width & Main & & & Experimental \\
{[MeV]} & {[MeV]} & {[MeV]} & channel & Sector & $J^P$ & states \\ \hline
600 & 4305.07 & 36.92 & $\bar{D}\Sigma_c$ & PB & $1/2^-$ & $P_c(4312)$ \\
700 & 4287.22 & 58.75 & $\bar{D}\Sigma_c$ & PB & $1/2^-$ &  \\
800 & 4248.12 & 65.69 & $\bar{D}\Sigma_c$ & PB & $1/2^-$ &  \\ \hline
600 & 4476.97 & 20.44 & $\bar{D}^*\Sigma_c$ & VB & $1/2^-$ &  \\
700 & 4453.30 & 15.00 & $\bar{D}^*\Sigma_c$ & VB & $1/2^-$ & $P_c(4440)$ \\
800 & 4413.54 & 11.96 & $\bar{D}^*\Sigma_c$ & VB & $1/2^-$ &  \\ \hline
600 & 4495.32 & 25.88 & $\bar{D}^*\Sigma_c^*$ & VB & $1/2^-$ & -- \\
700 & 4460.66 & 48.19 & $\bar{D}^*\Sigma_c^*$ & VB & $1/2^-$ & -- \\
800 & 4424.03 & 89.01 & $\bar{D}^*\Sigma_c^*$ & VB & $1/2^-$ & -- \\ \hline\hline
\end{tabular}
\end{table}

To check more details of these results, we split the coupled channel system into the PB and VB subsystems to see the coupled channel effect without the HQSS constraint, where the PB subsystem has three coupled channels, $\eta_c N$, $\bar{D} \Lambda_c$ and $\bar{D} \Sigma_c$, and there are four coupled channels in the VB subsystem, $J/\psi N$, $\bar{D}^* \Lambda_c$ $\bar{D}^* \Sigma_c$, and $\bar{D}^* \Sigma_c^*$. 
The results are shown in Fig.~\ref{fig:I1J1c7PV} and Table~\ref{tab:resI1J1c7PV}. 
From Fig.~\ref{fig:I1J1c7PV}, compared with Fig.~\ref{fig:I1J1c7}, it looks like that there is not much difference in the masses, where the line shape for the widths has been changed a lot. 
Now the width of the first pole mainly from the $\bar{D}\Sigma_c$ channel increases a lot from 20 MeV to 120 MeV. 
Whereas, the one for the second pole, contributed from the $\bar{D}^*\Sigma_c$ channel, becomes more reasonable values but decreases from 45 MeV to 10 MeV with some fluctuations, which is more consistent with the narrow width of the experimentally observed $P_c(4440)$ state. 
For the third pole of the $\bar{D}^* \Sigma_c^*$ channel, the width enhances with about 72 MeV just a little smaller than the one in the full coupled channel case before, but the mass is always below the threshold. 
From these results, one can find that the coupled channel effect under the HQSS constraint just affects the widths of these poles when varying the cutoff $q_{max}$, and does not change the bound properties of the strong interactions among the coupled channels. 

\begin{figure}[htbp]
\centering
\includegraphics[width=0.8\textwidth]{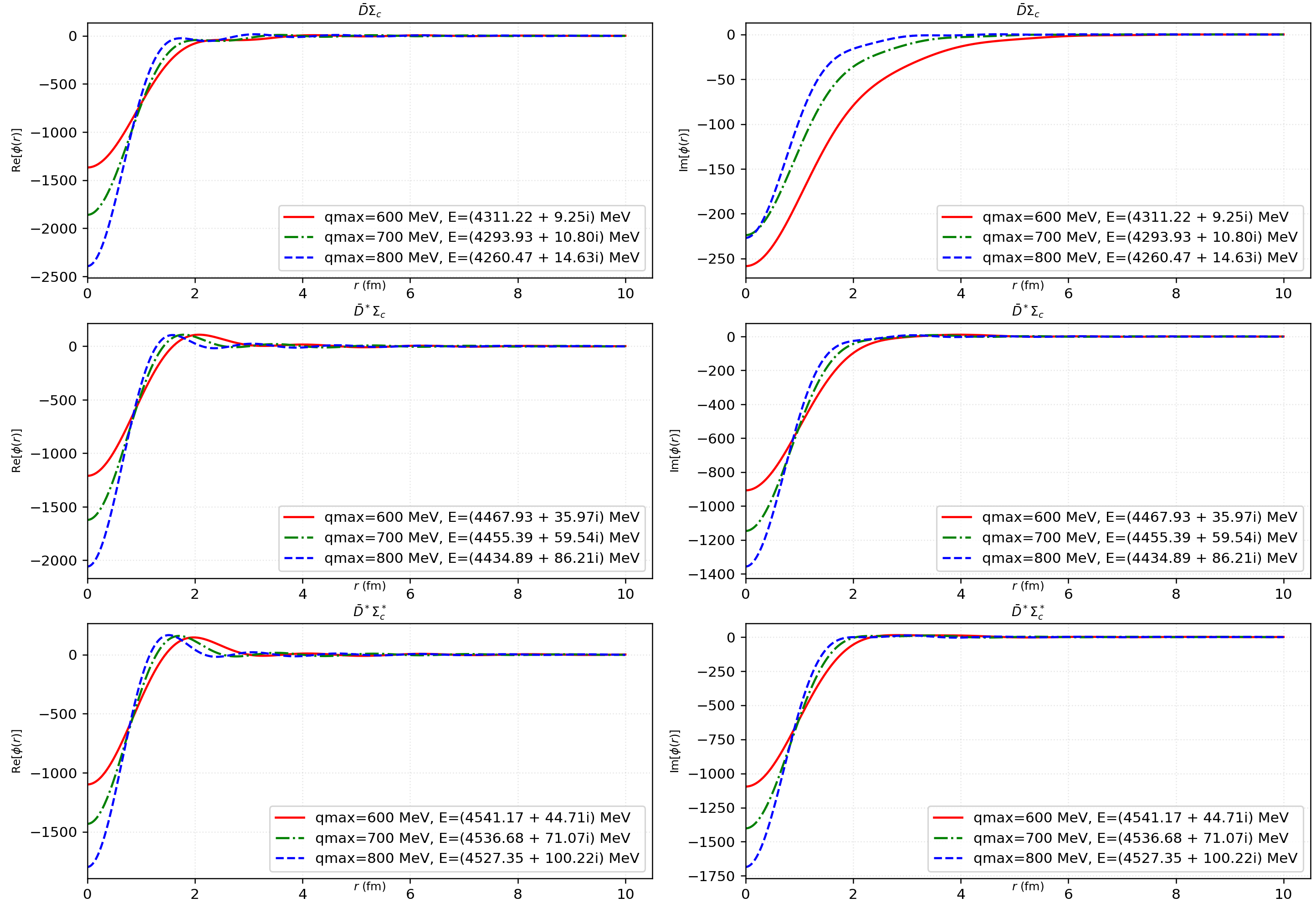}
\caption{Real (left) and imaginary (right) parts of the wave functions $\phi(r)$ of corresponding pole for the $I = 1/2,\, J^P = 1/2^-$ sector with different $q_{max}$ in the seven coupled-channel case.}
\label{fig:wfI1J1c7}
\end{figure}

\begin{figure}[htbp]
\centering
\includegraphics[width=0.8\textwidth]{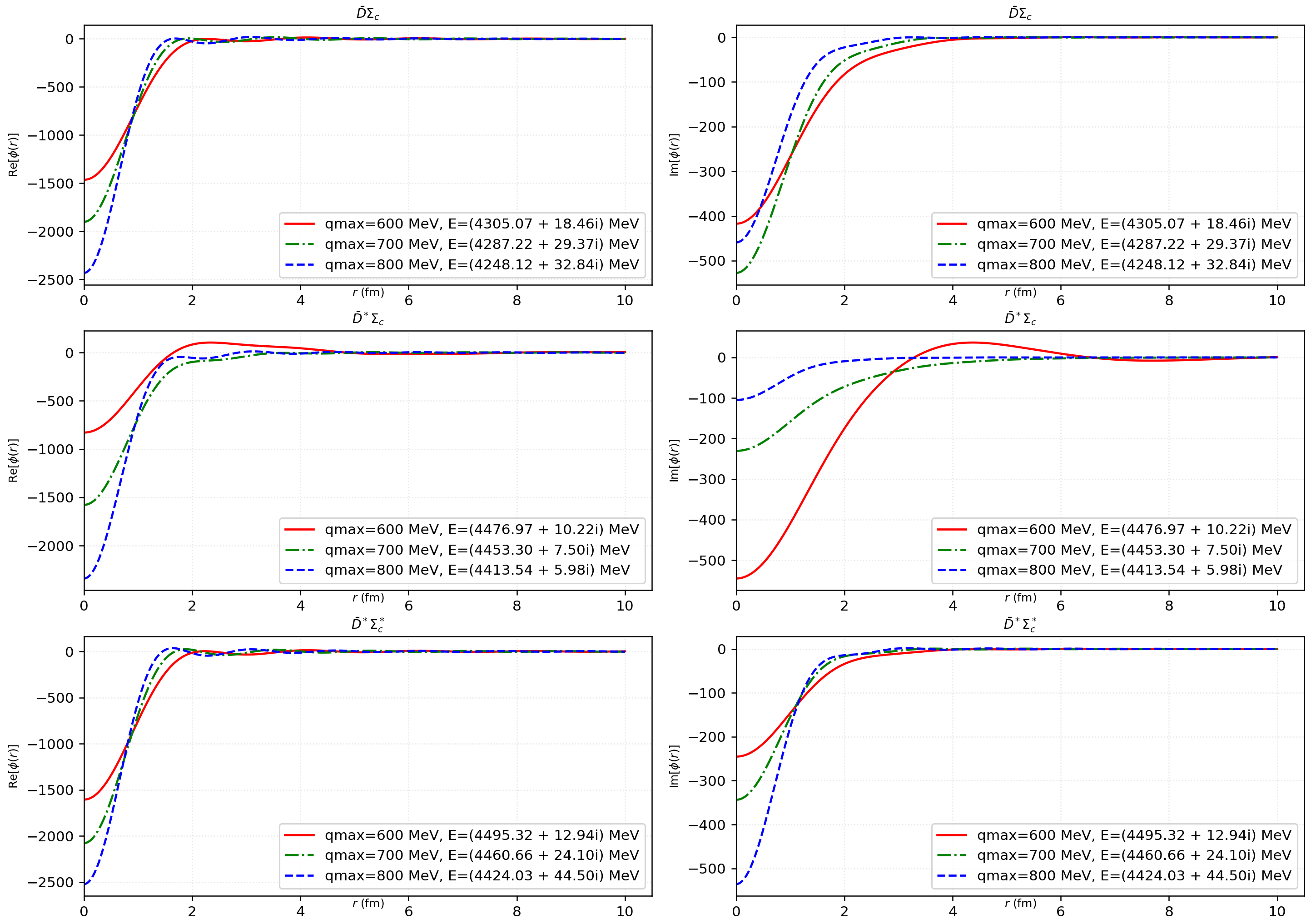}
\caption{Real (left) and imaginary (right) parts of the wave functions $\phi(r)$ of corresponding pole for the $I = 1/2,\, J^P = 1/2^-$ sector with different $q_{max}$ in the splitting PB and VB sectors.}
\label{fig:wfI1J1c7PV}
\end{figure}

As discussed in the last section, to reveal the molecular properties of the resonances, we continue to investigate the wave functions of these resonances at small distance. The results are shown in Figs.~\ref{fig:wfI1J1c7} and \ref{fig:wfI1J1c7PV}, for the seven coupled-channel case and the splitting PB and VB sectors, respectively. 
As the cutoff $q_{max}$ increases from $600 \, \text{MeV}$ to $800 \, \text{MeV}$, the real parts of the wave functions at the origin $r = 0$ increase significantly. 
From these results of Figs.~\ref{fig:wfI1J1c7} and \ref{fig:wfI1J1c7PV}, one can see that the wave functions nearly close to zero when $r > 4 \, \text{fm}$, at a reasonable size of the hadronic molecules, except for the second pole in the splitting PB and VB sectors, of which its imaginary parts can up to about 7 fm before going to zero.

\begin{figure}[htbp]
\centering
\includegraphics[width=0.4\textwidth,height=0.5\textheight,keepaspectratio]{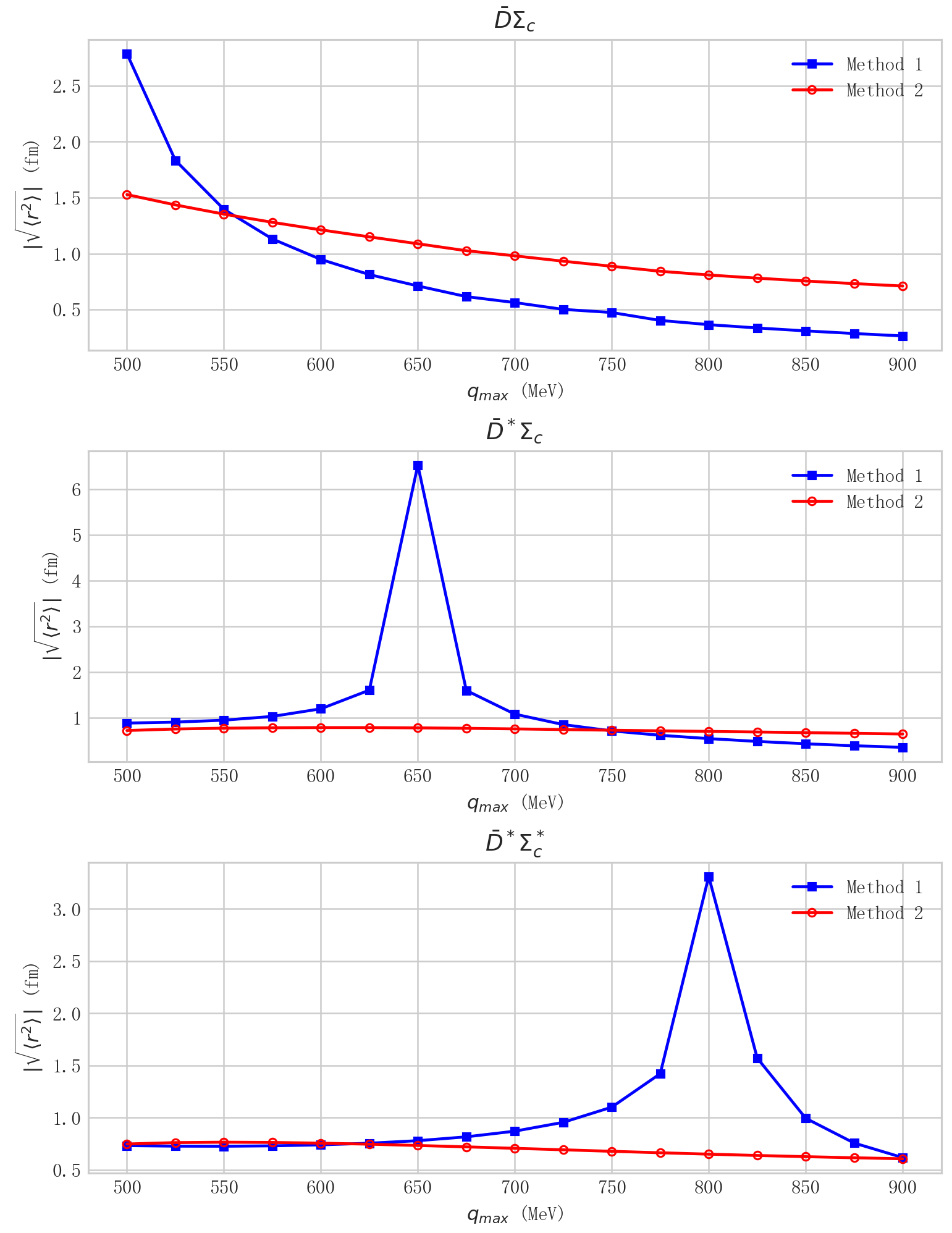}
\includegraphics[width=0.4\textwidth,height=0.5\textheight,keepaspectratio]{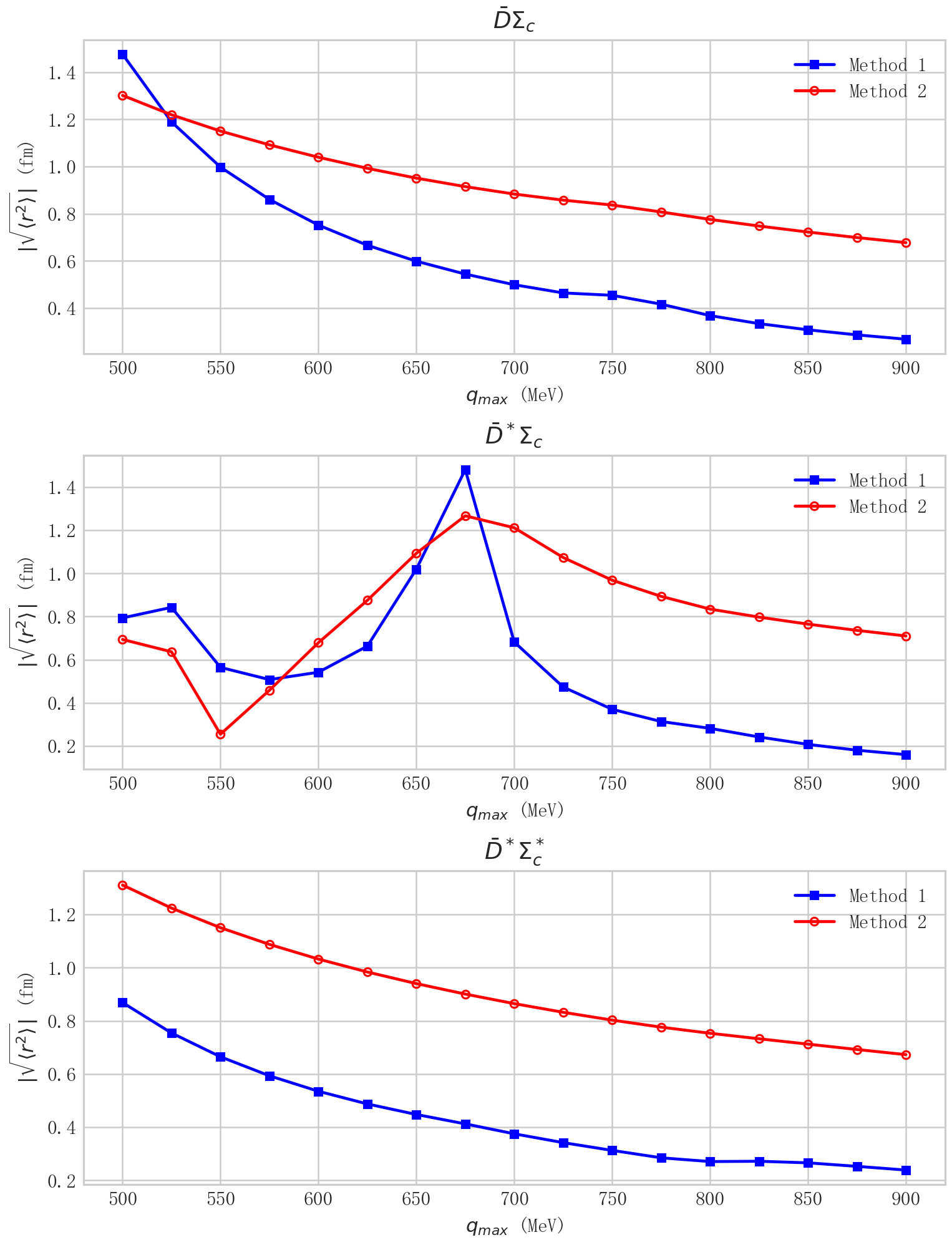}
\caption{RMS radii of the corresponding poles for the $I = 1/2,\, J^P = 1/2^-$ sector as a function of the cutoff $q_{max}$ in the seven coupled-channel case (left) and the splitting PB and VB sectors (right). Results from Method 1 (blue) and Method 2 (red) are compared.}
\label{fig:rmsI1J1c7PV}
\end{figure}

Next, using the wave functions obtained, we calculated the radii of these poles. As discussed in the last section, there are two ways to evaluate the radii. One way utilizes the derivative of the $G$ function with respect to the binding energy $B_{E,i}$, see Eq.~\eqref{eq:msr2}, named as ``Method 1". The other one explores the form factor evaluated from the wave functions, see Eq.~\eqref{eq:msr1}, labeled as ``Method 2". 
The results of the root-mean-square (RMS) radii obtained by varying the cutoffs are shown in Fig.~\ref{fig:rmsI1J1c7PV}, some of which are also listed in Tables~\ref{tab:rmsI1J1c7} and~\ref{tab:rmsI1J1c7PV} for the seven coupled-channel case and the splitting PB and VB sectors, respectively. 
From these results, one can see that the results with two methods are consistent with each other except for the situation of the pole closing to the threshold due to the binding energy $B_{E,i}$ going to zero. 
Indeed, the results of Method 2 are more numerical stability, with the cutoff $q_{max}$ smoothly changing. 
It is also found that the radii in most of the cases are less than 3 fm, which are consistent with the one estimated from the wave functions above. 
\begin{table}[ht]
\centering
\caption{RMS radii $\left| \sqrt{\langle r^2 \rangle} \right|$ of the corresponding poles for the $I = 1/2,\, J^P = 1/2^-$ sector using different methods in the seven coupled-channel case} \label{tab:rmsI1J1c7}
\begin{flushleft}
The radii of states calculated with Eq.~\eqref{eq:msr1}
\end{flushleft}
\renewcommand{\arraystretch}{0.7} 
\setlength{\tabcolsep}{2pt}
\begin{tabular}{ccccccc}
\hline\hline
Resonances & $q_{max} = 600 \, \text{MeV}$ & $\left| \sqrt{\langle r^2 \rangle} \right|_2$ & $q_{max} = 700 \, \text{MeV}$ & $\left| \sqrt{\langle r^2 \rangle} \right|_2$ & $q_{max} = 800 \, \text{MeV}$ & $\left| \sqrt{\langle r^2 \rangle} \right|_2$ \\ \hline
$\bar{D}\Sigma_c$ & $1.187 + 0.248i \text{ fm}$ & $1.212 \text{ fm}$ & $0.977 + 0.075i \text{ fm}$ & $0.980 \text{ fm}$ & $0.809 + 0.027i \text{ fm}$ & $0.809 \text{ fm}$ \\
$\bar{D}^*\Sigma_c$ & $0.774 + 0.095i \text{ fm}$ & $0.780 \text{ fm}$ & $0.748 + 0.070i \text{ fm}$ & $0.751 \text{ fm}$ & $0.693 + 0.057i \text{ fm}$ & $0.696 \text{ fm}$ \\
$\bar{D}^*\Sigma_c^*$ & $0.756 - 0.012i \text{ fm}$ & $0.756 \text{ fm}$ & $0.706 + 0.015i \text{ fm}$ & $0.706 \text{ fm}$ & $0.649 + 0.033i \text{ fm}$ & $0.650 \text{ fm}$ \\\hline\hline
\end{tabular}
\begin{flushleft}
The radii of states calculated with Eq.~\eqref{eq:msr2}
\end{flushleft}
\renewcommand{\arraystretch}{0.7} 
\setlength{\tabcolsep}{2pt}
\begin{tabular}{ccccccc}
\hline\hline
Resonances & $q_{max} = 600 \, \text{MeV}$ & $\left| \sqrt{\langle r^2 \rangle} \right|_1$ & $q_{max} = 700 \, \text{MeV}$ & $\left| \sqrt{\langle r^2 \rangle} \right|_1$ & $q_{max} = 800 \, \text{MeV}$ & $\left| \sqrt{\langle r^2 \rangle} \right|_1$ \\ \hline
$\bar{D}\Sigma_c$ & $0.948 - 0.003i \text{ fm}$ & $0.948 \text{ fm}$ & $0.562 - 0.027i \text{ fm}$ & $0.562 \text{ fm}$ & $0.364 - 0.035i \text{ fm}$ & $0.365 \text{ fm}$ \\
$\bar{D}^*\Sigma_c$ & $0.005 + 1.191i \text{ fm}$ & $1.191 \text{ fm}$ & $1.074 - 0.034i \text{ fm}$ & $1.075 \text{ fm}$ & $0.538 - 0.016i \text{ fm}$ & $0.538 \text{ fm}$ \\
$\bar{D}^*\Sigma_c^*$ & $0.008 + 0.739i \text{ fm}$ & $0.739 \text{ fm}$ & $0.012 + 0.871i \text{ fm}$ & $0.871 \text{ fm}$ & $0.028 + 3.311i \text{ fm}$ & $3.311 \text{ fm}$ \\\hline\hline
\end{tabular}
\end{table}
\begin{table}[ht]
\centering
\caption{RMS radii $\left| \sqrt{\langle r^2 \rangle} \right|$ of the corresponding poles for the $I = 1/2,\, J^P = 1/2^-$ sector using different methods  in the splitting PB and VB sectors.}\label{tab:rmsI1J1c7PV}
\begin{flushleft}
The radii of states calculated with Eq.~\eqref{eq:msr1}
\end{flushleft}
\renewcommand{\arraystretch}{0.7} 
\setlength{\tabcolsep}{2pt}
\begin{tabular}{ccccccc}
\hline\hline
Resonances & $q_{max} = 600 \, \text{MeV}$ & $\left| \sqrt{\langle r^2 \rangle} \right|_2$ & $q_{max} = 700 \, \text{MeV}$ & $\left| \sqrt{\langle r^2 \rangle} \right|_2$ & $q_{max} = 800 \, \text{MeV}$ & $\left| \sqrt{\langle r^2 \rangle} \right|_2$ \\ \hline
$\bar{D}\Sigma_c$ & $1.027 + 0.158i \text{ fm}$ & $1.039 \text{ fm}$ & $0.879 + 0.087i \text{ fm}$ & $0.883 \text{ fm}$ & $0.775 + 0.035i \text{ fm}$ & $0.776 \text{ fm}$ \\
$\bar{D}^*\Sigma_c$ & $0.429 + 0.526i \text{ fm}$ & $0.678 \text{ fm}$ & $1.181 + 0.269i \text{ fm}$ & $1.212 \text{ fm}$ & $0.834 + 0.017i \text{ fm}$ & $0.834 \text{ fm}$ \\
$\bar{D}^*\Sigma_c^*$ & $1.031 + 0.046i \text{ fm}$ & $1.032 \text{ fm}$ & $0.864 + 0.021i \text{ fm}$ & $0.865 \text{ fm}$ & $0.753 + 0.018i \text{ fm}$ & $0.753 \text{ fm}$ \\ \hline\hline
\end{tabular}
\begin{flushleft}
The radii of states calculated with Eq.~\eqref{eq:msr2}
\end{flushleft}
\renewcommand{\arraystretch}{0.7} 
\setlength{\tabcolsep}{2pt}
\begin{tabular}{ccccccc}
\hline\hline
Resonances & $q_{max} = 600 \, \text{MeV}$ & $\left| \sqrt{\langle r^2 \rangle} \right|_1$ & $q_{max} = 700 \, \text{MeV}$ & $\left| \sqrt{\langle r^2 \rangle} \right|_1$ & $q_{max} = 800 \, \text{MeV}$ & $\left| \sqrt{\langle r^2 \rangle} \right|_1$ \\ \hline
$\bar{D}\Sigma_c$ & $0.752 - 0.005i \text{ fm}$ & $0.752 \text{ fm}$ & $0.499 + 0.005i \text{ fm}$ & $0.499 \text{ fm}$ & $0.368 - 0.010i \text{ fm}$ & $0.368 \text{ fm}$ \\
$\bar{D}^*\Sigma_c$ & $0.192 + 0.506i \text{ fm}$ & $0.541 \text{ fm}$ & $0.658 - 0.177i \text{ fm}$ & $0.682 \text{ fm}$ & $0.271 - 0.076i \text{ fm}$ & $0.281 \text{ fm}$ \\
$\bar{D}^*\Sigma_c^*$ & $0.535 + 0.009i \text{ fm}$ & $0.535 \text{ fm}$ & $0.375 - 0.015i \text{ fm}$ & $0.375 \text{ fm}$ & $0.270 + 0.005i \text{ fm}$ & $0.271 \text{ fm}$ \\ \hline\hline
\end{tabular}
\end{table}
\vspace{0.5cm}

\begin{figure}[htbp]
\centering
\includegraphics[width=0.8\textwidth]{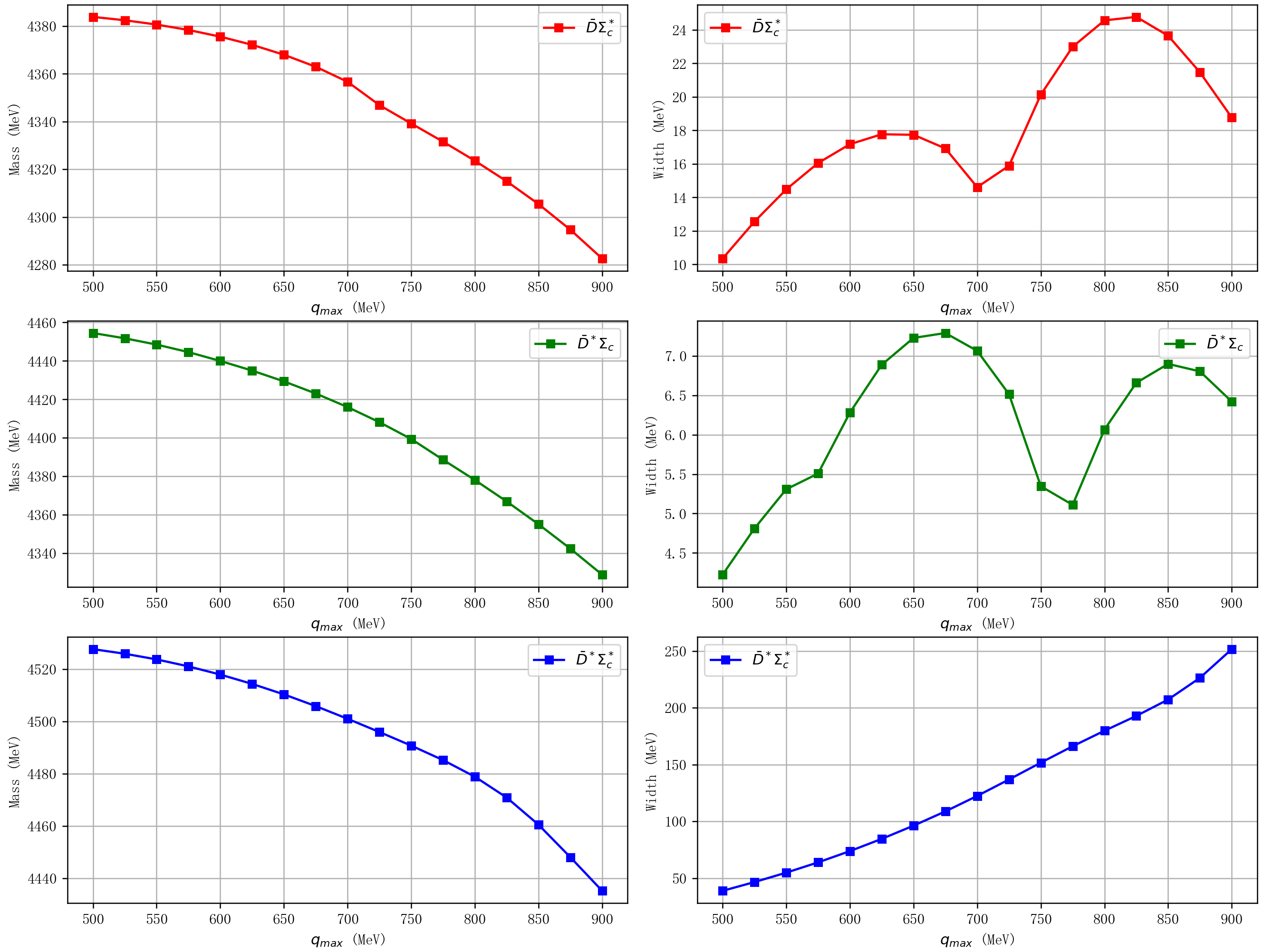}
\caption{Mass (left) and width (right) trajectories of the poles in the second Riemann sheets for the $I = 1/2,\, J^P = 3/2^-$ sector as a function of the cutoff $q_{max}$ in the five coupled-channel case.}
\label{fig:resI1J3c5}
\end{figure}

Furthermore, we examine the properties of the $P_c$ states in the $I = 1/2,\, J^P = 3/2^-$ sector. By varying the cutoff value $q_{max}$, the trajectories of the pole positions for the main channels $\bar{D}\Sigma_c^*$, $\bar{D}^*\Sigma_c$, and $\bar{D}^*\Sigma_c^*$ under the five coupled-channel case are calculated, as shown in Fig.~\ref{fig:resI1J3c5}, where some part results are presented in Table~\ref{tab:resI1J3c5}. The masses of these three poles exhibit a monotonic downward trend with the increasing of the cutoff $q_{max}$, which are all below the corresponding thresholds. Unlike the monotonic behaviours in the masses, the decay widths show the non-monotonic fluctuations when varying with $q_{max}$ for the first two poles. The fluctuations of the widths are due to the poles crossed the threshold of certain open channel. These results are also indicated different bound behaviours for three main channels. 
\begin{figure}[htbp]
\centering
\includegraphics[width=0.8\textwidth]{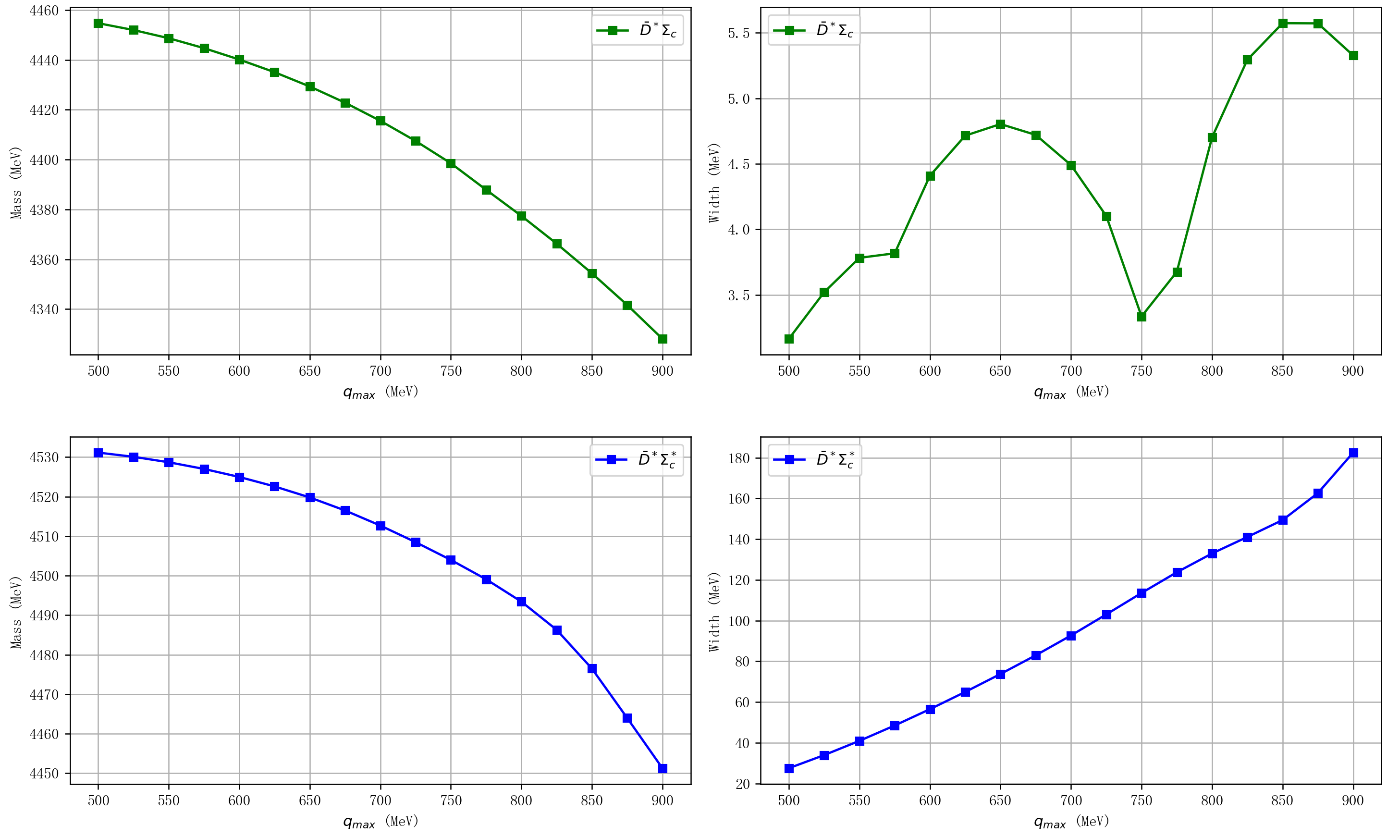}
\caption{Mass (left) and width (right) trajectories of the poles in the second Riemann sheets for the $I = 1/2,\, J^P = 3/2^-$ sector as a function of the cutoff $q_{max}$ in the splitting PB and VB sectors.}
\label{fig:resI1J3c5PV}
\end{figure}
\begin{table}[ht] 
\centering
\caption{Pole positions $(M, \Gamma)$ in the second Riemann sheets for the $I = 1/2,\, J^P = 3/2^-$ sector with different $q_{max}$ in the five coupled-channel case.}\label{tab:resI1J3c5}
\renewcommand{\arraystretch}{0.7}
\begin{tabular}{cccccc}
\hline\hline
$q_{max}$ & Mass & Width & Main & & Experimental \\
{[MeV]} & {[MeV]} & {[MeV]} & channel & $J^P$ & state \\ \hline
600 & 4375.58 & 17.18 & $\bar{D}\Sigma_c^*$ & $3/2^-$ & $P_c(4380)$ \\
700 & 4356.65 & 14.61 & $\bar{D}\Sigma_c^*$ & $3/2^-$ &  \\
800 & 4323.58 & 24.57 & $\bar{D}\Sigma_c^*$ & $3/2^-$ &  \\ \hline
600 & 4439.97 & 6.28 & $\bar{D}^*\Sigma_c$ & $3/2^-$ & $P_c(4457)$ \\
700 & 4416.00 & 7.07 & $\bar{D}^*\Sigma_c$ & $3/2^-$ &  \\
800 & 4378.06 & 6.07 & $\bar{D}^*\Sigma_c$ & $3/2^-$ &  \\ \hline
600 & 4517.93 & 73.76 & $\bar{D}^*\Sigma_c^*$ & $3/2^-$ & -- \\
700 & 4501.03 & 122.34 & $\bar{D}^*\Sigma_c^*$ & $3/2^-$ & -- \\
800 & 4478.88 & 179.73 & $\bar{D}^*\Sigma_c^*$ & $3/2^-$ & -- \\ \hline\hline
\end{tabular}
\end{table}
\begin{table}[ht]
\centering
\caption{Pole positions $(M, \Gamma)$ in the second Riemann sheets for the $I = 1/2,\, J^P = 3/2^-$ sector with different $q_{max}$ in the splitting VB sector.}\label{tab:resI1J3c5PV}
\renewcommand{\arraystretch}{0.7}
\begin{tabular}{ccccccc}
\hline\hline
$q_{max}$ & Mass & Width & Main & & & Experimental \\
{[MeV]} & {[MeV]} & {[MeV]} & channel & $J^P$ & Sector & state \\ \hline
600 & 4440.14 & 4.41 & $\bar{D}^*\Sigma_c$ & $3/2^-$ & VB & $P_c(4457)$ \\
700 & 4415.56 & 4.49 & $\bar{D}^*\Sigma_c$ & $3/2^-$ & VB &  \\
800 & 4377.40 & 4.70 & $\bar{D}^*\Sigma_c$ & $3/2^-$ & VB &  \\ \hline
600 & 4524.98 & 56.54 & $\bar{D}^*\Sigma_c^*$ & $3/2^-$ & VB & -- \\
700 & 4512.69 & 92.73 & $\bar{D}^*\Sigma_c^*$ & $3/2^-$ & VB & -- \\
800 & 4493.45 & 133.01 & $\bar{D}^*\Sigma_c^*$ & $3/2^-$ & VB & -- \\ \hline\hline
\end{tabular}
\end{table}

As done in the $I = 1/2,\, J^P = 1/2^-$ sector, we also split the five coupled channels into two PB and VB subsystems to check the coupled channel effect without the HQSS constraint, where the results are shown in Fig.~\ref{fig:resI1J3c5PV} and some part results are presented in Table~\ref{tab:resI1J3c5PV}. Note that in this case only one PB channel is removed from the five coupled channels. Thus, after retaining the VB coupled channels, once again the masses of the later two poles are not much changes, and their widths are significantly reduced. 
This indicates again that the coupled channels between the PB and VB sectors under the HQSS constraint only affect the decay widths of the poles and not much to their masses.
The width of the $\bar{D}^*\Sigma_c$ pole in full coupled channel case is more consistent with the narrow resonance of $P_c(4457)$ observed experimentally.

\begin{figure}[htbp]
\centering
\includegraphics[width=0.8\textwidth]{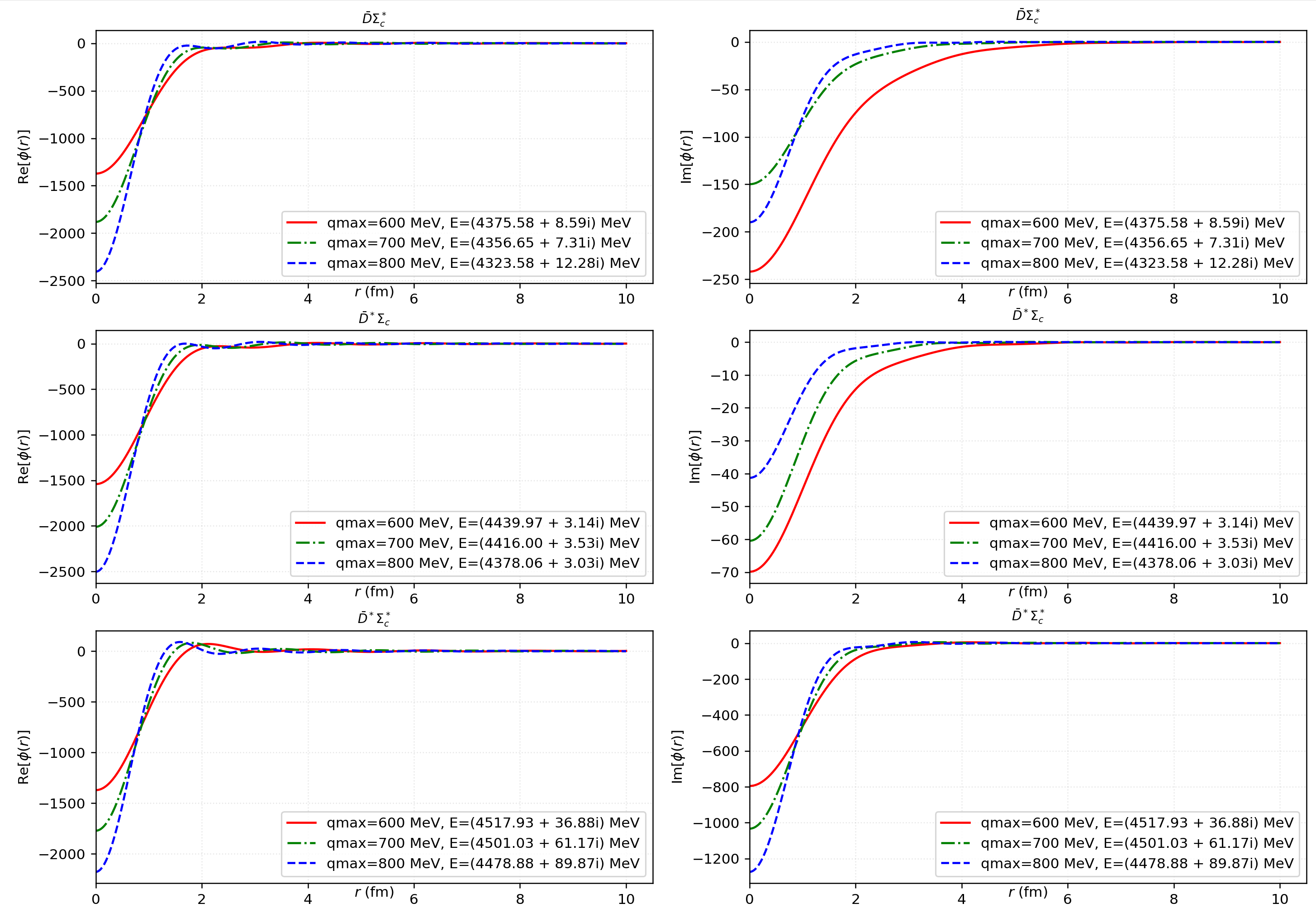}
\vspace{10pt}
\noindent\rule{0.6\linewidth}{0.5pt}
\vspace{10pt}
\includegraphics[width=0.8\textwidth]{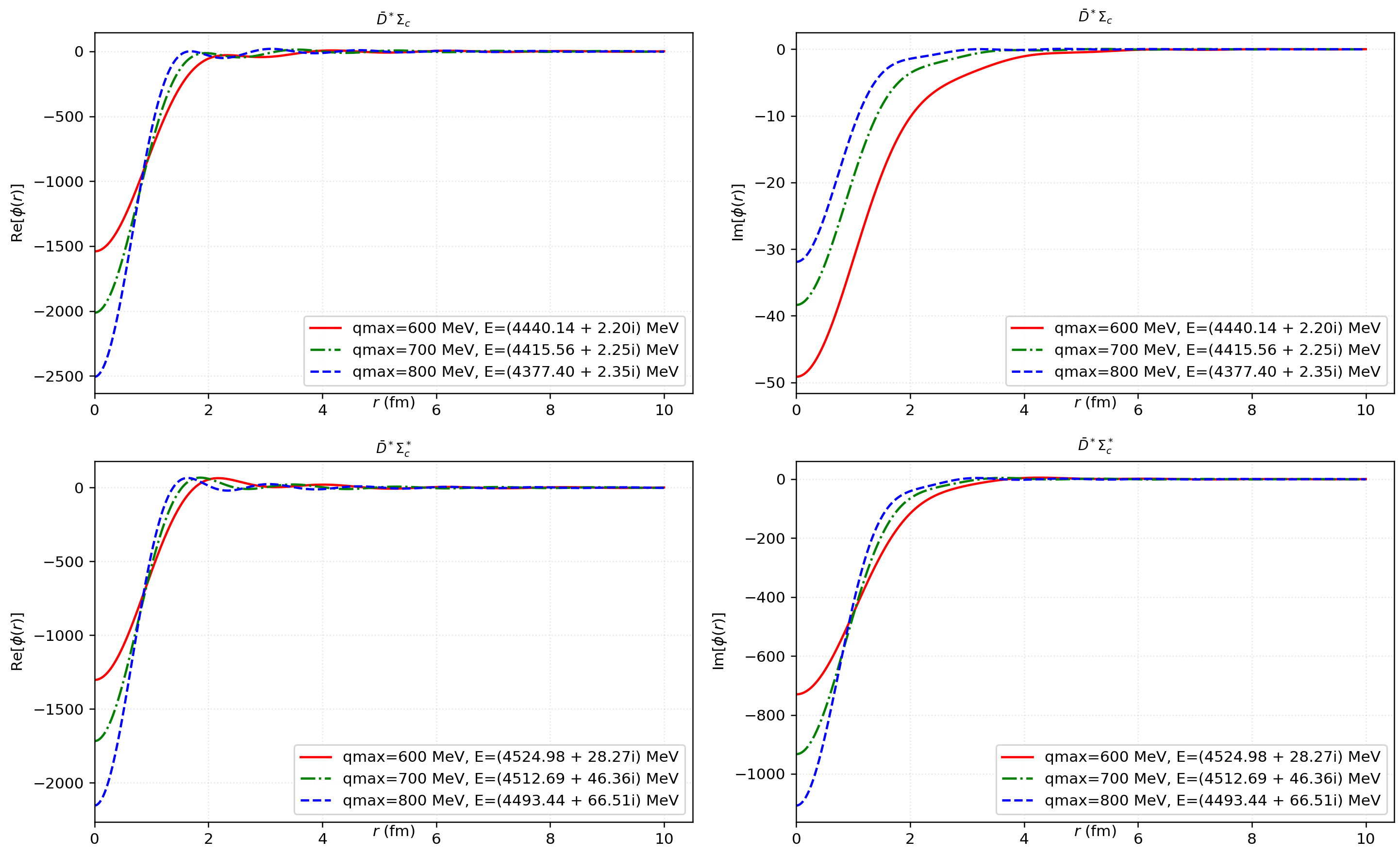}
\caption{Real (left) and imaginary (right) parts of the wave functions $\phi(r)$ of corresponding pole for the $I = 1/2,\, J^P = 3/2^-$ sector with different $q_{max}$. Results for the five coupled-channel case (top) and the splitting VB sector (bottom) are compared.}
\label{fig:wfI1J3c5VB}
\end{figure}

In Fig.~\ref{fig:wfI1J3c5VB}, we display the results of the wave functions $\phi(\vec{r})$ of the three poles $\bar{D}\Sigma_c^*$, $\bar{D}^*\Sigma_c$, and $\bar{D}^*\Sigma_c^*$ with different cutoff values $q_{max}$ in the $I = 1/2,\, J^P = 3/2^-$ sector. From Fig.~\ref{fig:wfI1J3c5VB}, one can see that the wave functions are mainly distributed within $0 \sim 6 \text{ fm}$ and go to zero rapidly after $r > 4 \text{ fm}$, which is consistent with the one found in the $I = 1/2,\, J^P = 1/2^-$ sector above. The wave functions after splitting into the VB subsystem do not change significantly, and thus, the results are not repeated here.

\begin{figure}[htbp]
\centering
\includegraphics[width=0.45\textwidth]{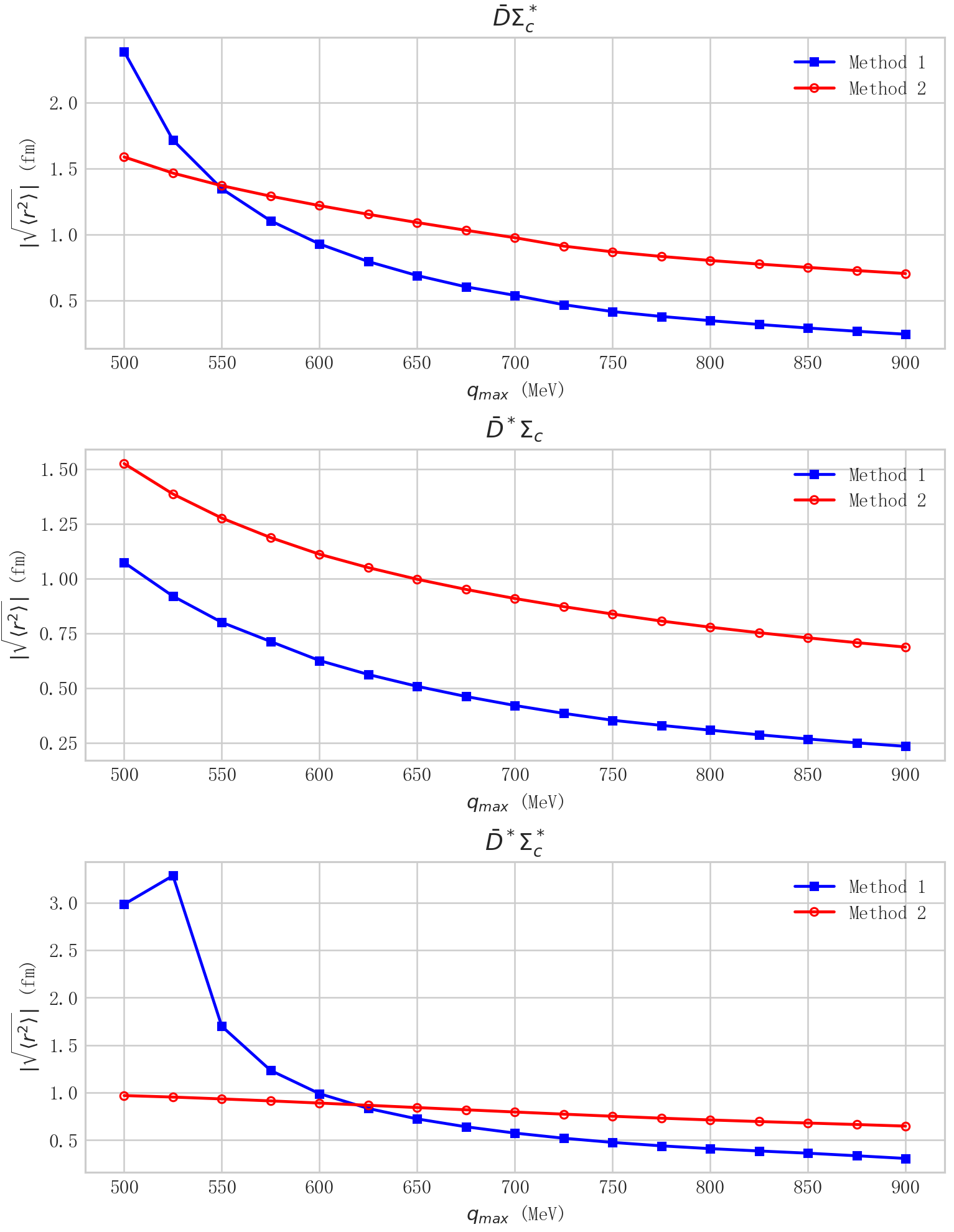}
\includegraphics[width=0.45\textwidth,height=0.3\textheight]{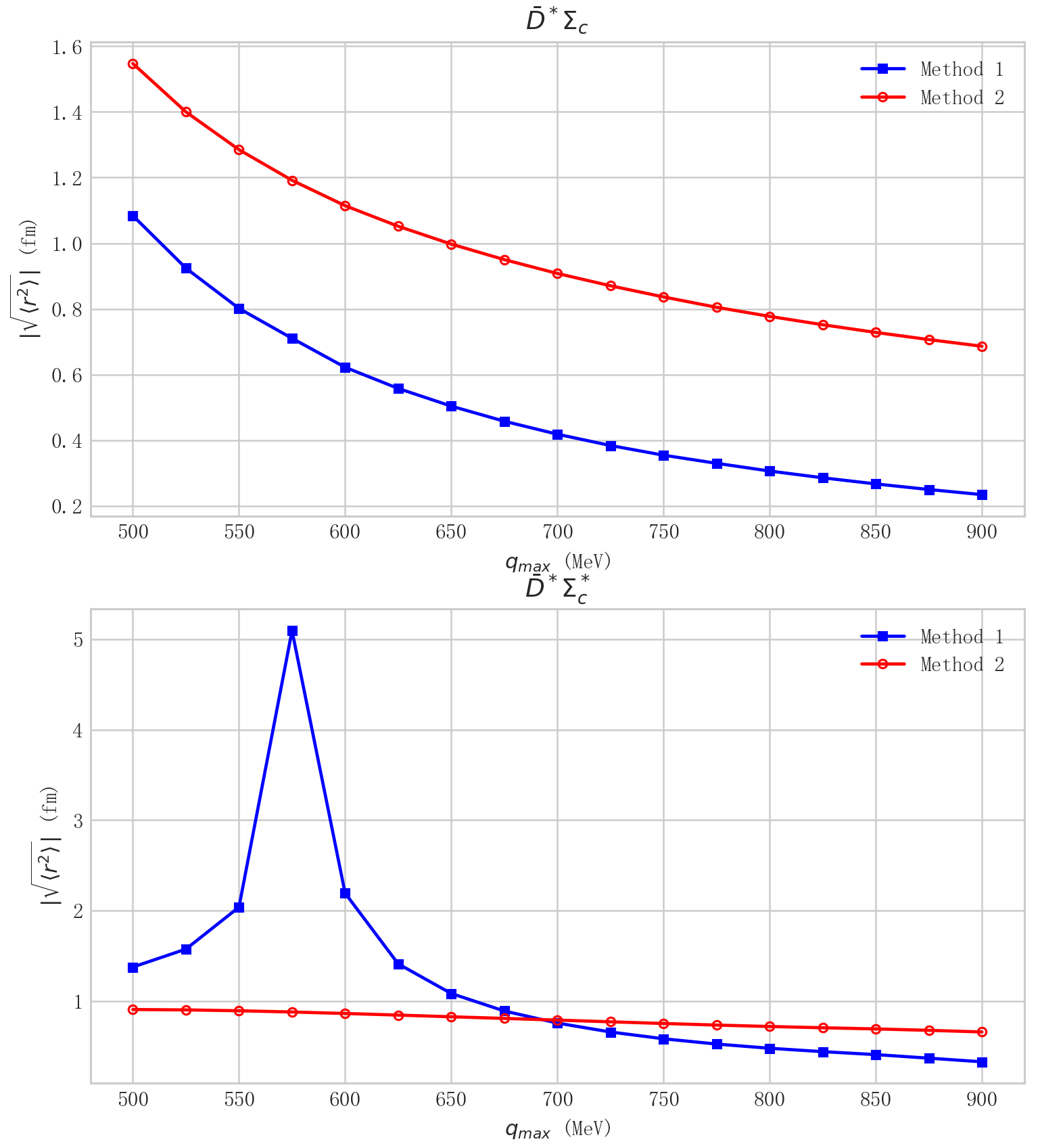}
\caption{RMS radii of the corresponding poles for the $I = 1/2,\, J^P = 3/2^-$ sector as a function of the cutoff $q_{max}$ in the five coupled-channel case (left) and the splitting VB sector (right). Results from Method 1 (blue) and Method 2 (red) are compared.}
\label{fig:rmsI1J3c5PV}
\end{figure}

With the wave functions obtained, we also evaluate the radii of these poles. 
The results of the RMS radii with varying the cutoffs are shown in Fig.~\ref{fig:rmsI1J3c5PV}, some of which are shown in Tables~\ref{tab:rmsI1J3c5} and~\ref{tab:rmsI1J3c5PV} for the five coupled-channel case and the splitting VB sector, respectively.  
From these results, one can see that the results with two methods are consistent with each other in most cases and the results of Method 2 are more stable. 
As found in the $I = 1/2,\, J^P = 1/2^-$ sector, the radii in most of the cases are less than 3 fm, consistent with the results of the wave functions above. 
From Fig.~\ref{fig:rmsI1J3c5PV}, also compared with Tables~\ref{tab:rmsI1J3c5} and~\ref{tab:rmsI1J3c5PV}, the coupled channel effect from the HQSS constraint has some influence on the radii, but not much. 

\begin{table}[ht]
\centering
\caption{RMS radii $\left| \sqrt{\langle r^2 \rangle} \right|$ of the corresponding poles for the $I = 1/2,\, J^P = 3/2^-$ sector using different methods in the five coupled-channel case.}
\label{tab:rmsI1J3c5}
\begin{flushleft}
The radii of states calculated with Eq.~\eqref{eq:msr1}
\end{flushleft}
\footnotesize
\renewcommand{\arraystretch}{0.7}
\setlength{\tabcolsep}{2pt}
\begin{tabular}{ccccccc}
\hline\hline
Resonances & $q_{max} = 600 \, \text{MeV}$ & $\left| \sqrt{\langle r^2 \rangle} \right|_2$ & $q_{max} = 700 \, \text{MeV}$ & $\left| \sqrt{\langle r^2 \rangle} \right|_2$ & $q_{max} = 800 \, \text{MeV}$ & $\left| \sqrt{\langle r^2 \rangle} \right|_2$ \\ \hline
$\bar{D}\Sigma_c^*$ & $1.198 + 0.236i \text{ fm}$ & $1.221 \text{ fm}$ & $0.976 + 0.048i \text{ fm}$ & $0.977 \text{ fm}$ & $0.804 + 0.022i \text{ fm}$ & $0.805 \text{ fm}$ \\
$\bar{D}^*\Sigma_c$ & $1.111 + 0.028i \text{ fm}$ & $1.111 \text{ fm}$ & $0.909 + 0.009i \text{ fm}$ & $0.909 \text{ fm}$ & $0.778 + 0.003i \text{ fm}$ & $0.778 \text{ fm}$ \\
$\bar{D}^*\Sigma_c^*$ & $0.885 + 0.095i \text{ fm}$ & $0.890 \text{ fm}$ & $0.794 + 0.056i \text{ fm}$ & $0.796 \text{ fm}$ & $0.711 + 0.044i \text{ fm}$ & $0.713 \text{ fm}$  \\ \hline\hline
\end{tabular}
\begin{flushleft}
The radii of states calculated with Eq.~\eqref{eq:msr2}
\end{flushleft}
\renewcommand{\arraystretch}{0.7} 
\setlength{\tabcolsep}{2pt}
\begin{tabular}{ccccccc}
\hline\hline
Resonances & $q_{max} = 600 \, \text{MeV}$ & $\left| \sqrt{\langle r^2 \rangle} \right|_1$ & $q_{max} = 700 \, \text{MeV}$ & $\left| \sqrt{\langle r^2 \rangle} \right|_1$ & $q_{max} = 800 \, \text{MeV}$ & $\left| \sqrt{\langle r^2 \rangle} \right|_1$ \\ \hline
$\bar{D}\Sigma_c^*$ & $0.930 - 0.021i \text{ fm}$ & $0.930 \text{ fm}$ & $0.540 - 0.001i \text{ fm}$ & $0.540 \text{ fm}$ & $0.347 - 0.035i \text{ fm}$ & $0.349 \text{ fm}$ \\
$\bar{D}^*\Sigma_c$ & $0.625 - 0.016i \text{ fm}$ & $0.625 \text{ fm}$ & $0.420 - 0.016i \text{ fm}$ & $0.420 \text{ fm}$ & $0.307 - 0.012i \text{ fm}$ & $0.307 \text{ fm}$ \\
$\bar{D}^*\Sigma_c^*$ & $0.989 - 0.014i \text{ fm}$ & $0.989 \text{ fm}$ & $0.574 - 0.024i \text{ fm}$ & $0.575 \text{ fm}$ & $0.410 - 0.018i \text{ fm}$ & $0.411 \text{ fm}$ \\ \hline\hline
\end{tabular}
\end{table}
\begin{table}[ht]
\centering
\caption{RMS radii $\left| \sqrt{\langle r^2 \rangle} \right|$ of the corresponding poles for the $I = 1/2,\, J^P = 3/2^-$ sector using different methods in the splitting VB sector.}
\label{tab:rmsI1J3c5PV}
\begin{flushleft}
The radii of states calculated with Eq.~\eqref{eq:msr1}
\end{flushleft}
\renewcommand{\arraystretch}{0.7} 
\setlength{\tabcolsep}{2pt}
\begin{tabular}{ccccccc}
\hline\hline
Resonances & $q_{max} = 600 \, \text{MeV}$ & $\left| \sqrt{\langle r^2 \rangle} \right|_2$ & $q_{max} = 700 \, \text{MeV}$ & $\left| \sqrt{\langle r^2 \rangle} \right|_2$ & $q_{max} = 800 \, \text{MeV}$ & $\left| \sqrt{\langle r^2 \rangle} \right|_2$ \\ \hline
$\bar{D}^*\Sigma_c$ & $1.114 + 0.020i \text{ fm}$ & $1.114 \text{ fm}$ & $0.908 + 0.006i \text{ fm}$ & $0.908 \text{ fm}$ & $0.777 + 0.002i \text{ fm}$ & $0.777 \text{ fm}$ \\
$\bar{D}^*\Sigma_c^*$ & $0.846 + 0.179i \text{ fm}$ & $0.865 \text{ fm}$ & $0.784 + 0.111i \text{ fm}$ & $0.792 \text{ fm}$ & $0.717 + 0.081i \text{ fm}$ & $0.721 \text{ fm}$ \\ \hline\hline
\end{tabular}
\begin{flushleft}
The radii of states calculated with Eq.~\eqref{eq:msr2}
\end{flushleft}
\renewcommand{\arraystretch}{0.7} 
\setlength{\tabcolsep}{2pt}
\begin{tabular}{ccccccc}
\hline\hline
Resonances & $q_{max} = 600 \, \text{MeV}$ & $\left| \sqrt{\langle r^2 \rangle} \right|_1$ & $q_{max} = 700 \, \text{MeV}$ & $\left| \sqrt{\langle r^2 \rangle} \right|_1$ & $q_{max} = 800 \, \text{MeV}$ & $\left| \sqrt{\langle r^2 \rangle} \right|_1$ \\ \hline
$\bar{D}^*\Sigma_c$ & $0.622 - 0.011i \text{ fm}$ & $0.622 \text{ fm}$ & $0.418 - 0.008i \text{ fm}$ & $0.418 \text{ fm}$ & $0.306 - 0.009i \text{ fm}$ & $0.306 \text{ fm}$ \\
$\bar{D}^*\Sigma_c^*$ & $2.194 - 0.017i \text{ fm}$ & $2.194 \text{ fm}$ & $0.759 - 0.011i \text{ fm}$ & $0.759 \text{ fm}$ & $0.481 - 0.001i \text{ fm}$ & $0.481 \text{ fm}$ \\ \hline\hline
\end{tabular}
\end{table}

\subsection{Results of coupled-channel interactions in hidden charm strange sector}

\begin{figure}[htbp]
\centering
\includegraphics[width=0.8\textwidth]{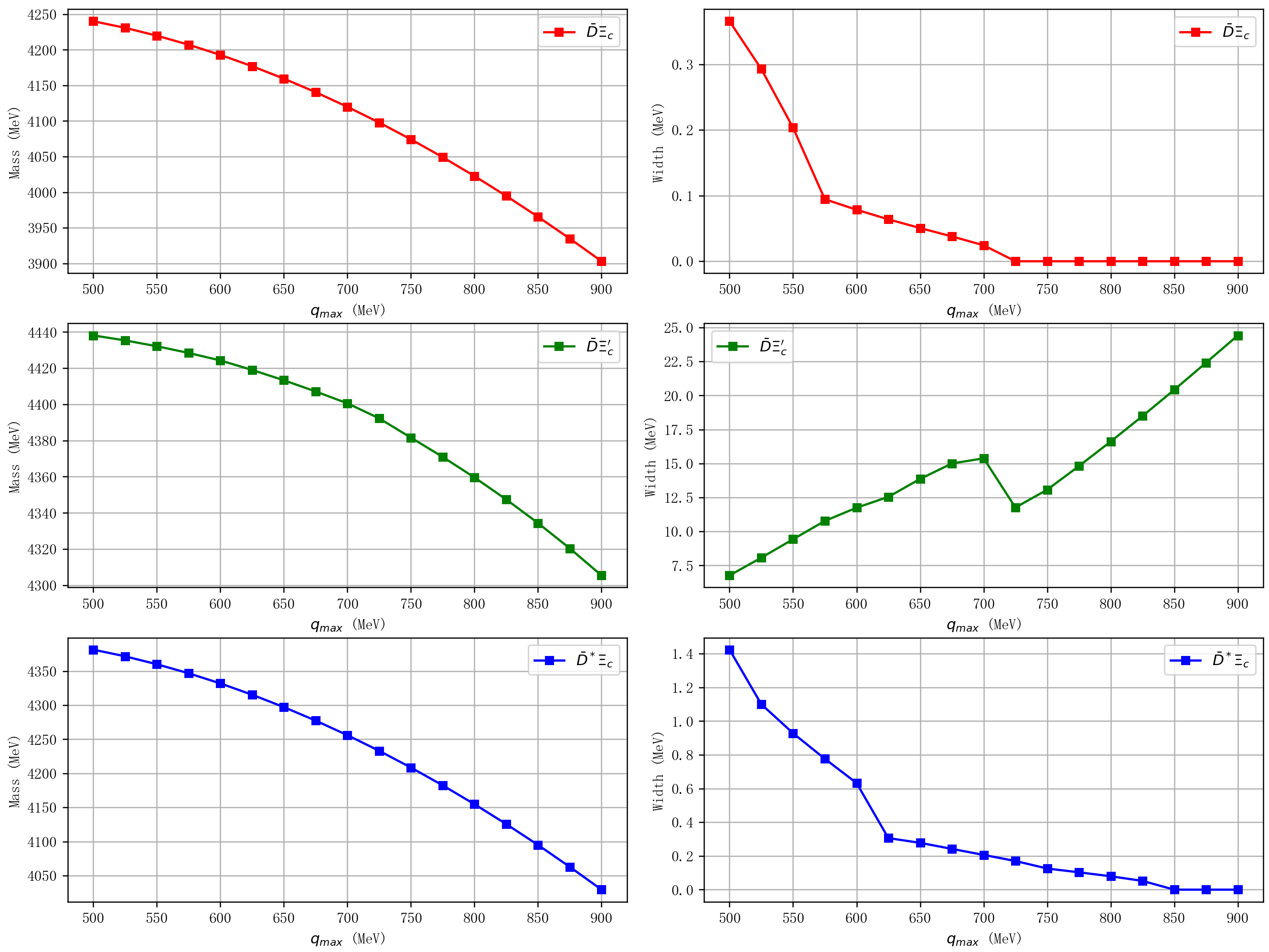}
\includegraphics[width=0.8\textwidth]{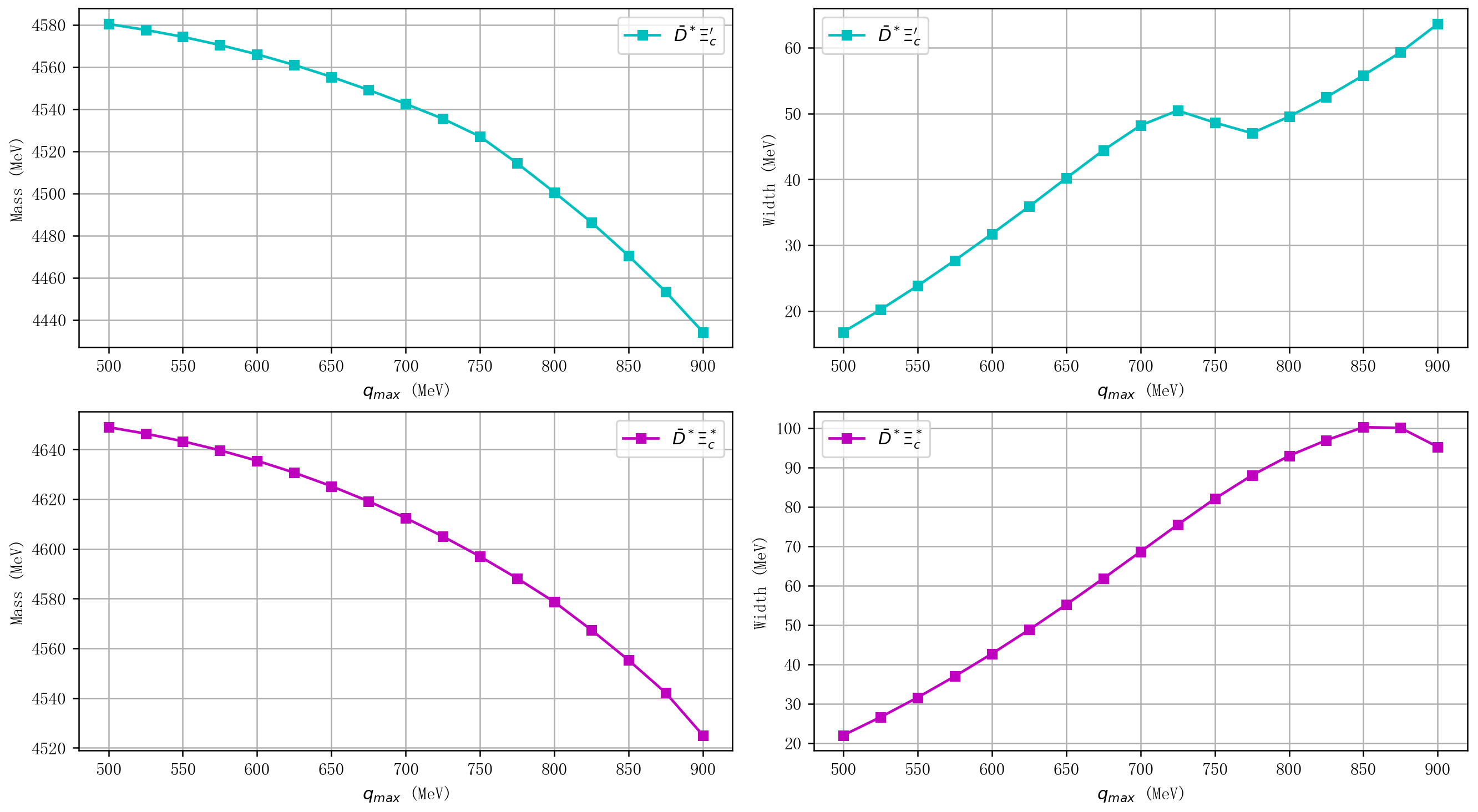}
\caption{Trajectories for the masses (left) and widths (right) of the poles in the second Riemann sheets for the $I=0,\, J^P = 1/2^-$ sector in the nine coupled-channel case.}
\label{fig:resI0J1cs9}
\end{figure}

Now we start to investigate the hidden charm strange sector to compare with what we have in the hidden charm sector obtained above. From the results of Ref.~\cite{Xiao:2019gjd}, we know that there are nine coupled channels under the HQSS constraint, as shown in Table~\ref{tab:I0J1cs9} for the $I=0,\, J^P = 1/2^-$ sector, where the bound systems are the $\bar{D} \Xi_c$, $\bar{D} \Xi'_c$, $\bar{D}^* \Xi_c$, $\bar{D}^* \Xi'_c$, and $\bar{D}^* \Xi^*_c$ channels. 
The trajectories of the masses and widths are shown in Fig.~\ref{fig:resI0J1cs9}, where some results are presented in Table~\ref{tab:resI0J1cs9}. From Fig.~\ref{fig:resI0J1cs9}, one can see that the masses of all the poles decrease monotonically with the increase of the cutoff $q_{max}$, which indicate that the strong attractive interactions among these systems dominate the generation of these poles. 
The poles corresponding to $\bar{D} \Xi_c$ and $\bar{D}^* \Xi_c$ exhibit extremely narrow widths and are located very close to the real axis on the complex energy plane, reflecting their relatively weak couplings to the low-energy open channels $\eta_c \Lambda$ and $J/\psi \Lambda$. And the binding energies of these two poles, more than 100 MeV, are much bigger than the other poles, which indicate the strong interactions for these two channels leading to deeply bound systems. 
Indeed, from the results of Refs.~\cite{Xiao:2019gjd,Xiao:2021rgp}, the poles of the $\bar{D} \Xi_c$ and $\bar{D}^* \Xi_c$ channels are strongly couple to the $\bar{D}_s \Lambda_c$ and $\bar{D}_s^* \Lambda_c$ channels, respectively, which lead to these system become more bound. 
Due to their large binding energy, as shown in detail in Table~\ref{tab:resI0J1cs9}, the $\bar{D}^* \Xi_c$ pole is assigned as the $P_{cs}(4338)$ state, while the $\bar{D} \Xi'_c$ pole as the $P_{cs}(4459)$ state, as found in Ref.~\cite{Feijoo:2022rxf}. This conclusion is different from the general views~\cite{Xiao:2021rgp,Wang:2022neq,Meng:2022wgl,Zhu:2022wpi,Wang:2019nvm}, the $P_{cs}(4338)$ as the $\bar{D} \Xi_c$ state, the $P_{cs}(4459)$ as the $\bar{D}^* \Xi_c$ molecule, which are analogous to the case of hidden charm sector, the $P_c(4312)$ as the $\bar{D} \Sigma$ state, the $P_c(4459)$ [$P_c(4440)$ and $P_c(4457)$] as the $\bar{D}^* \Sigma$ molecule. In Ref.~\cite{Wang:2019nvm}, it was predicted that there were two $P_{cs}(4459)$ states with different spins, which had 6 MeV for the masses differences, similar to the one of the $P_c(4459)$ state in the hidden charm system.  
\begin{figure}[htbp]
\centering
\includegraphics[width=0.8\textwidth]{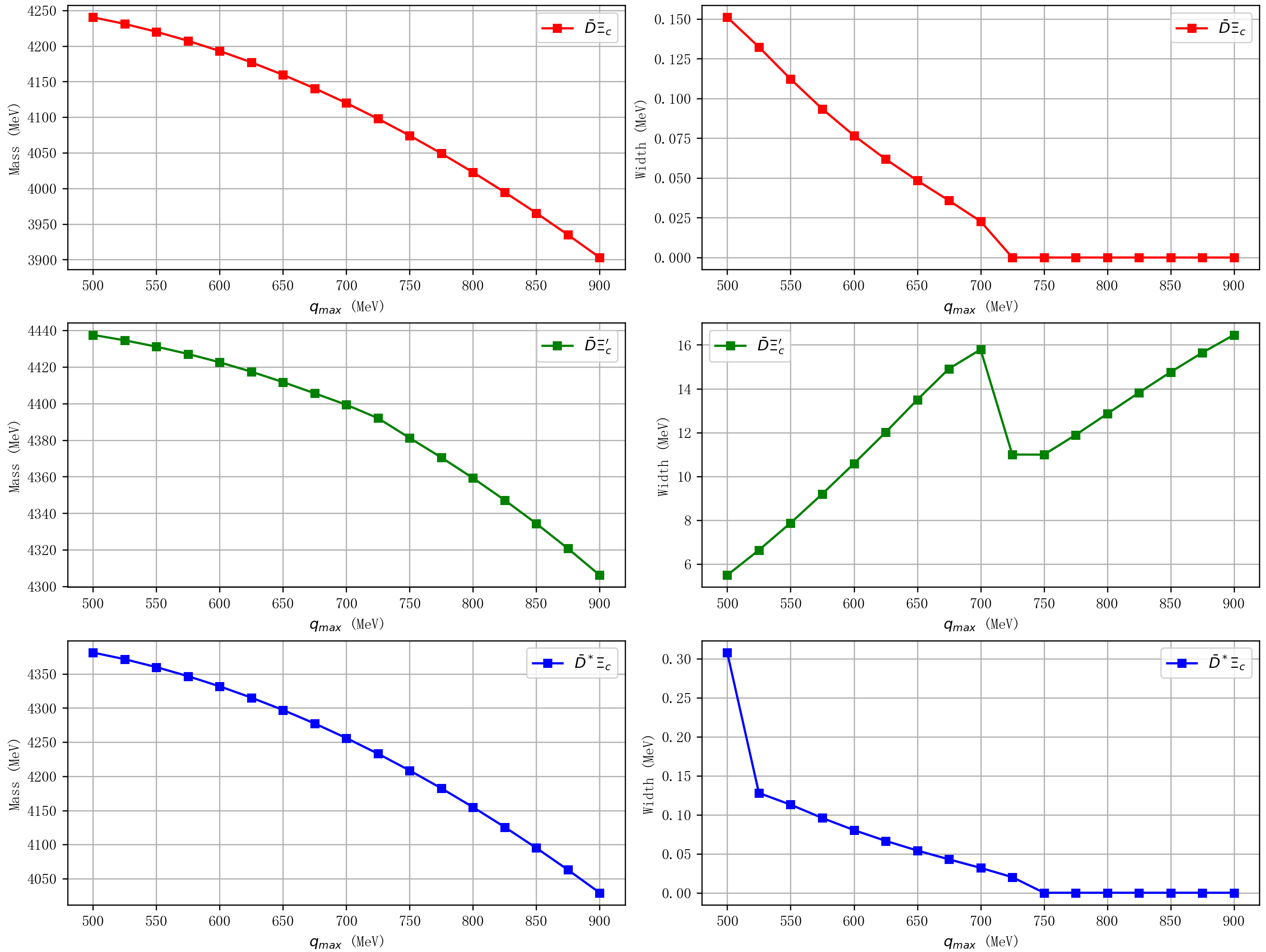}
\includegraphics[width=0.8\textwidth]{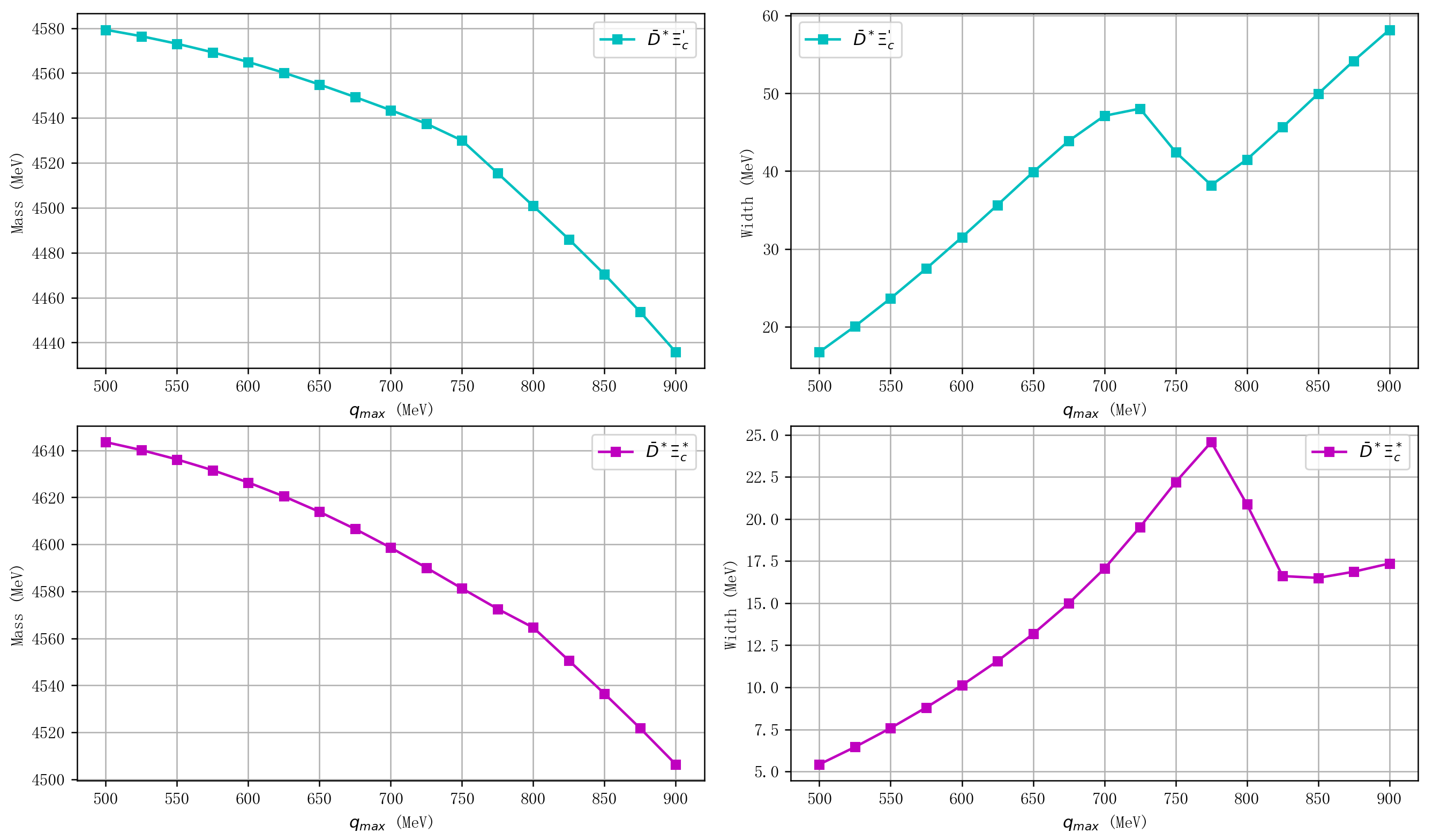}
\caption{Trajectories for the masses (left) and widths (right) of the poles in the second Riemann sheets for the $I=0,\, J^P = 1/2^-$ sector in the splitting PB and VB sectors.}
\label{fig:resI0J1cs9PV}
\end{figure}
\begin{table}[ht]
\centering
\caption{Pole positions $(M, \Gamma)$ in the second Riemann sheets for the $I=0,\, J^P = 1/2^-$ sector with the nine coupled-channel case.}
\label{tab:resI0J1cs9}
\renewcommand{\arraystretch}{0.7}
\begin{tabular}{cccccc}
\hline\hline
$q_{max}$ & Mass & Width & Main & & Experimental \\
{[MeV]} & {[MeV]} & {[MeV]} & channel & $J^P$ & states \\ \hline
600 & 4192.91 & 0.08 & $\bar{D} \Xi_c$ & $1/2^-$ & -- \\
700 & 4119.91 & 0.02 & $\bar{D} \Xi_c$ & $1/2^-$ & -- \\
800 & 4022.66 & 0.00 & $\bar{D} \Xi_c$ & -- \\ \hline
600 & 4331.89 & 0.63 & $\bar{D}^* \Xi_c$ & $1/2^-$ & $P_{cs}(4338)$ \\
700 & 4255.98 & 0.21 & $\bar{D}^* \Xi_c$ & $1/2^-$ & -- \\
800 & 4154.76 & 0.08 & $\bar{D}^* \Xi_c$ & $1/2^-$ & -- \\ \hline
600 & 4424.27 & 11.75 & $\bar{D} \Xi'_c$ & $1/2^-$ & $P_{cs}(4459)$ \\
700 & 4400.58 & 15.38 & $\bar{D} \Xi'_c$ & $1/2^-$ & -- \\
800 & 4359.59 & 16.61 & $\bar{D} \Xi'_c$ & $1/2^-$ & -- \\ \hline
600 & 4565.96 & 31.73 & $\bar{D}^* \Xi'_c$ & $1/2^-$ & -- \\
700 & 4542.46 & 48.20 & $\bar{D}^* \Xi'_c$ & $1/2^-$ & -- \\
800 & 4500.58 & 49.56 & $\bar{D}^* \Xi'_c$ & $1/2^-$ & -- \\ \hline
600 & 4635.46 & 42.75 & $\bar{D}^* \Xi^*_c$ & $1/2^-$ & -- \\
700 & 4612.43 & 68.66 & $\bar{D}^* \Xi^*_c$ & $1/2^-$ & -- \\
800 & 4578.71 & 93.02 & $\bar{D}^* \Xi^*_c$ & $1/2^-$ & -- \\ \hline\hline
\end{tabular}
\end{table}
\begin{table}[ht]
\centering
\caption{Pole positions $(M, \Gamma)$ in the second Riemann sheets for the $I=0,\, J^P = 1/2^-$ sector with the splitting PB and VB sectors.}
\label{tab:resI0J1cs9PV}
\renewcommand{\arraystretch}{0.7}
\begin{tabular}{ccccccc}
\hline\hline
$q_{max}$ & Mass & Width & Main & & & Experimental \\
{[MeV]} & {[MeV]} & {[MeV]} & channel & Sector & $J^P$ & state \\ \hline
600 & 4192.98 & 0.08 & $\bar{D} \Xi_c$ & PB & $1/2^-$ & -- \\
700 & 4119.94 & 0.02 & $\bar{D} \Xi_c$ & PB & $1/2^-$ & -- \\
800 & 4022.65 & 0.00 & $\bar{D} \Xi_c$ & PB & $1/2^-$ & -- \\ \hline
600 & 4331.75 & 0.08 & $\bar{D}^* \Xi_c$& VB & $1/2^-$ & $P_{cs}(4338)$  \\
700 & 4255.99 & 0.03 & $\bar{D}^* \Xi_c$ & VB & $1/2^-$ & -- \\
800 & 4154.78 & 0.00 & $\bar{D}^* \Xi_c$ & VB & $1/2^-$ & -- \\ \hline
600 & 4422.56 & 10.58 & $\bar{D} \Xi'_c$ & PB & $1/2^-$ & $P_{cs}(4459)$  \\
700 & 4399.34 & 15.79 & $\bar{D} \Xi'_c$ & PB & $1/2^-$ & -- \\
800 & 4359.24 & 12.85 & $\bar{D} \Xi'_c$ & PB & $1/2^-$ & -- \\ \hline
600 & 4564.95 & 31.49 & $\bar{D}^* \Xi'_c$ & VB & $1/2^-$ & -- \\
700 & 4543.50 & 47.12 & $\bar{D}^* \Xi'_c$ & VB & $1/2^-$ & -- \\
800 & 4500.82 & 41.50 & $\bar{D}^* \Xi'_c$ & VB & $1/2^-$ & -- \\ \hline
600 & 4626.34 & 10.12 & $\bar{D}^* \Xi^*_c$ & VB & $1/2^-$ & -- \\
700 & 4598.59 & 17.05 & $\bar{D}^* \Xi^*_c$ & VB & $1/2^-$ & -- \\
800 & 4564.59 & 20.87 & $\bar{D}^* \Xi^*_c$ & VB & $1/2^-$ & -- \\ \hline\hline
\end{tabular}
\end{table}

As done in the hidden charm sector, to check the coupled channel effect without the HQSS constraint, we also divide the coupled channel system into the PB and VB subsystems.
The trajectories of pole positions with varying $q_{max}$ are shown in Fig.~\ref{fig:resI0J1cs9PV}, where some of results are shown in Table~\ref{tab:resI0J1cs9PV}. 
Most of these results are not much different with the full coupled-channels case, see Fig.~\ref{fig:resI0J1cs9} and Table~\ref{tab:resI0J1cs9}, except for the width of the $\bar{D}^* \Xi^*_c$ pole having 4 time smaller, which indicate that the coupled channel effect with the HQSS constraint is not much important to this sector, different from the case in the hidden charm sector before. 
Besides, the width trajectories of the $\bar{D} \Xi'_c, \bar{D}^* \Xi'_c$ channels exhibit some fluctuations around $q_{max} = 700 \sim 850 \text{ MeV}$ due to the crossing the thresholds of some open channels.

\begin{figure}[htbp]
\centering
\includegraphics[width=0.8\textwidth]{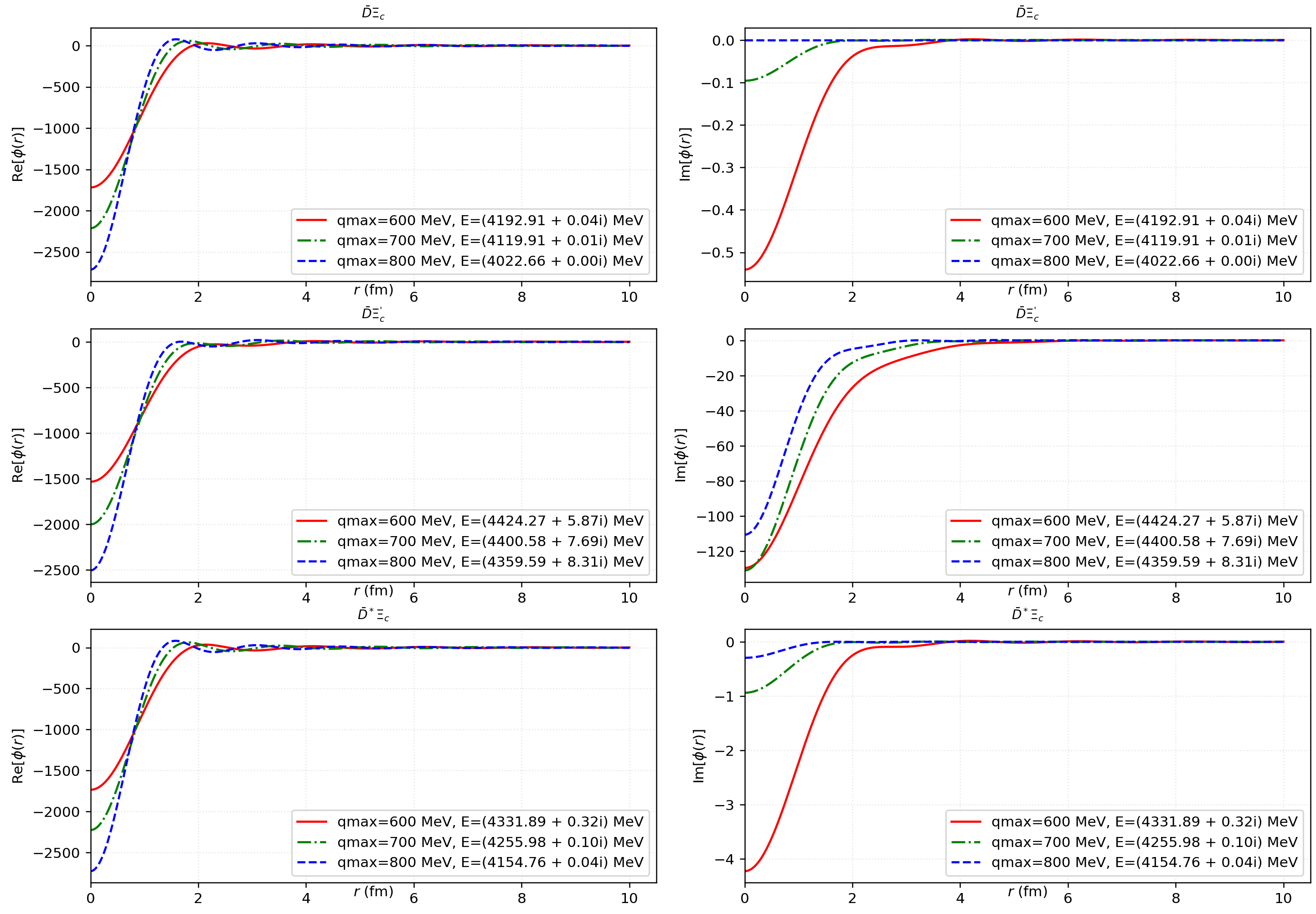}
\includegraphics[width=0.8\textwidth]{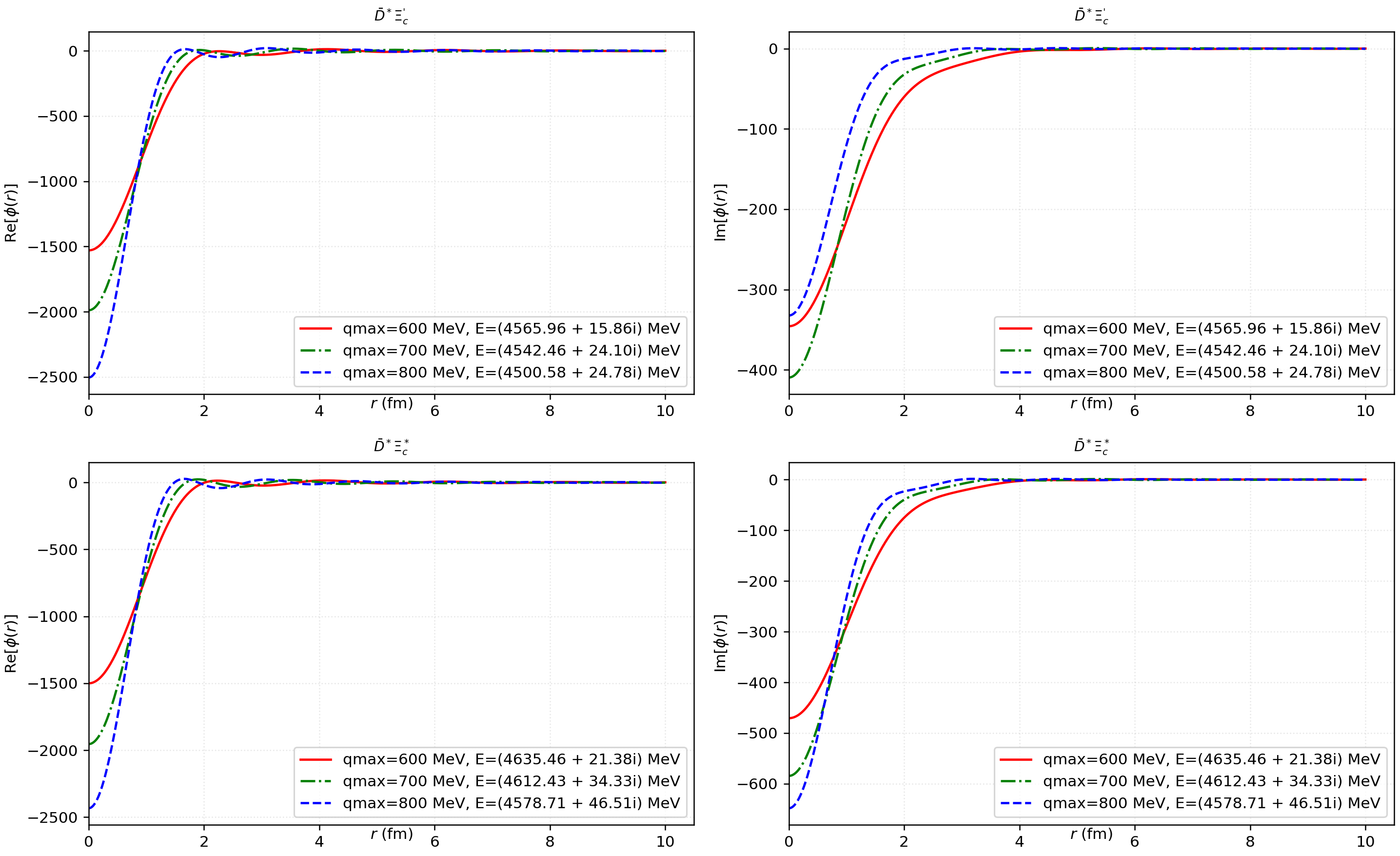}
\caption{Real (left) and imaginary (right) parts of the wave functions $\phi(r)$ of corresponding pole for the $I=0,\, J^P = 1/2^-$ sector with different $q_{max}$ in the nine coupled-channel case.}
\label{fig:wfI0J1cs9}
\end{figure}

Next, we show the results of the real and imaginary parts of the wave functions for each pole obtained in Fig.~\ref{fig:wfI0J1cs9}, where one can see once again that the wave functions are mainly contributed within $0 \sim 6 \text{ fm}$ and go to zero rapidly after $r > 4 \text{ fm}$, as found in the hidden charm sector above. 
The results for the splitting PB and VB sectors are not repeated to show here due to the similar results. 

\begin{figure}[htbp]
\centering
\includegraphics[width=0.4\textwidth]{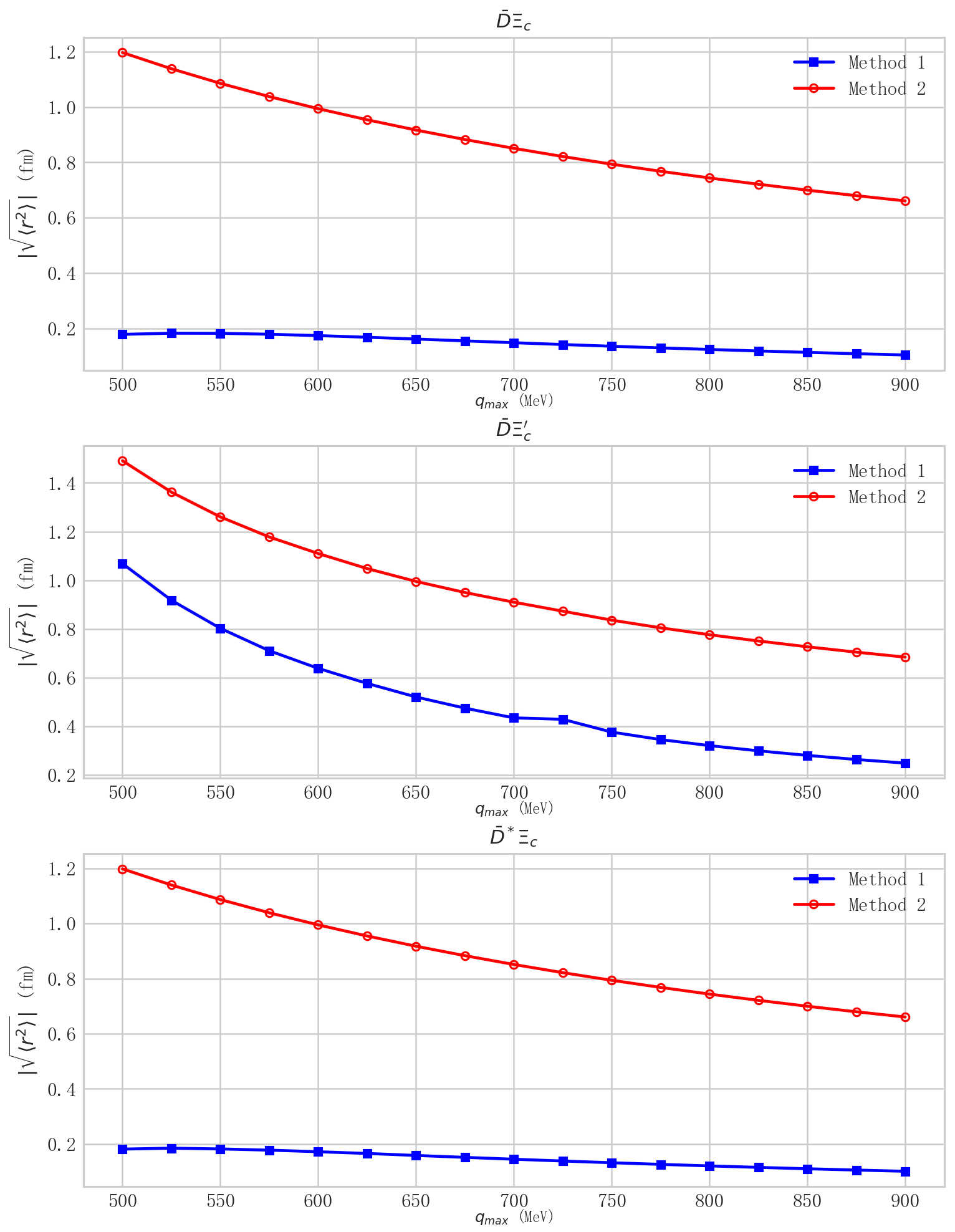}
\includegraphics[width=0.4\textwidth]{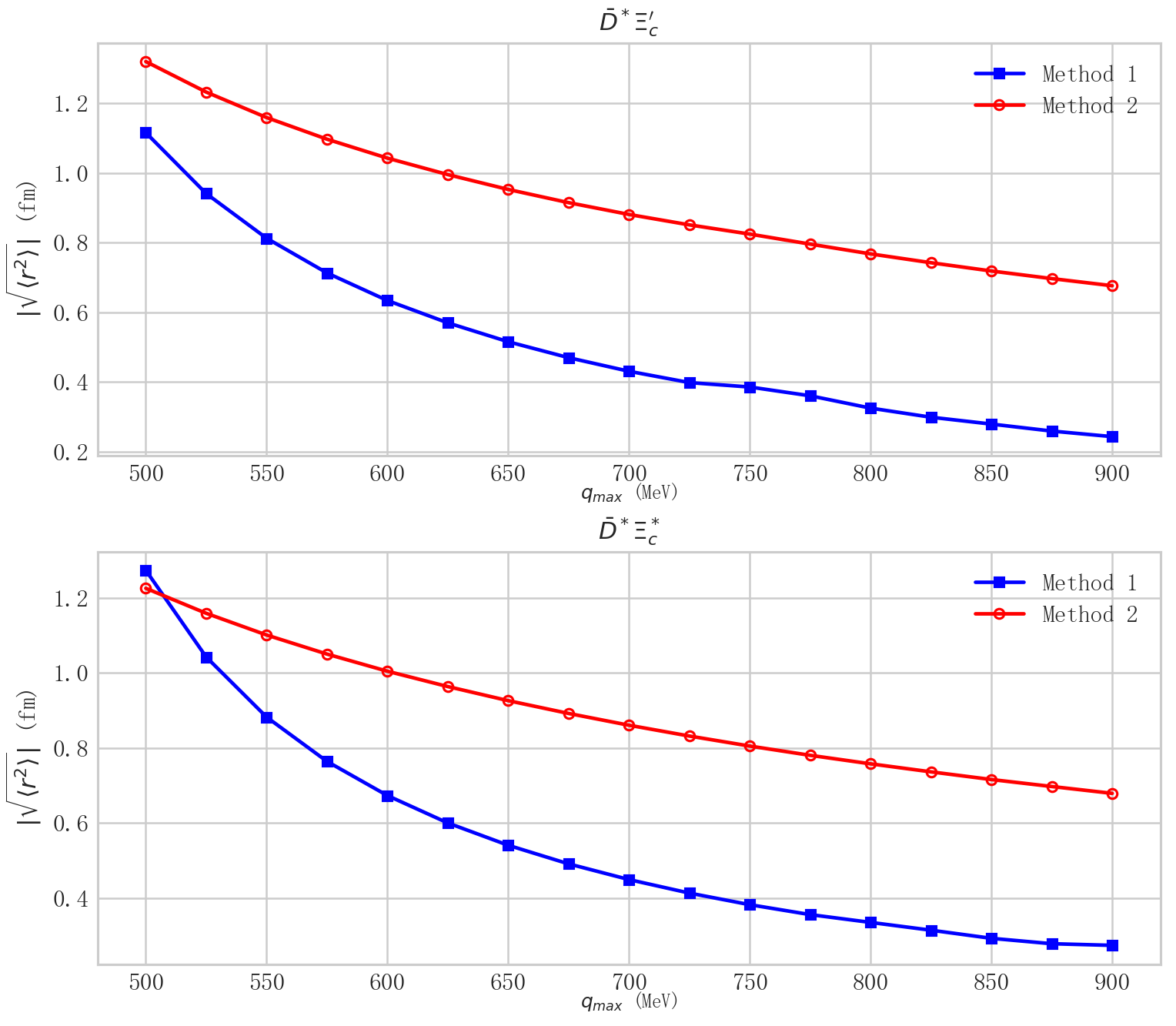}
\caption{RMS radii of the corresponding poles for the $I=0,\, J^P = 1/2^-$ sector as a function of the cutoff $q_{max}$ in the nine coupled-channel case. Results from Method 1 (blue) and Method 2 (red) are compared.}
\label{fig:rmsI0J1cs9}
\end{figure}

\begin{figure}[htbp]
\centering
\includegraphics[width=0.4\textwidth]{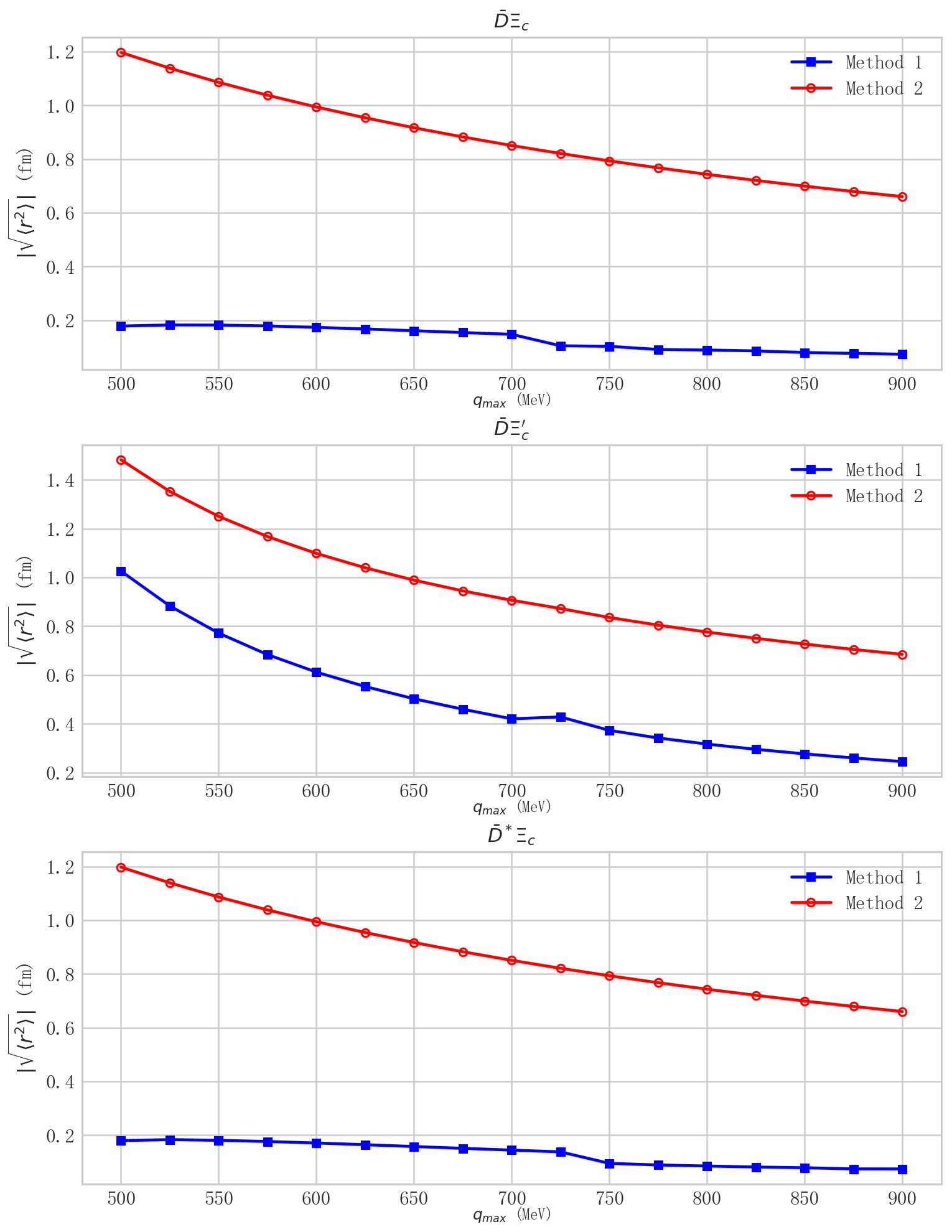}
\includegraphics[width=0.4\textwidth]{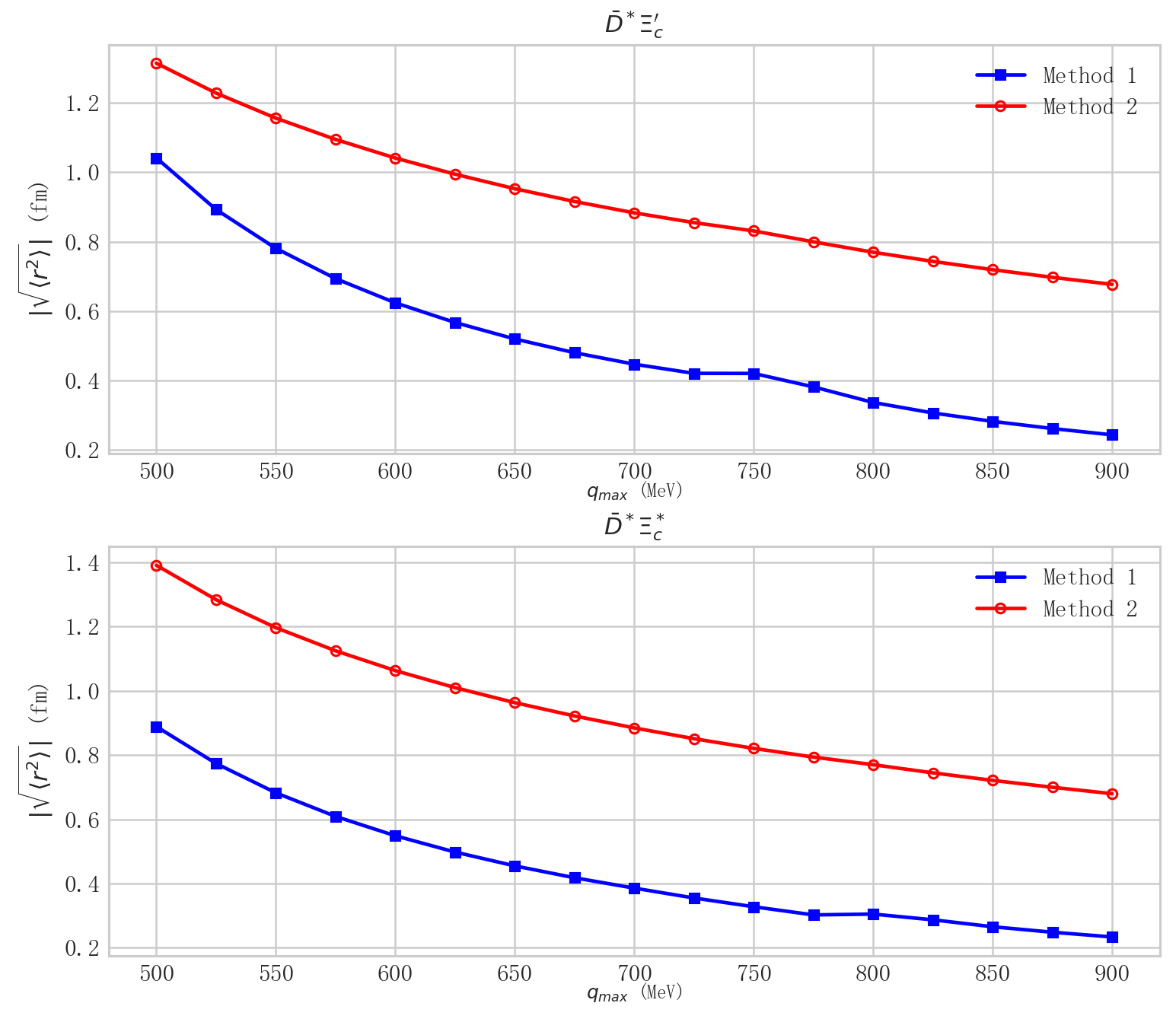}
\caption{RMS radii of the corresponding poles for the $I=0,\, J^P = 1/2^-$ sector as a function of the cutoff $q_{max}$ in the splitting PB and VB sectors. Results from Method 1 (blue) and Method 2 (red) are compared.}
\label{fig:rmsI0J1cs9PV}
\end{figure}

With the wave functions obtained, we continue to calculate the radii of these poles and show the results of the RMS radii with varying the cutoffs are shown in Fig.~\ref{fig:rmsI0J1cs9}, where some of them are listed in detail in Table~\ref{tab:rmsI0J1cs9}. 
As found above, the results of Method 2 are more stable than the results obtained with Method 1, all of which are less than 1.5 fm. 
From Fig.~\ref{fig:rmsI0J1cs9}, one can see that the results of Method 1 for the channels $\bar{D} \Xi$, $\bar{D}^* \Xi$ are small, less than 0.2 fm, due to the large binding energies, see Eq.~\eqref{eq:msr2}, meaning that Eq.~\eqref{eq:msr2} is not suit for the deep bound systems.
As show in Fig.~\ref{fig:rmsI0J1cs9PV},  one can see that the results in the splitting PB and VB sectors are not much different with the results of Fig.~\ref{fig:rmsI0J1cs9}, which once again indicate that the coupled channel effect from the HQSS constraint has little influence on the radii as found in the hidden charm sector. 
\begin{table}[ht]
\centering
\caption{RMS radii $\left| \sqrt{\langle r^2 \rangle} \right|$ of the corresponding poles for the $I=0,\, J^P = 1/2^-$ sector using different methods in the nine coupled-channel case.}
\label{tab:rmsI0J1cs9}
\begin{flushleft}
The radii of states calculated with Eq.~\eqref{eq:msr1}
\end{flushleft}
\renewcommand{\arraystretch}{0.7} 
\setlength{\tabcolsep}{2pt}
\begin{tabular}{ccccccc}
\hline\hline
Resonances & $q_{max} = 600 \, \text{MeV}$ & $\left| \sqrt{\langle r^2 \rangle} \right|_2$ & $q_{max} = 700 \, \text{MeV}$ & $\left| \sqrt{\langle r^2 \rangle} \right|_2$ & $q_{max} = 800 \, \text{MeV}$ & $\left| \sqrt{\langle r^2 \rangle} \right|_2$ \\ \hline
$\bar{D}\Xi_c$ & $0.994 - 0.000i \text{ fm}$ & $0.994 \text{ fm}$ & $0.851 - 0.000i \text{ fm}$ & $0.851 \text{ fm}$ & $0.744 + 0.000i \text{ fm}$ & $0.744 \text{ fm}$ \\
$\bar{D}\Xi_c'$ & $1.109 + 0.054i \text{ fm}$ & $1.110 \text{ fm}$ & $0.910 + 0.020i \text{ fm}$ & $0.910 \text{ fm}$ & $0.776 + 0.007i \text{ fm}$ & $0.776 \text{ fm}$ \\
$\bar{D}^*\Xi_c$ & $0.995 - 0.000i \text{ fm}$ & $0.995 \text{ fm}$ & $0.851 - 0.000i \text{ fm}$ & $0.851 \text{ fm}$ & $0.744 - 0.000i \text{ fm}$ & $0.744 \text{ fm}$ \\
$\bar{D}^*\Xi_c'$ & $1.038 + 0.100i \text{ fm}$ & $1.043 \text{ fm}$ & $0.880 + 0.048i \text{ fm}$ & $0.881 \text{ fm}$ & $0.768 + 0.018i \text{ fm}$ & $0.768 \text{ fm}$ \\
$\bar{D}^*\Xi_c^*$ & $0.998 + 0.117i \text{ fm}$ & $1.004 \text{ fm}$ & $0.858 + 0.058i \text{ fm}$ & $0.860 \text{ fm}$ & $0.757 + 0.034i \text{ fm}$ & $0.757 \text{ fm}$ \\ 
 \hline\hline
\end{tabular}
\begin{flushleft}
The radii of states calculated with Eq.~\eqref{eq:msr2}
\end{flushleft}
\renewcommand{\arraystretch}{0.7} 
\setlength{\tabcolsep}{2pt}
\begin{tabular}{ccccccc}
\hline\hline
Resonances & $q_{max} = 600 \, \text{MeV}$ & $\left| \sqrt{\langle r^2 \rangle} \right|_1$ & $q_{max} = 700 \, \text{MeV}$ & $\left| \sqrt{\langle r^2 \rangle} \right|_1$ & $q_{max} = 800 \, \text{MeV}$ & $\left| \sqrt{\langle r^2 \rangle} \right|_1$ \\ \hline
$\bar{D}\Xi_c$ & $0.174 - 0.000i \text{ fm}$ & $0.174 \text{ fm}$ & $0.148 + 0.000i \text{ fm}$ & $0.148 \text{ fm}$ & $0.123 + 0.000i \text{ fm}$ & $0.123 \text{ fm}$ \\
$\bar{D}\Xi_c'$ & $0.638 + 0.011i \text{ fm}$ & $0.638 \text{ fm}$ & $0.435 + 0.009i \text{ fm}$ & $0.435 \text{ fm}$ & $0.321 + 0.001i \text{ fm}$ & $0.321 \text{ fm}$ \\
$\bar{D}^*\Xi_c$ & $0.172 + 0.001i \text{ fm}$ & $0.172 \text{ fm}$ & $0.145 + 0.000i \text{ fm}$ & $0.145 \text{ fm}$ & $0.120 + 0.000i \text{ fm}$ & $0.120 \text{ fm}$ \\
$\bar{D}^*\Xi_c'$ & $0.634 + 0.009i \text{ fm}$ & $0.634 \text{ fm}$ & $0.431 + 0.013i \text{ fm}$ & $0.431 \text{ fm}$ & $0.325 + 0.009i \text{ fm}$ & $0.325 \text{ fm}$ \\
$\bar{D}^*\Xi_c^*$ & $0.673 + 0.010i \text{ fm}$ & $0.673 \text{ fm}$ & $0.449 + 0.012i \text{ fm}$ & $0.449 \text{ fm}$ & $0.334 + 0.019i \text{ fm}$ & $0.334 \text{ fm}$ \\  \hline\hline
\end{tabular}
\end{table}

\begin{figure}[htbp]
\centering
\includegraphics[width=0.8\textwidth]{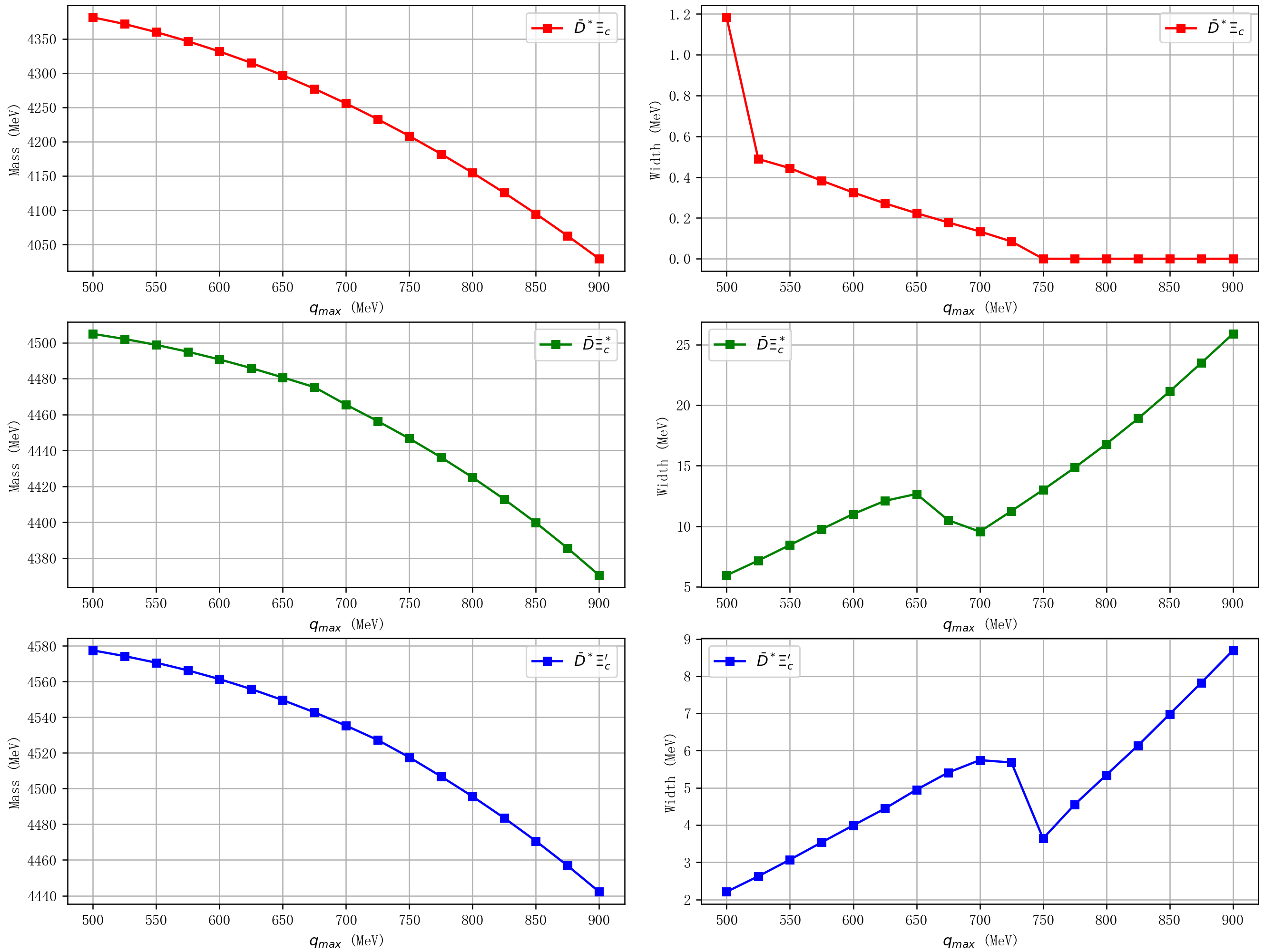}
\includegraphics[width=0.8\textwidth]{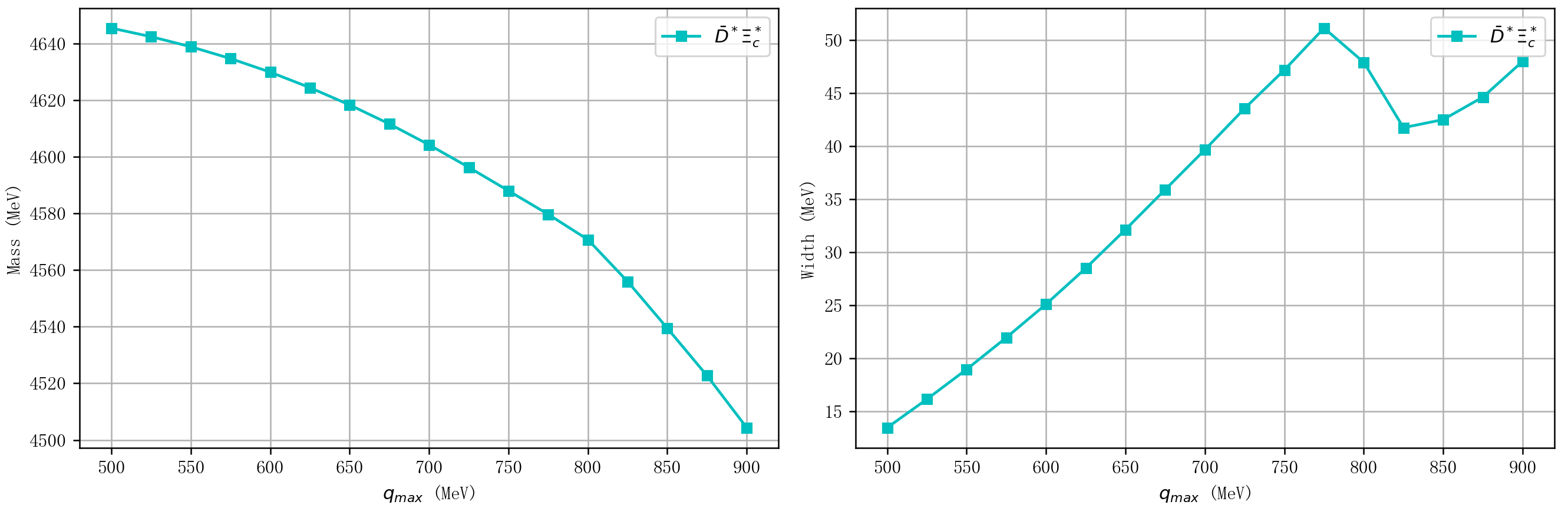}
\caption{Trajectories for the masses (left) and widths (right) of the poles in the second Riemann sheets for the $I=0,\, J^P = 3/2^-$ sector in the six coupled-channel case.}
\label{fig:resI0J3cs6}
\end{figure}

Moreover, we examine the properties of the interactions of the $I=0,\, J^P = 3/2^-$ sector in the hidden charm strange system, where the main bound systems are the $\bar{D}^* \Xi_c$, $\bar{D} \Xi_c^*$, $\bar{D}^* \Xi'_c$, $\bar{D}^* \Xi^*_c$ channels. 
For the six full coupled channels, the trajectories of the pole positions are shown in Fig.~\ref{fig:resI0J3cs6}, some of which are presented in Table~\ref{tab:resI0J3cs6}. 
As shown in Fig.~\ref{fig:resI0J3cs6}, the masses of these poles reduce for the increasing of the cutoff $q_{max}$, all of which are below the corresponding thresholds. 
Unlike the monotonic behaviours in the masses, the decay widths show the different line shapes for these poles. The fluctuations of the widths are due to the pole crossed the thresholds of certain open channels. These results are also indicated different bound behaviours for three main channels, where the $\bar{D}^* \Xi_c$ channel is deeply bound compared to the others, as in the $I=0,\, J^P = 1/2^-$ sector. 
Indeed, the $\bar{D}^* \Xi_c$ pole has an extremely narrow decay width, and as the mass decreases, the pole moves rapidly below all the thresholds with the zero width. This also indicate that the $\bar{D}^* \Xi_c$ couples to the low-energy open channel $J/\psi \Lambda$ weakly. 
As found in Refs.~\cite{Xiao:2019gjd,Xiao:2021rgp}, the poles of the $\bar{D}^* \Xi_c$ channel is strongly couple to the $\bar{D}_s^* \Lambda_c$ channels, which leads to the system become more bound. 
Note that, in this sector, there is only one PB channel, $\bar{D} \Xi_c^*$, and the results without it are not much different with Fig.~\ref{fig:resI0J3cs6} and not shown here, meaning that the influence of this channel can be ignored. 
\begin{table}[ht]
\centering
\caption{Pole positions $(M, \Gamma)$ in the second Riemann sheets for the $I=0,\, J^P = 3/2^-$ sector with the six coupled-channel case.}
\label{tab:resI0J3cs6}
\renewcommand{\arraystretch}{0.7}
\begin{tabular}{cccccc}
\hline\hline
$q_{max}$ & Mass & Width & Main & & Experimental \\
{[MeV]} & {[MeV]} & {[MeV]} & channel & $J^P$ & state \\ \hline
600 & 4331.73 & 0.32 & $\bar{D}^* \Xi_c$ & $3/2^-$ & $P_{cs}(4338)$ \\
700 & 4255.93 & 0.13 & $\bar{D}^* \Xi_c$ & $3/2^-$ & -- \\
800 & 4154.78 & 0.00 & $\bar{D}^* \Xi_c$ & $3/2^-$ & -- \\ \hline
600 & 4490.65 & 11.01 & $\bar{D} \Xi^*_c$ & $3/2^-$ & -- \\
700 & 4465.48 & 9.55 & $\bar{D} \Xi^*_c$ & $3/2^-$ & -- \\
800 & 4424.98 & 16.80 & $\bar{D} \Xi^*_c$ & $3/2^-$ & -- \\ \hline
600 & 4561.40 & 3.99 & $\bar{D}^* \Xi'_c$ & $3/2^-$ & -- \\
700 & 4535.31 & 5.74 & $\bar{D}^* \Xi'_c$ & $3/2^-$ & -- \\
800 & 4495.59 & 5.35 & $\bar{D}^* \Xi'_c$ & $3/2^-$ & -- \\ \hline
600 & 4629.98 & 25.11 & $\bar{D}^* \Xi^*_c$ & $3/2^-$ & -- \\
700 & 4604.30 & 39.67 & $\bar{D}^* \Xi^*_c$ & $3/2^-$ & -- \\
800 & 4570.78 & 47.92 & $\bar{D}^* \Xi^*_c$ & $3/2^-$ & -- \\ \hline\hline
\end{tabular}
\end{table}

\begin{figure}[htbp]
\centering
\includegraphics[width=0.8\textwidth]{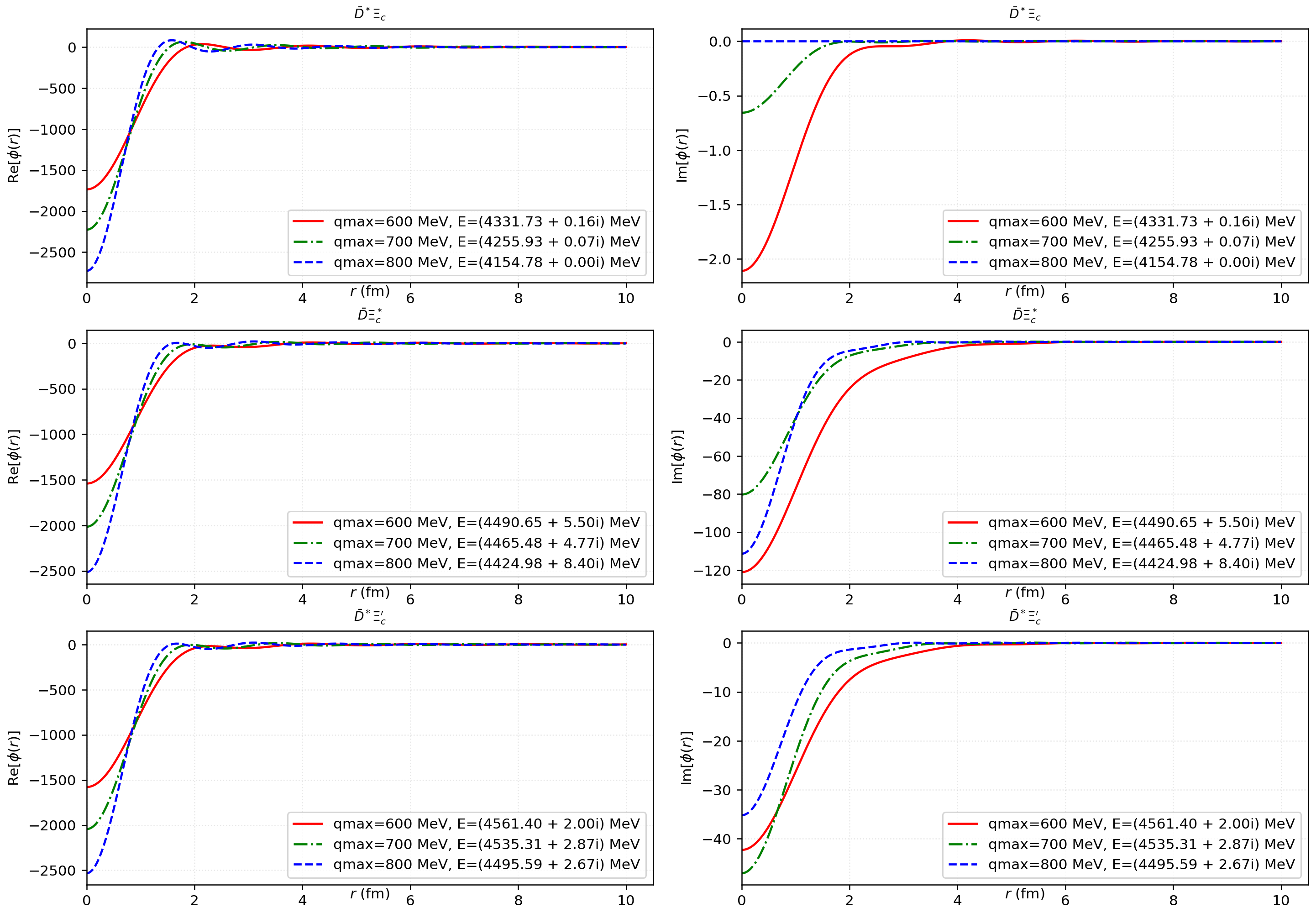}
\includegraphics[width=0.8\textwidth,height=0.1\textheight]{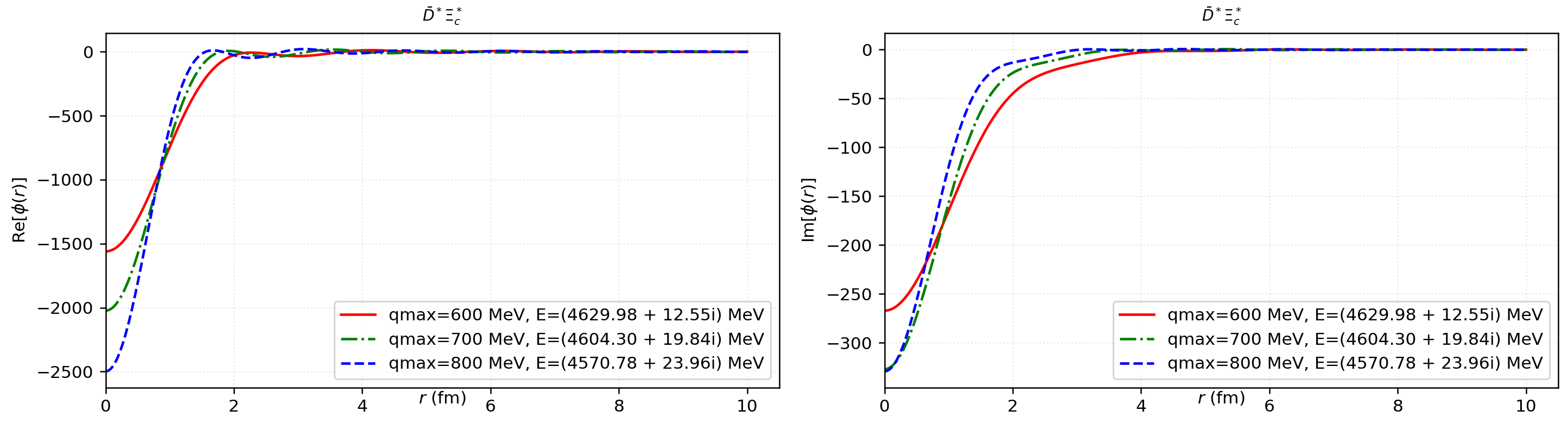}
\caption{Real (left) and imaginary (right) parts of the wave functions $\phi(r)$ of corresponding pole for the $I = 0,\, J^P = 3/2^-$ sector with different $q_{max}$ in the six coupled-channel case.}
\label{fig:wfI0J3cs6}
\end{figure}

In Fig.~\ref{fig:wfI0J3cs6}, we present the results of the wave functions $\phi(\vec{r})$ of the four primary pole components in the six coupled-channel case with different cutoffs. 
From Fig.~\ref{fig:wfI0J3cs6}, as found in the $I = 0,\, J^P = 1/2^-$ sector above, the wave functions are mainly distributed within $0 \sim 6 \text{ fm}$ and go to zero rapidly after $r > 4 \text{ fm}$. 
It should be mentioned that the results with only splitting VB sector are similar, and thus, not shown them here.

\begin{figure}[htbp]
\centering
\includegraphics[width=0.5\textwidth]{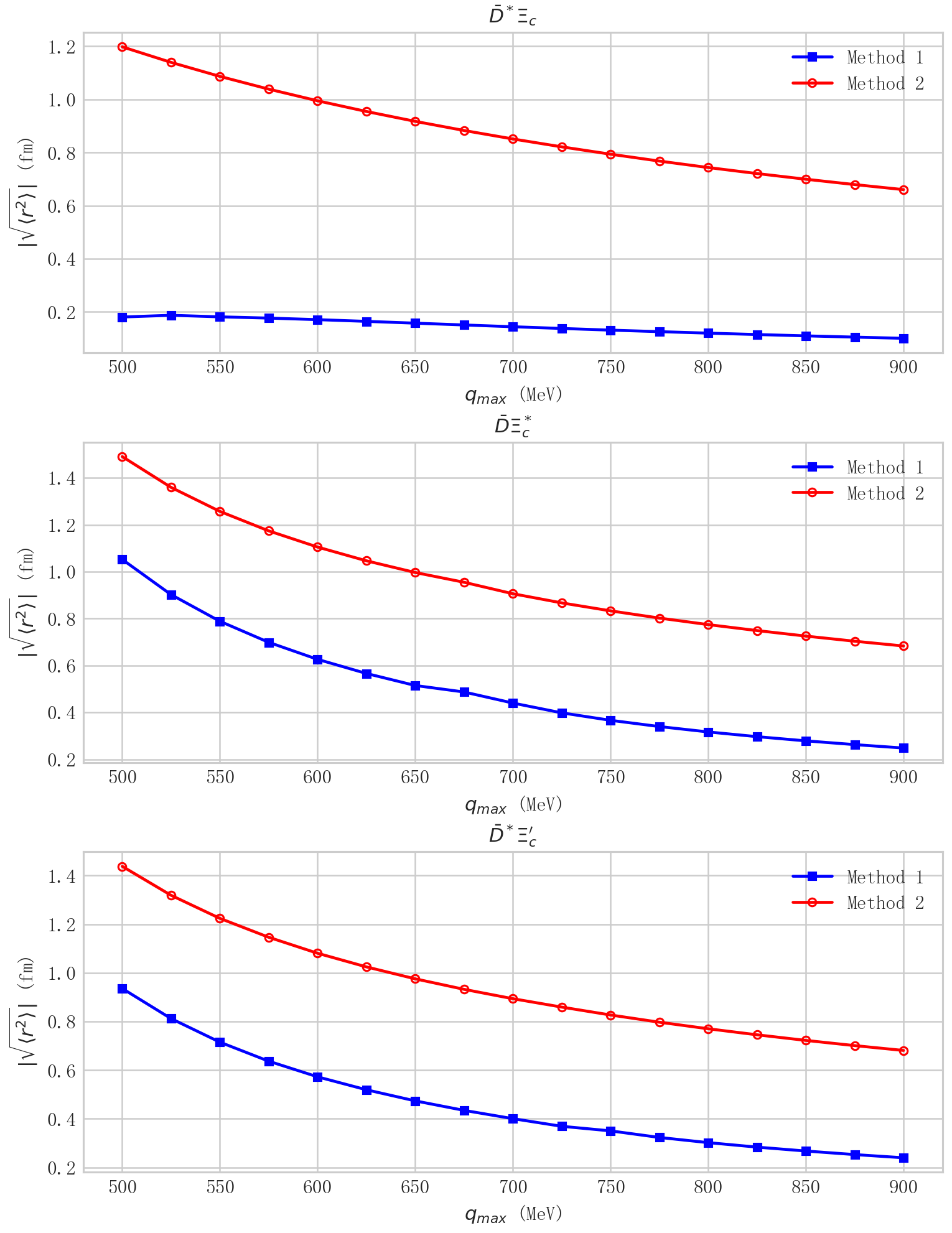}
\includegraphics[width=0.5\textwidth]{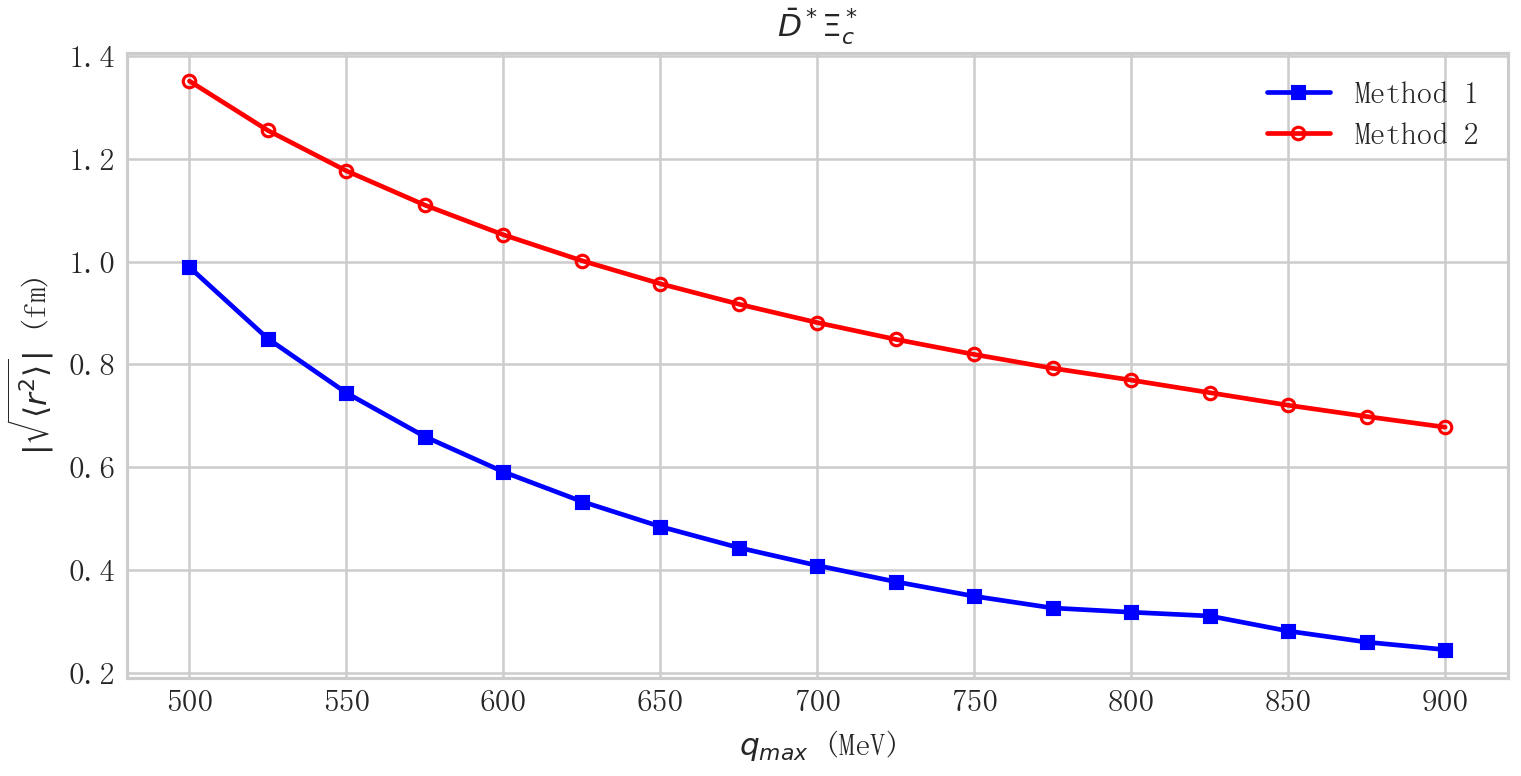}
\caption{RMS radii of the corresponding poles for the $I=0,\, J^P = 3/2^-$ sector as a function of the cutoff $q_{max}$ in the six coupled-channel case. Results from Method 1 (blue) and Method 2 (red) are compared.}
\label{fig:rmsI0J3cs6}
\end{figure}

Then, using the wave functions calculated, we further evaluate the radii of these poles. 
The results of the RMS radii with varying the cutoffs are shown in Fig.~\ref{fig:rmsI0J3cs6}, some of which are shown in Table~\ref{tab:rmsI0J3cs6}. 
From Fig.~\ref{fig:rmsI0J3cs6}, it is found that the results with two methods are not much different and the results of Method 2 are more stable. 
As found in the $I = 0,\, J^P = 1/2^-$ sector, most of the radii are less than 1.5 fm, and the results of Method 1 for the $\bar{D}^* \Xi_c$ are smaller than the others for its deep bound with large binding energy. 
Note that, the similar results of the splitting VB sector are not shown any more, where only the radii of the $\bar{D}^* \Xi_c$ pole with Method 1 are a bit smaller. 
\begin{table}[ht]
\centering
\caption{RMS radii $\left| \sqrt{\langle r^2 \rangle} \right|$ of the corresponding poles for the $I=0,\, J^P = 3/2^-$ sector using different methods in the six coupled-channel case.}
\label{tab:rmsI0J3cs6}
\begin{flushleft}
The radii of states calculated with Eq.~\eqref{eq:msr1}
\end{flushleft}
\footnotesize
\renewcommand{\arraystretch}{0.7}
\setlength{\tabcolsep}{2pt}
\begin{tabular}{ccccccc}
\hline\hline
Resonances & $q_{max} = 600 \, MeV$ & $\left| \sqrt{\langle r^2 \rangle} \right|_2$ & $q_{max} = 700 \, MeV$ & $\left| \sqrt{\langle r^2 \rangle} \right|_2$ & $q_{max} = 800 \, MeV$ & $\left| \sqrt{\langle r^2 \rangle} \right|_2$ \\ \hline
$\bar{D}^*\Xi_c$ & $0.995 - 0.000i \text{ fm}$ & $0.995 \text{ fm}$ & $0.851 - 0.000i \text{ fm}$ & $0.851 \text{ fm}$ & $0.744 + 0.000i \text{ fm}$ & $0.744 \text{ fm}$ \\
$\bar{D}\Xi_c^*$ & $1.104 + 0.048i \text{ fm}$ & $1.105 \text{ fm}$ & $0.906 + 0.011i \text{ fm}$ & $0.906 \text{ fm}$ & $0.774 + 0.007i \text{ fm}$ & $0.774 \text{ fm}$ \\
$\bar{D}^*\Xi_c'$ & $1.081 + 0.013i \text{ fm}$ & $1.081 \text{ fm}$ & $0.894 + 0.005i \text{ fm}$ & $0.894 \text{ fm}$ & $0.770 + 0.002i \text{ fm}$ & $0.770 \text{ fm}$ \\
$\bar{D}^*\Xi_c^*$ & $1.050 + 0.073i \text{ fm}$ & $1.052 \text{ fm}$ & $0.881 + 0.034i \text{ fm}$ & $0.881 \text{ fm}$ & $0.769 + 0.019i \text{ fm}$ & $0.769 \text{ fm}$ \\ 
\hline\hline
\end{tabular}
\begin{flushleft}
The radii of states calculated with Eq.~\eqref{eq:msr2}
\end{flushleft}
\begin{tabular}{ccccccc}
\hline\hline
Resonances & $q_{max} = 600 \, MeV$ & $\left| \sqrt{\langle r^2 \rangle} \right|_1$ & $q_{max} = 700 \, MeV$ & $\left| \sqrt{\langle r^2 \rangle} \right|_1$ & $q_{max} = 800 \, MeV$ & $\left| \sqrt{\langle r^2 \rangle} \right|_1$ \\ \hline
$\bar{D}^*\Xi_c$ & $0.171 + 0.001i \text{ fm}$ & $0.171 \text{ fm}$ & $0.144 + 0.000i \text{ fm}$ & $0.144 \text{ fm}$ & $0.120 + 0.000i \text{ fm}$ & $0.120 \text{ fm}$ \\
$\bar{D}\Xi_c^*$ & $0.626 + 0.002i \text{ fm}$ & $0.626 \text{ fm}$ & $0.440 - 0.003i \text{ fm}$ & $0.440 \text{ fm}$ & $0.317 + 0.001i \text{ fm}$ & $0.317 \text{ fm}$ \\
$\bar{D}^*\Xi_c'$ & $0.573 + 0.000i \text{ fm}$ & $0.573 \text{ fm}$ & $0.401 - 0.002i \text{ fm}$ & $0.401 \text{ fm}$ & $0.302 - 0.002i \text{ fm}$ & $0.302 \text{ fm}$ \\
$\bar{D}^*\Xi_c^*$ & $0.591 + 0.008i \text{ fm}$ & $0.591 \text{ fm}$ & $0.408 + 0.007i \text{ fm}$ & $0.408 \text{ fm}$ & $0.316 + 0.032i \text{ fm}$ & $0.317 \text{ fm}$ \\  
\hline\hline
\end{tabular}
\end{table}

\subsection{Results of single-channel interactions}

In the last two subsections, firstly we utilize the coupled-channel formalism for the hidden charm system, where the main bound channels in the $I = 1/2, J^P = 1/2^-$ sector ($\bar{D}\Sigma_c$, $\bar{D}^*\Sigma_c$, $\bar{D}^*\Sigma_c^*$) and in the $I = 1/2, J^P = 3/2^-$ sector ($\bar{D}\Sigma_c^*$, $\bar{D}^*\Sigma_c$, $\bar{D}^*\Sigma_c^*$) are studied in detail. 
For the hidden charm strange system, the main bound channels in the $I = 0, J^P = 1/2^-$ sector ($\bar{D}\Xi_c$, $\bar{D}^*\Xi_c$, $\bar{D}\Xi'_c$, $\bar{D}^*\Xi'_c$, $\bar{D}^*\Xi^*_c$) and in the $I = 0, J^P = 3/2^-$ sector ($\bar{D}^*\Xi_c$, $\bar{D}\Xi^*_c$, $\bar{D}^*\Xi'_c$, $\bar{D}^*\Xi^*_c$) are also investigated. 
The strong interactions of these channels and their coupled channels with/without the HQSS constraint are studied carefully. 
In this subsection, we make further investigation on the properties of these bound systems with the single channel interaction to check the coupled channel effect in detail. 

\begin{figure}[htbp]
\centering
\includegraphics[width=0.8\textwidth]{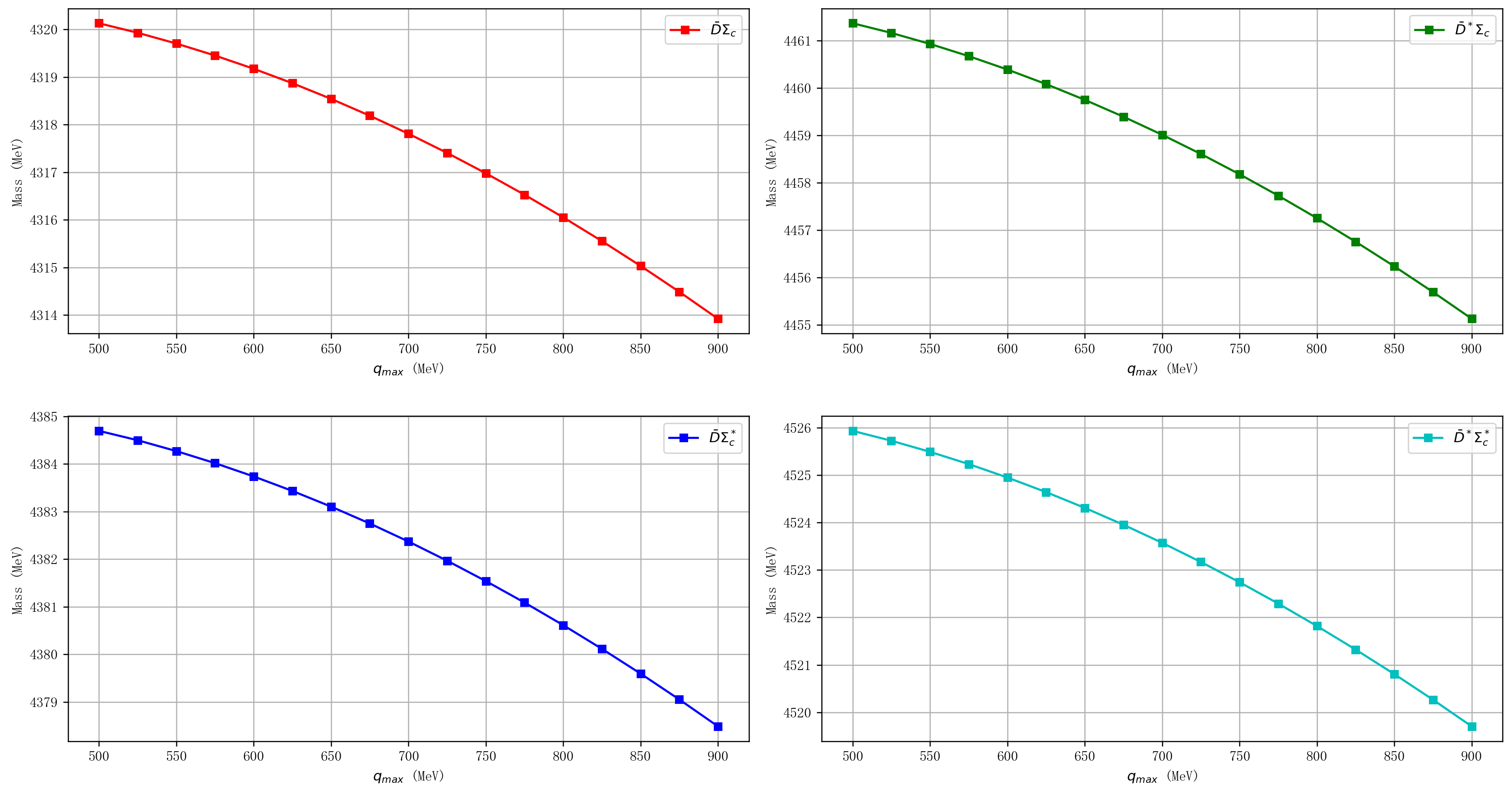}
\caption{Trajectories for the masses of the poles in the first Riemann sheet for the single channel interactions of the hidden charm system.}
\label{fig:reshc1}
\end{figure}

First, we investigate the hidden charm system. Note that, for the single channel interaction, the isospin and spin-parity are not necessary to specify for different sectors. As one can see from the coupled channel interactions before and the interaction potentials of Table~\ref{tab:I1J1c7}, in the hidden charm system, there are only four bound channels, $\bar{D}\Sigma_c$, $\bar{D}\Sigma_c^*$, $\bar{D}^*\Sigma_c$, $\bar{D}^*\Sigma_c^*$, where the trajectories of the corresponding poles are shown in Fig.~\ref{fig:reshc1}. 
As seen from Fig.~\ref{fig:reshc1}, the masses of the four poles show a monotonic downward trend as $q_{max}$ increases. 
Note that, all these poles locate on the real axis of the first Riemann sheet below the corresponding thresholds, becoming pure bound states with zero decay width, since there is no coupled channel to decay for the single channel interaction. 
Moreover, one also can easily find that the binding energies of all these poles are similar, and the mass differences of these poles are nearly the same, about 6 MeV, for the cutoff $q_{max}$ varying from 500 MeV to 900 MeV, due the similar interaction potentials, see the value of $\mu_3$ in Eq.~\eqref{eq:lechc}. 
When the coupled channel effect taken into account, the poles' trajectories are quite different, as found from the results before.

\begin{figure}[htbp]
\centering
\includegraphics[width=0.8\textwidth]{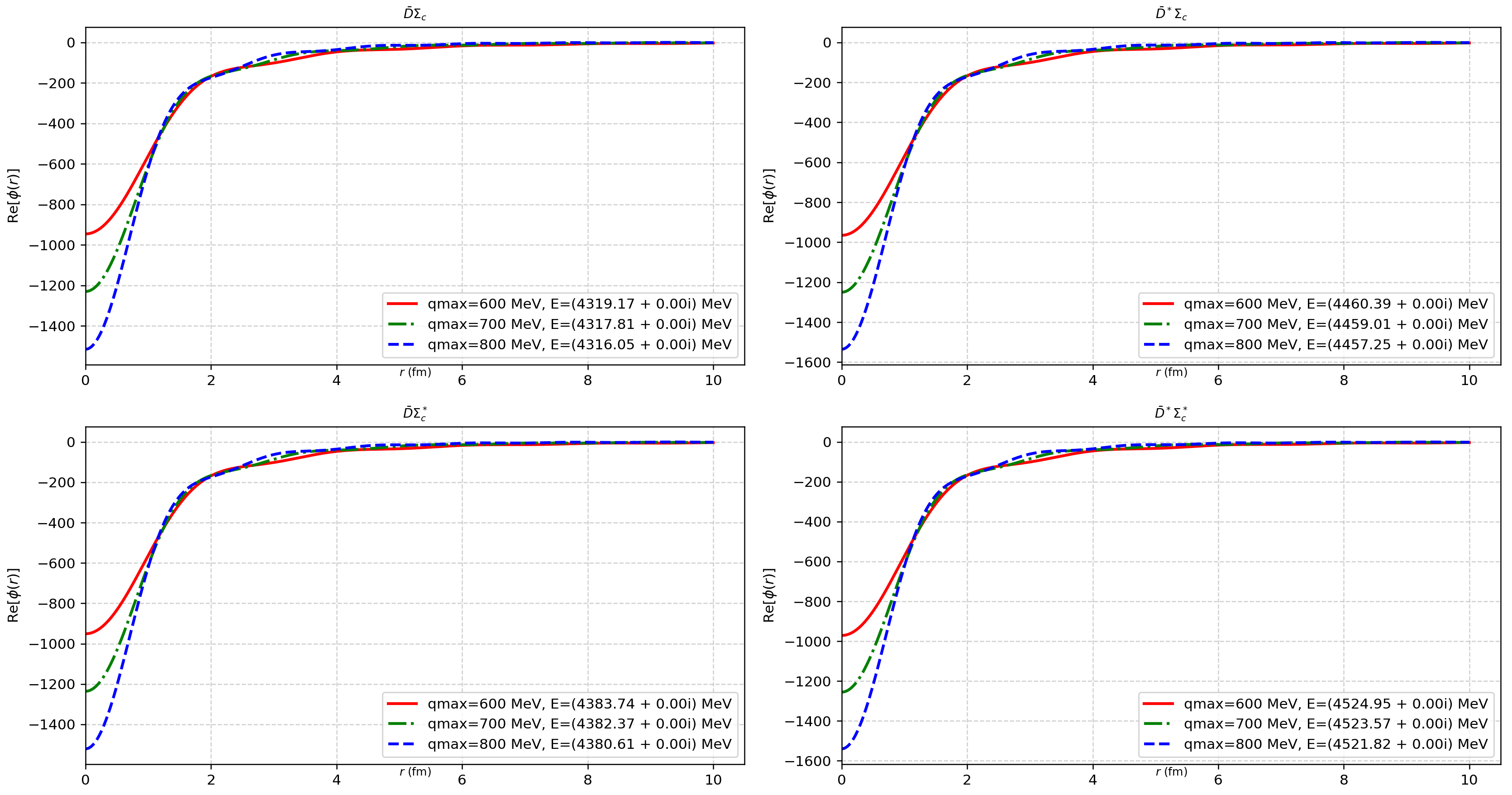}
\caption{Wave functions $\phi(r)$ of corresponding pole for the single channel interactions of the hidden charm system with different $q_{max}$.}
\label{fig:wfhc1}
\end{figure}

Next, the results of the wave functions $\phi(\vec{r})$ of the four poles are presented in Fig.~\ref{fig:wfhc1} with different cutoffs. 
Note that, now the wave functions $\phi(\vec{r})$ are also real due to the real bound poles with no width, see the results of Fig.~\ref{fig:reshc1}. 
From Fig.~\ref{fig:wfhc1}, the wave functions are distributed within $0 \sim 6 \text{ fm}$ and go to zero rapidly after $r > 4 \text{ fm}$ as normal. 
Furthermore, one also see that the line shapes of these wave functions are similar for the same interaction dynamics just under different channels. 

\begin{figure}[htbp]
\centering
\includegraphics[width=0.8\textwidth]{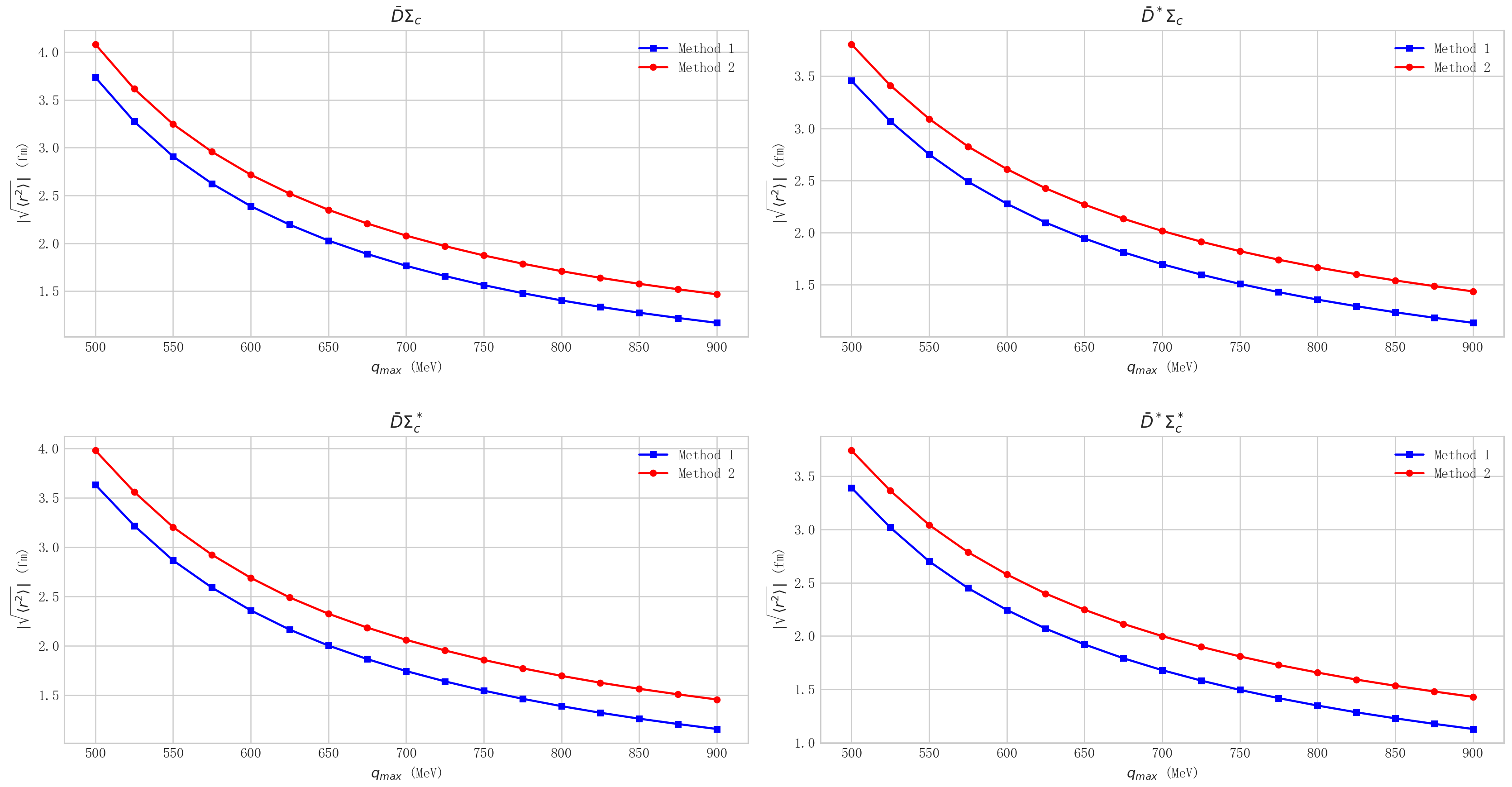}
\caption{RMS radii of the corresponding poles for the single channel interactions of the hidden charm system as a function of the cutoff $q_{max}$. Results from Method 1 (blue) and Method 2 (red) are compared.}
\label{fig:rmshc1}
\end{figure}

With the wave functions calculated, the radii of these poles are obtained, where the results of the RMS radii with varying the cutoffs are shown in Fig.~\ref{fig:rmshc1}. 
As shown in Fig.~\ref{fig:rmshc1}, unlike the unstable fluctuations observed in the coupled channel cases above, the curves of Method 1 and Method 2 for the poles of the single-channel interactions appear extremely smooth and monotonic, since all of these poles are pure bound states with zero width below the corresponding threshold. 
Thus, the results with two methods are not much different and their line shapes are similar, where only the values reduce a little bit for the heavier channels. 
One also can see that all these radii are in the range from about 4 fm deceasing to 1 fm for the varying cutoffs, bigger than what we have in the coupled channel cases. 

\begin{figure}[htbp]
\centering
\includegraphics[width=0.8\textwidth]{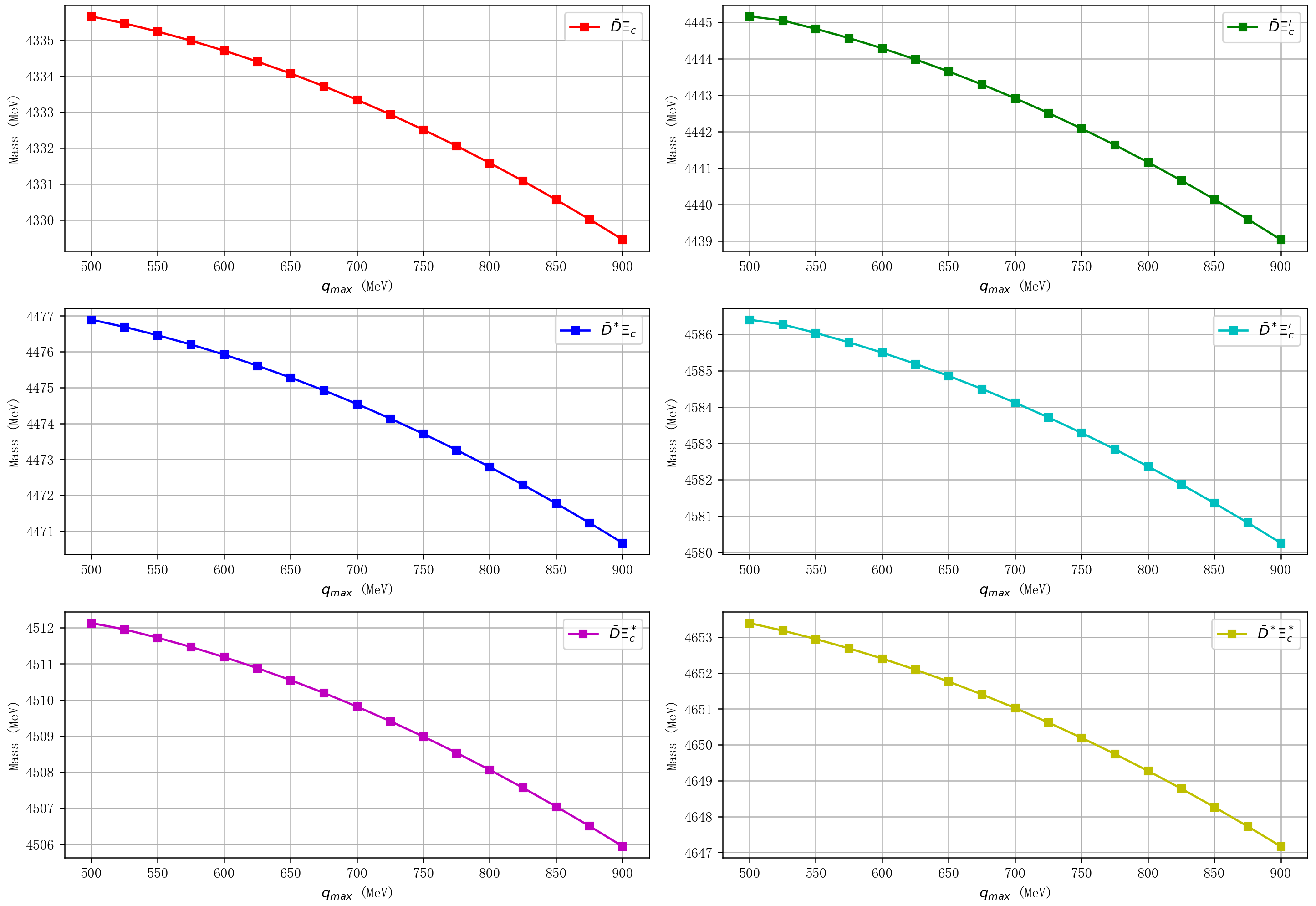}
\caption{Trajectories for the masses of the poles in the first Riemann sheet for the single channel interactions of the hidden charm strange system.}
\label{fig:reshcs1}
\end{figure}

Second, we study the hidden charm strange system. In this system, there are six bound channels as found in the coupled channel cases and seen from the interaction potentials, see Table~\ref{tab:I0J1cs9}, which are the channels $\bar{D}\Xi_c$, $\bar{D}\Xi'_c$, $\bar{D}^*\Xi_c$, $\bar{D}^*\Xi'_c$, $\bar{D} \Xi^*_c$, and $\bar{D}^*\Xi^*_c$. 
The trajectories of their poles in the first Riemann sheet are shown in Fig.~\ref{fig:reshcs1}, which are analogous to the results of the hidden charm sector above with zero width, and also have 6 MeV differences of the masses for these poles when the cutoff varying. 
Indeed, the binding energies of these poles are similar too, for the same potentials, see the values of $\mu_2$, $\mu_4$ and $\lambda$ in Eq.~\eqref{eq:lechcs}. 

\begin{figure}[htbp]
\centering
\includegraphics[width=0.8\textwidth]{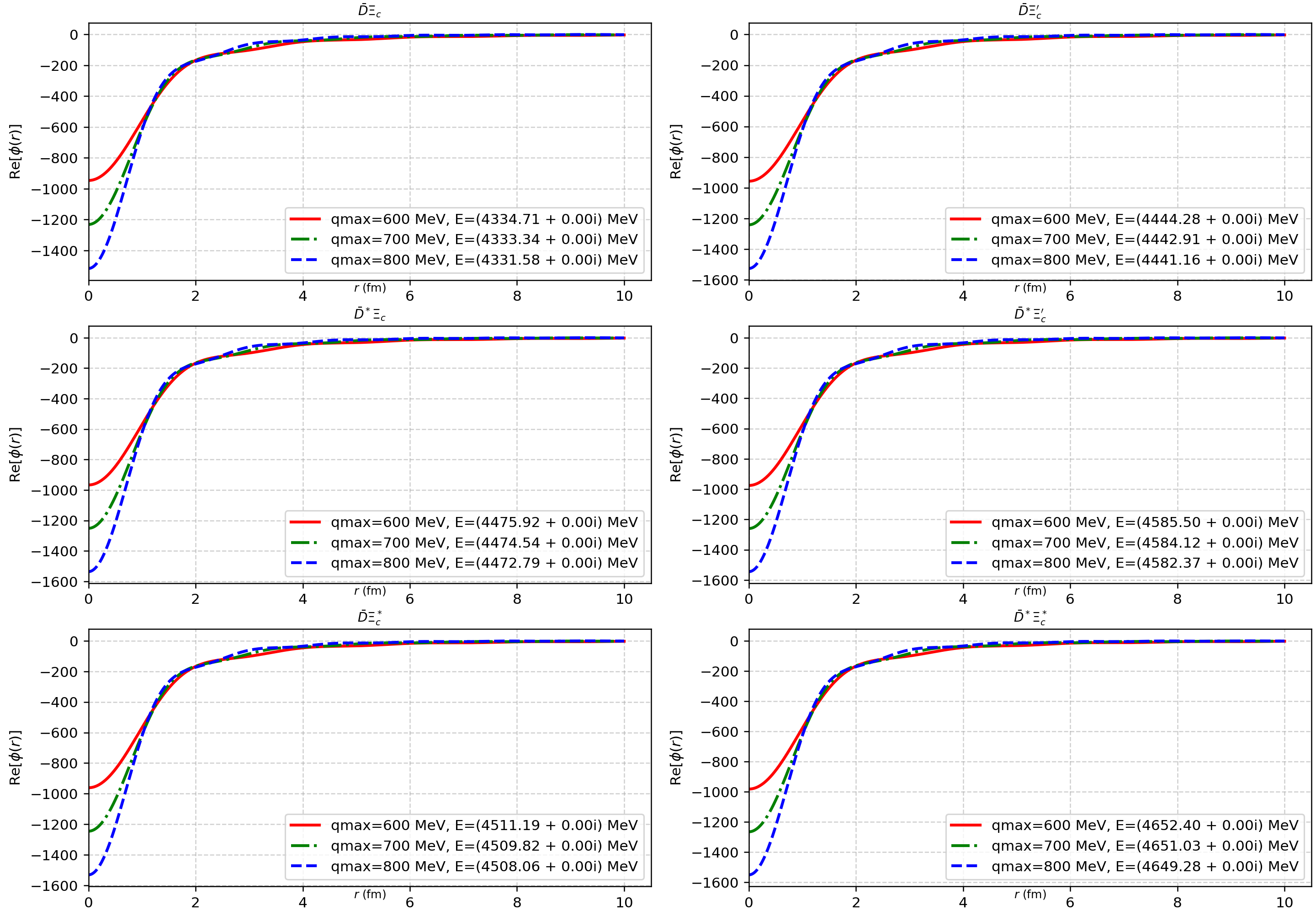}
\caption{Wave functions $\phi(r)$ of corresponding pole for the single channel interactions of the hidden charm strange system with different $q_{max}$.}
\label{fig:wfhcs1}
\end{figure}

Then, we show the results of the wave functions $\phi(\vec{r})$ of the six poles in Fig.~\ref{fig:wfhcs1} with different cutoffs, which are real too for the pure bound states with zero width. 
From the results of Fig.~\ref{fig:wfhcs1}, one can see that these wave functions go to zero quickly after $r > 4 \text{ fm}$ as the others above. 

\begin{figure}[htbp]
\centering
\includegraphics[width=0.8\textwidth]{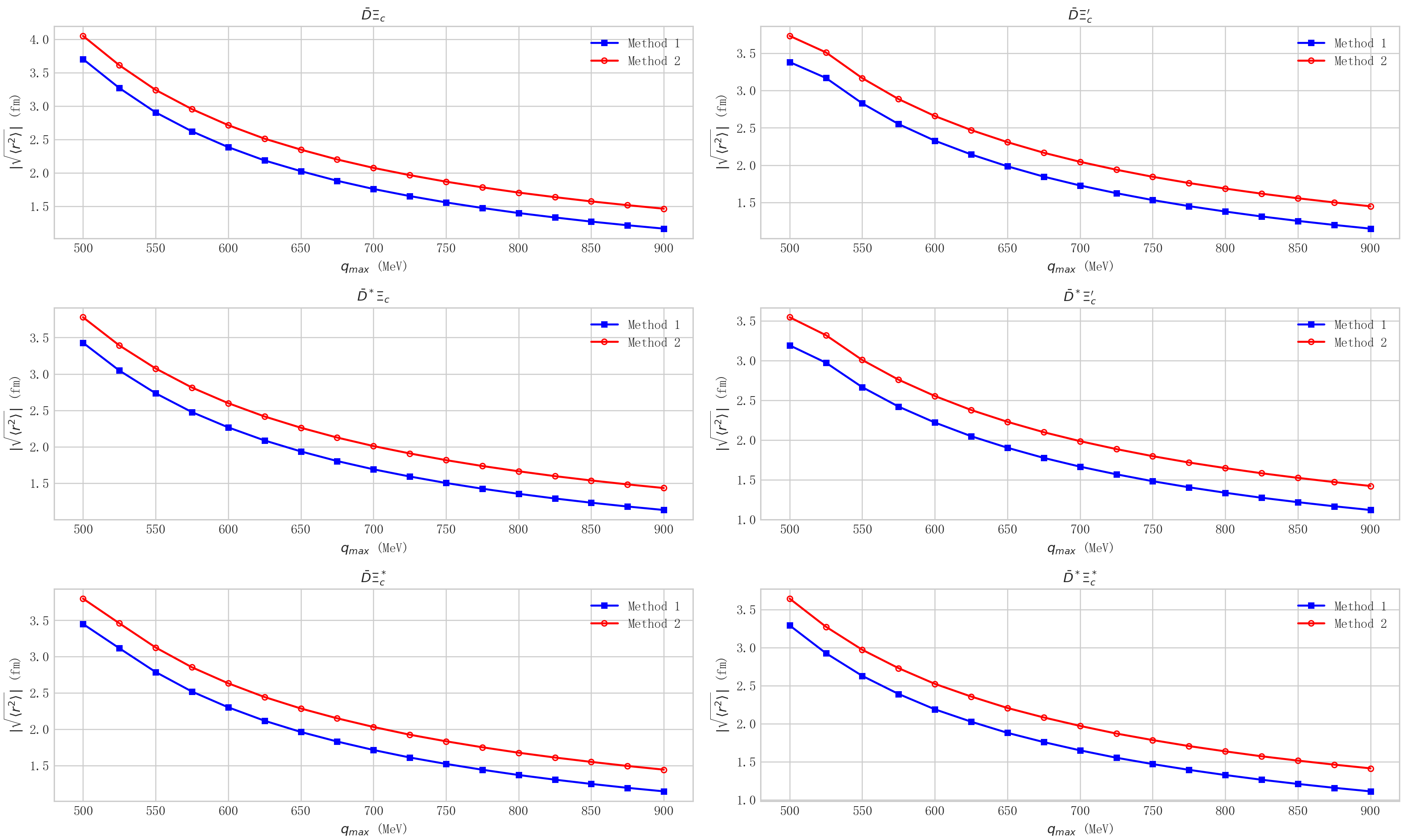}
\caption{RMS radii of the corresponding poles for the single channel interactions of the hidden charm strange system as a function of the cutoff $q_{max}$. Results from Method 1 (blue) and Method 2 (red) are compared.}
\label{fig:rmshcs1}
\end{figure}

Furthermore, the radii of these poles are shown in Fig.~\ref{fig:rmshcs1}, where the results of two methods are consistent with each other. From Fig.~\ref{fig:rmshcs1}, it can be found that all of these radii are around the range of 1 to 4 fm, which are also larger than the results of the coupled channel cases before and consistent with the results obtained in the hidden charm system, see the results of Fig.~\ref{fig:rmshc1}.  

\section{Conclusions}

In the present work, we systematically investigate the meson-baryon interactions for the molecular properties of the pentaquark states $P_c$ and $P_{cs}$ in the hidden charm and hidden charm strange systems using a coupled channel approach, based on our former work that combined the heavy quark spin symmetry and the local hidden gauge formalism. 
By solving the Bethe-Salpeter equation with the momentum cutoff method, we obtain the poles' trajectories, wave functions, and root-mean-square radii. 

For both of the hidden charm and hidden charm strange systems, we investigate the interactions of the full coupled channel systems under the constraint of the heavy quark spin symmetry, the splitting PB and VB sectors, and the single channels for varying the cutoffs, to understand more about the properties of the bound systems. 
To generate the $P_c$ states, the full coupled channel interactions with the heavy quark spin symmetry are important, which affect seriously the widths of the corresponding poles. 
Indeed, in the hidden charm system, the main bound channels are the ones $\bar{D} \Sigma_c$, $\bar{D}^* \Sigma_c$, which couple to the lower decay channels strongly too. 
Whereas, it is not so necessary to reproduce the $P_{cs}$ state using the full coupled channel interactions with the heavy quark spin symmetry, showing that the results of the splitting PB and VB sectors are not much differences. 
For the hidden charm strange system, the main bound channels $\bar{D} \Xi_c$, $\bar{D}^* \Xi_c$ couple strongly to the $\bar{D}_s \Lambda_c$, $\bar{D}_s^* \Lambda_c$, respectively, and not much to the lower decay channels, which are different from the case of the hidden charm system. 
The widths of the corresponding poles for the loose bound channels, $\bar{D} \Xi'_c$, $\bar{D}^* \Xi'_c$ and $\bar{D}^* \Xi_c^*$ exhibit different behaviours of the trajectories. 
Note that, all the main bound channels have similar binding energies for the single channel interactions, since they have the same attractive interaction potentials. 

Furthermore, we also calculated the wave functions and root-mean-square radii of the corresponding poles. 
The wave functions exhibit the effective range around $0\sim 6$ fm and go to zero fast for $r > 4$ fm. 
We use two methods to evaluate the root-mean-square radii, the results of which are consistent with each other in most of the cases. 
The root-mean-square radii are mostly typical size between $0.5 \sim 2$ fm, which is comparable to the characteristic scale of the molecular states. 
Indeed, the root-mean-square radii are dependent on the pole trajectories, and have different results for the full coupled channel case, the splitting PB and VB sectors, and the single channel interactions. From these results, we can understand more about these bound hidden charm and hidden charm strange systems to hint the molecular nature of the $P_c$ and $P_{cs}$ states.

\section*{Acknowledgements}

We acknowledge Profs. Kazem Azizi and Mao-Jun Yan for useful comments. 
This work is supported by the Natural Science Foundation of Guangxi province under Grant No. 2023JJA110076, the Natural Science Foundation of Hunan province under Grant No. 2023JJ30647, and the National Natural Science Foundation of China under Grants No. 12365019 and No. 12575081.

\bibliography{paper}

@article{Gell-Mann:1964ewy,
    author = "Gell-Mann, Murray",
    title = "{A Schematic Model of Baryons and Mesons}",
    doi = "10.1016/S0031-9163(64)92001-3",
    journal = "Phys. Lett.",
    volume = "8",
    pages = "214--215",
    year = "1964"
}

@article{Zweig:1964ruk,
    author = "Zweig, G.",
    title = "{An SU(3) model for strong interaction symmetry and its breaking. Version 1}",
    reportNumber = "CERN-TH-401",
    doi = "10.17181/CERN-TH-401",
    month = "1",
    year = "1964"
}

@article{Klempt:2009pi,
    author = "Klempt, Eberhard and Richard, Jean-Marc",
    title = "{Baryon spectroscopy}",
    eprint = "0901.2055",
    archivePrefix = "arXiv",
    primaryClass = "hep-ph",
    doi = "10.1103/RevModPhys.82.1095",
    journal = "Rev. Mod. Phys.",
    volume = "82",
    pages = "1095--1153",
    year = "2010"
}

@article{Richard:2016eis,
    author = "Richard, Jean-Marc",
    title = "{Exotic hadrons: review and perspectives}",
    eprint = "1606.08593",
    archivePrefix = "arXiv",
    primaryClass = "hep-ph",
    doi = "10.1007/s00601-016-1159-0",
    journal = "Few Body Syst.",
    volume = "57",
    number = "12",
    pages = "1185--1212",
    year = "2016"
}

@article{Esposito:2016noz,
    author = "Esposito, A. and Pilloni, A. and Polosa, A. D.",
    title = "{Multiquark Resonances}",
    eprint = "1611.07920",
    archivePrefix = "arXiv",
    primaryClass = "hep-ph",
    reportNumber = "JLAB-THY-16-2301",
    doi = "10.1016/j.physrep.2016.11.002",
    journal = "Phys. Rept.",
    volume = "668",
    pages = "1--97",
    year = "2017"
}

@article{Guo:2017jvc,
    author = "Guo, Feng-Kun and Hanhart, Christoph and Mei{\ss}ner, Ulf-G. and Wang, Qian and Zhao, Qiang and Zou, Bing-Song",
    title = "{Hadronic molecules}",
    eprint = "1705.00141",
    archivePrefix = "arXiv",
    primaryClass = "hep-ph",
    doi = "10.1103/RevModPhys.90.015004",
    journal = "Rev. Mod. Phys.",
    volume = "90",
    number = "1",
    pages = "015004",
    year = "2018",
    note = "[Erratum: Rev.Mod.Phys. 94, 029901 (2022)]"
}

@article{Karliner:2017qhf,
    author = "Karliner, Marek and Rosner, Jonathan L. and Skwarnicki, Tomasz",
    title = "{Multiquark States}",
    eprint = "1711.10626",
    archivePrefix = "arXiv",
    primaryClass = "hep-ph",
    doi = "10.1146/annurev-nucl-101917-020902",
    journal = "Ann. Rev. Nucl. Part. Sci.",
    volume = "68",
    pages = "17--44",
    year = "2018"
}

@article{Chen:2016spr,
    author = "Chen, Hua-Xing and Chen, Wei and Liu, Xiang and Liu, Yan-Rui and Zhu, Shi-Lin",
    title = "{A review of the open charm and open bottom systems}",
    eprint = "1609.08928",
    archivePrefix = "arXiv",
    primaryClass = "hep-ph",
    doi = "10.1088/1361-6633/aa6420",
    journal = "Rept. Prog. Phys.",
    volume = "80",
    number = "7",
    pages = "076201",
    year = "2017"
}

@article{Lebed:2016hpi,
    author = "Lebed, Richard F. and Mitchell, Ryan E. and Swanson, Eric S.",
    title = "{Heavy-Quark QCD Exotica}",
    eprint = "1610.04528",
    archivePrefix = "arXiv",
    primaryClass = "hep-ph",
    doi = "10.1016/j.ppnp.2016.11.003",
    journal = "Prog. Part. Nucl. Phys.",
    volume = "93",
    pages = "143--194",
    year = "2017"
}

@article{Ali:2017jda,
    author = {Ali, Ahmed and Lange, Jens S{\"o}ren and Stone, Sheldon},
    title = "{Exotics: Heavy Pentaquarks and Tetraquarks}",
    eprint = "1706.00610",
    archivePrefix = "arXiv",
    primaryClass = "hep-ph",
    reportNumber = "DESY-17-071",
    doi = "10.1016/j.ppnp.2017.08.003",
    journal = "Prog. Part. Nucl. Phys.",
    volume = "97",
    pages = "123--198",
    year = "2017"
}

@article{Olsen:2017bmm,
    author = "Olsen, Stephen Lars and Skwarnicki, Tomasz and Zieminska, Daria",
    title = "{Nonstandard heavy mesons and baryons: Experimental evidence}",
    eprint = "1708.04012",
    archivePrefix = "arXiv",
    primaryClass = "hep-ph",
    doi = "10.1103/RevModPhys.90.015003",
    journal = "Rev. Mod. Phys.",
    volume = "90",
    number = "1",
    pages = "015003",
    year = "2018"
}

@article{Yuan:2018inv,
    author = "Yuan, Chang-Zheng",
    title = "{The XYZ states revisited}",
    eprint = "1808.01570",
    archivePrefix = "arXiv",
    primaryClass = "hep-ex",
    doi = "10.1142/S0217751X18300181",
    journal = "Int. J. Mod. Phys. A",
    volume = "33",
    number = "21",
    pages = "1830018",
    year = "2018"
}

@article{Liu:2019zoy,
    author = "Liu, Yan-Rui and Chen, Hua-Xing and Chen, Wei and Liu, Xiang and Zhu, Shi-Lin",
    title = "{Pentaquark and Tetraquark states}",
    eprint = "1903.11976",
    archivePrefix = "arXiv",
    primaryClass = "hep-ph",
    doi = "10.1016/j.ppnp.2019.04.003",
    journal = "Prog. Part. Nucl. Phys.",
    volume = "107",
    pages = "237--320",
    year = "2019"
}

@article{Brambilla:2019esw,
    author = "Brambilla, Nora and Eidelman, Simon and Hanhart, Christoph and Nefediev, Alexey and Shen, Cheng-Ping and Thomas, Christopher E. and Vairo, Antonio and Yuan, Chang-Zheng",
    title = "{The $XYZ$ states: experimental and theoretical status and perspectives}",
    eprint = "1907.07583",
    archivePrefix = "arXiv",
    primaryClass = "hep-ex",
    reportNumber = "TUM-EFT 125/19",
    doi = "10.1016/j.physrep.2020.05.001",
    journal = "Phys. Rept.",
    volume = "873",
    pages = "1--154",
    year = "2020"
}

@article{Chen:2019asm,
    author = "Chen, Rui and Sun, Zhi-Feng and Liu, Xiang and Zhu, Shi-Lin",
    title = "{Strong LHCb evidence supporting the existence of the hidden-charm molecular pentaquarks}",
    eprint = "1903.11013",
    archivePrefix = "arXiv",
    primaryClass = "hep-ph",
    doi = "10.1103/PhysRevD.100.011502",
    journal = "Phys. Rev. D",
    volume = "100",
    number = "1",
    pages = "011502",
    year = "2019"
}

@article{Liu:2019tjn,
    author = "Liu, Ming-Zhu and Pan, Ya-Wen and Peng, Fang-Zheng and S{\'a}nchez S{\'a}nchez, Mario and Geng, Li-Sheng and Hosaka, Atsushi and Pavon Valderrama, Manuel",
    title = "{Emergence of a complete heavy-quark spin symmetry multiplet: seven molecular pentaquarks in light of the latest LHCb analysis}",
    eprint = "1903.11560",
    archivePrefix = "arXiv",
    primaryClass = "hep-ph",
    doi = "10.1103/PhysRevLett.122.242001",
    journal = "Phys. Rev. Lett.",
    volume = "122",
    number = "24",
    pages = "242001",
    year = "2019"
}

@article{Xiao:2019mvs,
    author = "Xiao, Cheng-Jian and Huang, Yin and Dong, Yu-Bing and Geng, Li-Sheng and Chen, Dian-Yong",
    title = "{Exploring the molecular scenario of Pc(4312) , Pc(4440) , and Pc(4457)}",
    eprint = "1904.00872",
    archivePrefix = "arXiv",
    primaryClass = "hep-ph",
    doi = "10.1103/PhysRevD.100.014022",
    journal = "Phys. Rev. D",
    volume = "100",
    number = "1",
    pages = "014022",
    year = "2019"
}

@article{Guo:2019kdc,
    author = "Guo, Zhi-Hui and Oller, J. A.",
    title = "{Anatomy of the newly observed hidden-charm pentaquark states: $P_c(4312)$, $P_c(4440)$ and $P_c(4457)$}",
    eprint = "1904.00851",
    archivePrefix = "arXiv",
    primaryClass = "hep-ph",
    doi = "10.1016/j.physletb.2019.04.053",
    journal = "Phys. Lett. B",
    volume = "793",
    pages = "144--149",
    year = "2019"
}

@article{Xiao:2019aya,
    author = "Xiao, C. W. and Nieves, J. and Oset, E.",
    title = "{Heavy quark spin symmetric molecular states from ${\bar D}^{(*)}\Sigma_c^{(*)}$ and other coupled channels in the light of the recent LHCb pentaquarks}",
    eprint = "1904.01296",
    archivePrefix = "arXiv",
    primaryClass = "hep-ph",
    doi = "10.1103/PhysRevD.100.014021",
    journal = "Phys. Rev. D",
    volume = "100",
    number = "1",
    pages = "014021",
    year = "2019"
}

@article{Burns:2019iih,
    author = "Burns, T. J. and Swanson, E. S.",
    title = "{Molecular interpretation of the $P_c$(4440) and $P_c$(4457) states}",
    eprint = "1908.03528",
    archivePrefix = "arXiv",
    primaryClass = "hep-ph",
    doi = "10.1103/PhysRevD.100.114033",
    journal = "Phys. Rev. D",
    volume = "100",
    number = "11",
    pages = "114033",
    year = "2019"
}

@article{Zhu:2019iwm,
    author = "Zhu, Ruilin and Liu, Xuejie and Huang, Hongxia and Qiao, Cong-Feng",
    title = "{Analyzing doubly heavy tetra- and penta-quark states by variational method}",
    eprint = "1904.10285",
    archivePrefix = "arXiv",
    primaryClass = "hep-ph",
    doi = "10.1016/j.physletb.2019.134869",
    journal = "Phys. Lett. B",
    volume = "797",
    pages = "134869",
    year = "2019"
}

@article{Chen:2020uif,
    author = "Chen, Hua-Xing and Chen, Wei and Liu, Xiang and Liu, Xiao-Hai",
    title = "{Establishing the first hidden-charm pentaquark with strangeness}",
    eprint = "2011.01079",
    archivePrefix = "arXiv",
    primaryClass = "hep-ph",
    doi = "10.1140/epjc/s10052-021-09196-4",
    journal = "Eur. Phys. J. C",
    volume = "81",
    number = "5",
    pages = "409",
    year = "2021"
}

@article{Wang:2020eep,
    author = "Wang, Zhi-Gang",
    title = "{Analysis of the $P_{cs}(4459)$ as the hidden-charm pentaquark state with QCD sum rules}",
    eprint = "2011.05102",
    archivePrefix = "arXiv",
    primaryClass = "hep-ph",
    doi = "10.1142/S0217751X21500718",
    journal = "Int. J. Mod. Phys. A",
    volume = "36",
    number = "10",
    pages = "2150071",
    year = "2021"
}

@article{Chen:2020kco,
    author = "Chen, Rui",
    title = "{Can the newly reported $P_{cs}(4459)$ be a strange hidden-charm $\Xi_c\bar D^*$ molecular pentaquark?}",
    eprint = "2011.07214",
    archivePrefix = "arXiv",
    primaryClass = "hep-ph",
    doi = "10.1103/PhysRevD.103.054007",
    journal = "Phys. Rev. D",
    volume = "103",
    number = "5",
    pages = "054007",
    year = "2021"
}

@article{Liu:2020hcv,
    author = "Liu, Ming-Zhu and Pan, Ya-Wen and Geng, Li-Sheng",
    title = "{Can discovery of hidden charm strange pentaquark states help determine the spins of $P_c(4440)$ and $P_c(4457)$ ?}",
    eprint = "2011.07935",
    archivePrefix = "arXiv",
    primaryClass = "hep-ph",
    doi = "10.1103/PhysRevD.103.034003",
    journal = "Phys. Rev. D",
    volume = "103",
    number = "3",
    pages = "034003",
    year = "2021"
}

@article{Zhu:2021lhd,
    author = "Zhu, Jun-Tao and Song, Lin-Qing and He, Jun",
    title = "{$P_{cs}(4459)$ and other possible molecular states from $\Xi_{c}^{(*)}\bar{D}^{(*)}$ and $\Xi'_c\bar{D}^{(*)}$ interactions}",
    eprint = "2101.12441",
    archivePrefix = "arXiv",
    primaryClass = "hep-ph",
    doi = "10.1103/PhysRevD.103.074007",
    journal = "Phys. Rev. D",
    volume = "103",
    number = "7",
    pages = "074007",
    year = "2021"
}

@article{Xiao:2021rgp,
    author = "Xiao, C. W. and Wu, J. J. and Zou, B. S.",
    title = "{Molecular nature of $P_{cs} (4459)$ and its heavy quark spin partners}",
    eprint = "2102.02607",
    archivePrefix = "arXiv",
    primaryClass = "hep-ph",
    doi = "10.1103/PhysRevD.103.054016",
    journal = "Phys. Rev. D",
    volume = "103",
    number = "5",
    pages = "054016",
    year = "2021"
}

@article{Wang:2022neq,
    author = "Wang, Xiu-Wu and Wang, Zhi-Gang",
    title = "{Analysis of P $_{cs}$(4338) and related pentaquark molecular states via QCD sum rules*}",
    eprint = "2207.06060",
    archivePrefix = "arXiv",
    primaryClass = "hep-ph",
    doi = "10.1088/1674-1137/ac9aab",
    journal = "Chin. Phys. C",
    volume = "47",
    number = "1",
    pages = "013109",
    year = "2023"
}

@article{Meng:2022wgl,
    author = "Meng, Lu and Wang, Bo and Zhu, Shi-Lin",
    title = "{Double thresholds distort the line shapes of the P{\ensuremath{\psi}}s{\ensuremath{\Lambda}}(4338)0 resonance}",
    eprint = "2208.03883",
    archivePrefix = "arXiv",
    primaryClass = "hep-ph",
    doi = "10.1103/PhysRevD.107.014005",
    journal = "Phys. Rev. D",
    volume = "107",
    number = "1",
    pages = "014005",
    year = "2023"
}

@article{Zhu:2022wpi,
    author = "Zhu, Jun-Tao and Kong, Shu-Yi and He, Jun",
    title = "{P{\ensuremath{\psi}}s{\ensuremath{\Lambda}}(4459) and P{\ensuremath{\psi}}s{\ensuremath{\Lambda}}(4338) as molecular states in J/{\ensuremath{\psi}}{\ensuremath{\Lambda}} invariant mass spectra}",
    eprint = "2211.06232",
    archivePrefix = "arXiv",
    primaryClass = "hep-ph",
    doi = "10.1103/PhysRevD.107.034029",
    journal = "Phys. Rev. D",
    volume = "107",
    number = "3",
    pages = "034029",
    year = "2023"
}

@article{Feijoo:2022rxf,
    author = "Feijoo, Albert and Wang, Wen-Fei and Xiao, Chu-Wen and Wu, Jia-Jun and Oset, Eulogio and Nieves, Juan and Zou, Bing-Song",
    title = "{A new look at the Pcs states from a molecular perspective}",
    eprint = "2212.12223",
    archivePrefix = "arXiv",
    primaryClass = "hep-ph",
    doi = "10.1016/j.physletb.2023.137760",
    journal = "Phys. Lett. B",
    volume = "839",
    pages = "137760",
    year = "2023"
}

@article{LHCb:2015yax,
    author = "Aaij, Roel and others",
    collaboration = "LHCb",
    title = "{Observation of $J/\psi p$ Resonances Consistent with Pentaquark States in $\Lambda_b^0 \to J/\psi K^- p$ Decays}",
    doi = "10.1103/PhysRevLett.115.072001",
    journal = "Phys. Rev. Lett.",
    volume = "115",
    pages = "072001",
    year = "2015"
}

@article{LHCb:2019kea,
    author = "Aaij, Roel and others",
    collaboration = "LHCb",
    title = "{Observation of a narrow pentaquark state, $P_c(4312)^+$, and of two-peak structure of the $P_c(4450)^+$}",
    doi = "10.1103/PhysRevLett.122.222001",
    journal = "Phys. Rev. Lett.",
    volume = "122",
    pages = "222001",
    year = "2019"
}

@article{LHCb:2020pdt,
    author = "Aaij, Roel and others",
    collaboration = "LHCb",
    title = "{Evidence of a $J/\psi\Lambda$ structure and observation of excited $\Xi^-$ states in the $\Xi_b^- \to J/\psi\Lambda K^-$ decay}",
    doi = "10.1016/j.scib.2021.04.009",
    journal = "Sci. Bull.",
    volume = "66",
    pages = "1278--1287",
    year = "2021",
    eprint = "2012.10380",
    archivePrefix = "arXiv",
    primaryClass = "hep-ex"
}

@article{LHCb:2022ogu,
    author = "Aaij, Roel and others",
    collaboration = "LHCb",
    title = "{Observation of a $J/\psi\Lambda$ Resonance Consistent with a Strange Pentaquark Candidate in $B^- \to J/\psi\Lambda\bar{p}$ Decays}",
    doi = "10.1103/PhysRevLett.131.031901",
    journal = "Phys. Rev. Lett.",
    volume = "131",
    number = "3",
    pages = "031901",
    year = "2023"
}

@article{Du:2019pij,
    author = "Du, Meng-Lin and Baru, Vadim and Guo, Feng-Kun and Hanhart, Christoph and Mei{\ss}ner, Ulf-G and Oller, Jos{\'e} A. and Wang, Qian",
    title = "{Interpretation of the LHCb $P_c$ States as Hadronic Molecules and Hints of a Narrow $P_c(4380)$}",
    eprint = "1910.11846",
    archivePrefix = "arXiv",
    primaryClass = "hep-ph",
    doi = "10.1103/PhysRevLett.124.072001",
    journal = "Phys. Rev. Lett.",
    volume = "124",
    number = "7",
    pages = "072001",
    year = "2020"
}

@article{Dong:2020hxe,
    author = "Dong, Xiang-Kun and Guo, Feng-Kun and Zou, Bing-Song",
    title = "{Explaining the Many Threshold Structures in the Heavy-Quark Hadron Spectrum}",
    eprint = "2011.14517",
    archivePrefix = "arXiv",
    primaryClass = "hep-ph",
    doi = "10.1103/PhysRevLett.126.152001",
    journal = "Phys. Rev. Lett.",
    volume = "126",
    number = "15",
    pages = "152001",
    year = "2021"
}

@article{Dong:2021juy,
    author = "Dong, Xiang-Kun and Guo, Feng-Kun and Zou, Bing-Song",
    title = "{A survey of heavy-antiheavy hadronic molecules}",
    eprint = "2101.01021",
    archivePrefix = "arXiv",
    primaryClass = "hep-ph",
    doi = "10.13725/j.cnki.pip.2021.02.001",
    journal = "Progr. Phys.",
    volume = "41",
    pages = "65--93",
    year = "2021"
}

@article{Dong:2021bvy,
    author = "Dong, Xiang-Kun and Guo, Feng-Kun and Zou, Bing-Song",
    title = "{A survey of heavy{\textendash}heavy hadronic molecules}",
    eprint = "2108.02673",
    archivePrefix = "arXiv",
    primaryClass = "hep-ph",
    doi = "10.1088/1572-9494/ac27a2",
    journal = "Commun. Theor. Phys.",
    volume = "73",
    number = "12",
    pages = "125201",
    year = "2021"
}

@article{Chen:2022asf,
    author = "Chen, Hua-Xing and Chen, Wei and Liu, Xiang and Liu, Yan-Rui and Zhu, Shi-Lin",
    title = "{An updated review of the new hadron states}",
    eprint = "2204.02649",
    archivePrefix = "arXiv",
    primaryClass = "hep-ph",
    doi = "10.1088/1361-6633/aca3b6",
    journal = "Rept. Prog. Phys.",
    volume = "86",
    number = "2",
    pages = "026201",
    year = "2023"
}

@article{Zou:2021sha,
    author = "Zou, Bing-Song",
    title = "{Building up the spectrum of pentaquark states as hadronic molecules}",
    eprint = "2103.15273",
    archivePrefix = "arXiv",
    primaryClass = "hep-ph",
    doi = "10.1016/j.scib.2021.04.023",
    journal = "Sci. Bull.",
    volume = "66",
    pages = "1258",
    year = "2021"
}

@article{Meng:2022ozq,
    author = "Meng, Lu and Wang, Bo and Wang, Guang-Juan and Zhu, Shi-Lin",
    title = "{Chiral perturbation theory for heavy hadrons and chiral effective field theory for heavy hadronic molecules}",
    eprint = "2204.08716",
    archivePrefix = "arXiv",
    primaryClass = "hep-ph",
    doi = "10.1016/j.physrep.2023.04.003",
    journal = "Phys. Rept.",
    volume = "1019",
    pages = "1--149",
    year = "2023"
}

@article{Yamaguchi:2019seo,
    author = "Yamaguchi, Yasuhiro and Garc{\'\i}a-Tecocoatzi, Hugo and Giachino, Alessandro and Hosaka, Atsushi and Santopinto, Elena and Takeuchi, Sachiko and Takizawa, Makoto",
    title = "{$P_c$ pentaquarks with chiral tensor and quark dynamics}",
    eprint = "1907.04684",
    archivePrefix = "arXiv",
    primaryClass = "hep-ph",
    doi = "10.1103/PhysRevD.101.091502",
    journal = "Phys. Rev. D",
    volume = "101",
    number = "9",
    pages = "091502",
    year = "2020"
}

@article{Giannuzzi:2019esi,
    author = "Giannuzzi, Floriana",
    title = "{Heavy pentaquark spectroscopy in the diquark model}",
    eprint = "1903.04430",
    archivePrefix = "arXiv",
    primaryClass = "hep-ph",
    reportNumber = "BARI-TH/719-19",
    doi = "10.1103/PhysRevD.99.094006",
    journal = "Phys. Rev. D",
    volume = "99",
    number = "9",
    pages = "094006",
    year = "2019"
}

@article{Ali:2019npk,
    author = "Ali, Ahmed and Parkhomenko, Alexander Ya.",
    title = "{Interpretation of the narrow $J/\psi p$ Peaks in $\Lambda_b \to J/\psi p K^-$ decay in the compact diquark model}",
    eprint = "1904.00446",
    archivePrefix = "arXiv",
    primaryClass = "hep-ph",
    reportNumber = "DESY-19-051",
    doi = "10.1016/j.physletb.2019.05.002",
    journal = "Phys. Lett. B",
    volume = "793",
    pages = "365--371",
    year = "2019"
}

@article{Eides:2019tgv,
    author = "Eides, Michael I. and Petrov, Victor Yu and Polyakov, Maxim V.",
    title = "{New LHCb pentaquarks as hadrocharmonium states}",
    eprint = "1904.11616",
    archivePrefix = "arXiv",
    primaryClass = "hep-ph",
    doi = "10.1142/S0217732320501515",
    journal = "Mod. Phys. Lett. A",
    volume = "35",
    number = "18",
    pages = "2050151",
    year = "2020"
}

@article{Shi:2021wyt,
    author = "Shi, Pan-Pan and Huang, Fei and Wang, Wen-Ling",
    title = "{Hidden charm pentaquark states in a diquark model}",
    eprint = "2107.08680",
    archivePrefix = "arXiv",
    primaryClass = "hep-ph",
    doi = "10.1140/epja/s10050-021-00542-4",
    journal = "Eur. Phys. J. A",
    volume = "57",
    number = "7",
    pages = "237",
    year = "2021"
}

@article{Giron:2021fnl,
    author = "Giron, Jesse F. and Lebed, Richard F.",
    title = "{Fine structure of pentaquark multiplets in the dynamical diquark model}",
    eprint = "2110.05557",
    archivePrefix = "arXiv",
    primaryClass = "hep-ph",
    reportNumber = "LA-UR-21-30038",
    doi = "10.1103/PhysRevD.104.114028",
    journal = "Phys. Rev. D",
    volume = "104",
    number = "11",
    pages = "114028",
    year = "2021"
}

@article{Mohan:2026blk,
    author = "Mohan, Binesh and Dhir, Rohit",
    title = "{A baryon-calibrated unified quark-diquark effective mass formalism for heavy multiquarks}",
    eprint = "2603.04175",
    archivePrefix = "arXiv",
    primaryClass = "hep-ph",
    month = "3",
    year = "2026"
}

@article{Hanhart:2025bun,
    author = "Hanhart, C.",
    title = "{Hadronic molecules and multiquark states}",
    eprint = "2504.06043",
    archivePrefix = "arXiv",
    primaryClass = "hep-ph",
    month = "4",
    year = "2025"
}

@article{Gross:2022hyw,
    author = "Gross, Franz and others",
    title = "{50 Years of Quantum Chromodynamics}",
    eprint = "2212.11107",
    archivePrefix = "arXiv",
    primaryClass = "hep-ph",
    doi = "10.1140/epjc/s10052-023-11949-2",
    journal = "Eur. Phys. J. C",
    volume = "83",
    pages = "1125",
    year = "2023"
}

@article{Guo:2015umn,
    author = "Guo, Feng-Kun and Mei{\ss}ner, Ulf-G. and Wang, Wei and Yang, Zhi",
    title = "{How to reveal the exotic nature of the P$_c$(4450)}",
    eprint = "1507.04950",
    archivePrefix = "arXiv",
    primaryClass = "hep-ph",
    doi = "10.1103/PhysRevD.92.071502",
    journal = "Phys. Rev. D",
    volume = "92",
    number = "7",
    pages = "071502",
    year = "2015"
}

@article{Liu:2015fea,
    author = "Liu, Xiao-Hai and Wang, Qian and Zhao, Qiang",
    title = "{Understanding the newly observed heavy pentaquark candidates}",
    eprint = "1507.05359",
    archivePrefix = "arXiv",
    primaryClass = "hep-ph",
    doi = "10.1016/j.physletb.2016.03.089",
    journal = "Phys. Lett. B",
    volume = "757",
    pages = "231--236",
    year = "2016"
}

@article{Mikhasenko:2015vca,
    author = "Mikhasenko, Mikhail",
    title = "{A triangle singularity and the LHCb pentaquarks}",
    eprint = "1507.06552",
    archivePrefix = "arXiv",
    primaryClass = "hep-ph",
    month = "7",
    year = "2015"
}

@article{Bayar:2016ftu,
    author = "Bayar, Melahat and Aceti, Francesca and Guo, Feng-Kun and Oset, Eulogio",
    title = "{A Discussion on Triangle Singularities in the $\Lambda_b \to J/\psi K^{-} p$ Reaction}",
    eprint = "1609.04133",
    archivePrefix = "arXiv",
    primaryClass = "hep-ph",
    doi = "10.1103/PhysRevD.94.074039",
    journal = "Phys. Rev. D",
    volume = "94",
    number = "7",
    pages = "074039",
    year = "2016"
}

@article{Guo:2019twa,
    author = "Guo, Feng-Kun and Liu, Xiao-Hai and Sakai, Shuntaro",
    title = "{Threshold cusps and triangle singularities in hadronic reactions}",
    eprint = "1912.07030",
    archivePrefix = "arXiv",
    primaryClass = "hep-ph",
    doi = "10.1016/j.ppnp.2020.103757",
    journal = "Prog. Part. Nucl. Phys.",
    volume = "112",
    pages = "103757",
    year = "2020"
}

@article{Shen:2020gpw,
    author = "Shen, Chao-Wei and Jing, Hao-Jie and Guo, Feng-Kun and Wu, Jia-Jun",
    title = "{Exploring Possible Triangle Singularities in the $\Xi^-_{b} \to K^- J/\psi \Lambda$ Decay}",
    eprint = "2008.09082",
    archivePrefix = "arXiv",
    primaryClass = "hep-ph",
    doi = "10.3390/sym12101611",
    journal = "Symmetry",
    volume = "12",
    number = "10",
    pages = "1611",
    year = "2020"
}

@article{Nakamura:2021qvy,
    author = "Nakamura, Satoshi X.",
    title = "{$P_c(4312)^+$, $P_c(4380)^+$, and $P_c(4457)^+$ as double triangle cusps}",
    eprint = "2103.06817",
    archivePrefix = "arXiv",
    primaryClass = "hep-ph",
    doi = "10.1103/PhysRevD.103.L111503",
    journal = "Phys. Rev. D",
    volume = "103",
    pages = "111503",
    year = "2021"
}

@article{Duan:2023dky,
    author = "Duan, Ming-Xiao and Qiu, Lin and Ling, Xi-Zhe and Zhao, Qiang",
    title = "{Predictions for feed-down enhancements at the {\ensuremath{\Lambda}}cD{\textasciimacron} and {\ensuremath{\Lambda}}cD{\textasciimacron}* thresholds via the triangle and box singularities}",
    eprint = "2303.13329",
    archivePrefix = "arXiv",
    primaryClass = "hep-ph",
    doi = "10.1103/PhysRevD.109.L031507",
    journal = "Phys. Rev. D",
    volume = "109",
    number = "3",
    pages = "L031507",
    year = "2024"
}

@article{Kuang:2020bnk,
    author = "Kuang, Shi-Qing and Dai, Ling-Yun and Kang, Xian-Wei and Yao, De-Liang",
    title = "{Pole analysis on the hadron spectroscopy of $\Lambda_b\to J/\Psi p K^-$}",
    eprint = "2002.11959",
    archivePrefix = "arXiv",
    primaryClass = "hep-ph",
    doi = "10.1140/epjc/s10052-020-8008-5",
    journal = "Eur. Phys. J. C",
    volume = "80",
    number = "5",
    pages = "433",
    year = "2020"
}

@article{Nakamura:2021dix,
    author = "Nakamura, S. X. and Hosaka, A. and Yamaguchi, Y.",
    title = "{$P_c(4312)^+$ and $P_c(4337)^+$ as interfering $\Sigma_c \bar{D}$ and $\Lambda_c \bar{D}^{*}$ threshold cusps}",
    eprint = "2109.15235",
    archivePrefix = "arXiv",
    primaryClass = "hep-ph",
    doi = "10.1103/PhysRevD.104.L091503",
    journal = "Phys. Rev. D",
    volume = "104",
    number = "9",
    pages = "L091503",
    year = "2021"
}

@article{Co:2024bfl,
    author = "Co, Darwin Alexander O. and Chavez, Vince Angelo A. and Sombillo, Denny Lane B.",
    title = "{Deep learning framework for disentangling triangle singularity and pole-based enhancements}",
    eprint = "2403.18265",
    archivePrefix = "arXiv",
    primaryClass = "hep-ph",
    doi = "10.1103/PhysRevD.110.114034",
    journal = "Phys. Rev. D",
    volume = "110",
    number = "11",
    pages = "114034",
    year = "2024"
}

@article{Zhang:2024qkg,
    author = "Zhang, Zhen-Hua and Guo, Feng-Kun",
    title = "{Classification of coupled-channel near-threshold structures}",
    eprint = "2407.10620",
    archivePrefix = "arXiv",
    primaryClass = "hep-ph",
    doi = "10.1016/j.physletb.2025.139387",
    journal = "Phys. Lett. B",
    volume = "863",
    pages = "139387",
    year = "2025"
}

@article{Sakthivasan:2024uwd,
    author = {Sakthivasan, Ajay S. and Mai, Maxim and Rusetsky, Akaki and D{\"o}ring, Michael},
    title = "{Effects of final state interactions on Landau singularities}",
    eprint = "2407.17969",
    archivePrefix = "arXiv",
    primaryClass = "hep-ph",
    doi = "10.1007/JHEP10(2024)246",
    journal = "JHEP",
    volume = "10",
    pages = "246",
    year = "2024"
}

@article{Oller:2000fj,
    author = "Oller, J. A. and Meissner, Ulf G.",
    title = "{Chiral dynamics in the presence of bound states: Kaon nucleon interactions revisited}",
    eprint = "hep-ph/0011146",
    archivePrefix = "arXiv",
    reportNumber = "FZJ-IKP-TH-2000-26",
    doi = "10.1016/S0370-2693(01)00078-8",
    journal = "Phys. Lett. B",
    volume = "500",
    pages = "263--272",
    year = "2001"
}

@article{Oller:1998zr,
    author = "Oller, J. A. and Oset, E.",
    title = "{N/D description of two meson amplitudes and chiral symmetry}",
    eprint = "hep-ph/9809337",
    archivePrefix = "arXiv",
    doi = "10.1103/PhysRevD.60.074023",
    journal = "Phys. Rev. D",
    volume = "60",
    pages = "074023",
    year = "1999"
}

@article{Xiao:2013yca,
    author = "Xiao, C. W. and Nieves, J. and Oset, E.",
    title = "{Combining heavy quark spin and local hidden gauge symmetries in the dynamical generation of hidden charm baryons}",
    eprint = "1304.5368",
    archivePrefix = "arXiv",
    primaryClass = "hep-ph",
    doi = "10.1103/PhysRevD.88.056012",
    journal = "Phys. Rev. D",
    volume = "88",
    pages = "056012",
    year = "2013"
}

@article{Xiao:2019gjd,
    author = "Xiao, C. W. and Nieves, J. and Oset, E.",
    title = "{Prediction of hidden charm strange molecular baryon states with heavy quark spin symmetry}",
    eprint = "1906.09010",
    archivePrefix = "arXiv",
    primaryClass = "hep-ph",
    doi = "10.1016/j.physletb.2019.135051",
    journal = "Phys. Lett. B",
    volume = "799",
    pages = "135051",
    year = "2019"
}

@article{Oset:1997it,
    author = "Oset, E. and Ramos, A.",
    title = "{Nonperturbative chiral approach to s wave anti-K N interactions}",
    eprint = "nucl-th/9711022",
    archivePrefix = "arXiv",
    doi = "10.1016/S0375-9474(98)00170-5",
    journal = "Nucl. Phys. A",
    volume = "635",
    pages = "99--120",
    year = "1998"
}

@article{Oller:1997ti,
    author = "Oller, J. A. and Oset, E.",
    title = "{Chiral symmetry amplitudes in the S wave isoscalar and isovector channels and the $\sigma$, f$_0$(980), a$_0$(980) scalar mesons}",
    eprint = "hep-ph/9702314",
    archivePrefix = "arXiv",
    doi = "10.1016/S0375-9474(97)00160-7",
    journal = "Nucl. Phys. A",
    volume = "620",
    pages = "438--456",
    year = "1997",
    note = "[Erratum: Nucl.Phys.A 652, 407--409 (1999)]"
}

@article{Guo:2005wp,
    author = "Guo, Feng-Kun and Ping, Rong-Gang and Shen, Peng-Nian and Chiang, Huan-Ching and Zou, Bing-Song",
    title = "{S wave K pi scattering and effects of kappa in J/psi ---{\ensuremath{>}} anti-K*0 (892) K+ pi-}",
    eprint = "hep-ph/0509050",
    archivePrefix = "arXiv",
    doi = "10.1016/j.nuclphysa.2006.04.008",
    journal = "Nucl. Phys. A",
    volume = "773",
    pages = "78--94",
    year = "2006"
}

@article{Guo:2006fu,
    author = "Guo, Feng-Kun and Shen, Peng-Nian and Chiang, Huan-Ching and Ping, Rong-Gang and Zou, Bing-Song",
    title = "{Dynamically generated 0+ heavy mesons in a heavy chiral unitary approach}",
    eprint = "hep-ph/0603072",
    archivePrefix = "arXiv",
    doi = "10.1016/j.physletb.2006.08.064",
    journal = "Phys. Lett. B",
    volume = "641",
    pages = "278--285",
    year = "2006"
}

@article{Oller:2004xm,
    author = "Oller, Jose A.",
    title = "{Final state interactions in hadronic D decays}",
    eprint = "hep-ph/0411105",
    archivePrefix = "arXiv",
    doi = "10.1103/PhysRevD.71.054030",
    journal = "Phys. Rev. D",
    volume = "71",
    pages = "054030",
    year = "2005"
}

@article{Ozpineci:2013zas,
    author = "Ozpineci, A. and Xiao, C. W. and Oset, E.",
    title = "{Hidden beauty molecules within the local hidden gauge approach and heavy quark spin symmetry}",
    eprint = "1306.3154",
    archivePrefix = "arXiv",
    primaryClass = "hep-ph",
    doi = "10.1103/PhysRevD.88.034018",
    journal = "Phys. Rev. D",
    volume = "88",
    pages = "034018",
    year = "2013"
}

@article{Yamagata-Sekihara:2010kpd,
    author = "Yamagata-Sekihara, J. and Nieves, J. and Oset, E.",
    title = "{Couplings in coupled channels versus wave functions in the case of resonances: application to the two $\Lambda(1405)$ states}",
    eprint = "1007.3923",
    archivePrefix = "arXiv",
    primaryClass = "hep-ph",
    doi = "10.1103/PhysRevD.83.014003",
    journal = "Phys. Rev. D",
    volume = "83",
    pages = "014003",
    year = "2011"
}

@article{Ahmed:2020kmp,
    author = "Ahmed, Hiwa A. and Xiao, C. W.",
    title = "{Study the molecular nature of $\sigma$, $f_{0}(980)$, and $a_{0}(980)$ states}",
    eprint = "2001.08141",
    archivePrefix = "arXiv",
    primaryClass = "hep-ph",
    doi = "10.1103/PhysRevD.101.094034",
    journal = "Phys. Rev. D",
    volume = "101",
    number = "9",
    pages = "094034",
    year = "2020",
    note = "[Erratum: Phys.Rev.D 112, 099902 (2025)]"
}

@article{Sekihara:2013wlq,
    author = "Sekihara, Takayasu and Hyodo, Tetsuo",
    title = "{Size measurement of dynamically generated hadronic resonances with finite boxes}",
    eprint = "1209.0577",
    archivePrefix = "arXiv",
    primaryClass = "nucl-th",
    doi = "10.1103/PhysRevC.87.045202",
    journal = "Phys. Rev. C",
    volume = "87",
    number = "4",
    pages = "045202",
    year = "2013"
}

@article{Wang:2019nvm,
    author = "Wang, Bo and Meng, Lu and Zhu, Shi-Lin",
    title = "{Spectrum of the strange hidden charm molecular pentaquarks in chiral effective field theory}",
    eprint = "1912.12592",
    archivePrefix = "arXiv",
    primaryClass = "hep-ph",
    doi = "10.1103/PhysRevD.101.034018",
    journal = "Phys. Rev. D",
    volume = "101",
    number = "3",
    pages = "034018",
    year = "2020"
}

@article{Isgur:1989vq,
    author = "Isgur, Nathan and Wise, Mark B.",
    title = "{Weak Decays of Heavy Mesons in the Static Quark Approximation}",
    reportNumber = "UTPT-89-27, CALT-68-1585",
    doi = "10.1016/0370-2693(89)90566-2",
    journal = "Phys. Lett. B",
    volume = "232",
    pages = "113--117",
    year = "1989"
}

@article{Neubert:1993mb,
    author = "Neubert, Matthias",
    title = "{Heavy quark symmetry}",
    eprint = "hep-ph/9306320",
    archivePrefix = "arXiv",
    reportNumber = "SLAC-PUB-6263",
    doi = "10.1016/0370-1573(94)90091-4",
    journal = "Phys. Rept.",
    volume = "245",
    pages = "259--396",
    year = "1994"
}

@article{Bando:1984ej,
    author = "Bando, M. and Kugo, T. and Uehara, S. and Yamawaki, K. and Yanagida, T.",
    title = "{Is rho Meson a Dynamical Gauge Boson of Hidden Local Symmetry?}",
    reportNumber = "RRK 84-22",
    doi = "10.1103/PhysRevLett.54.1215",
    journal = "Phys. Rev. Lett.",
    volume = "54",
    pages = "1215",
    year = "1985"
}

@article{Bando:1987br,
    author = "Bando, Masako and Kugo, Taichiro and Yamawaki, Koichi",
    title = "{Nonlinear Realization and Hidden Local Symmetries}",
    reportNumber = "DPNU-87-63, AICHI-1, KUNS-903",
    doi = "10.1016/0370-1573(88)90019-1",
    journal = "Phys. Rept.",
    volume = "164",
    pages = "217--314",
    year = "1988"
}

@article{Meissner:1987ge,
    author = "Meissner, Ulf G.",
    title = "{Low-Energy Hadron Physics from Effective Chiral Lagrangians with Vector Mesons}",
    reportNumber = "MIT-CTP-1471",
    doi = "10.1016/0370-1573(88)90090-7",
    journal = "Phys. Rept.",
    volume = "161",
    pages = "213",
    year = "1988"
}

@article{Nagahiro:2008cv,
    author = "Nagahiro, H. and Roca, L. and Hosaka, A. and Oset, E.",
    title = "{Hidden gauge formalism for the radiative decays of axial-vector mesons}",
    eprint = "0809.0943",
    archivePrefix = "arXiv",
    primaryClass = "hep-ph",
    doi = "10.1103/PhysRevD.79.014015",
    journal = "Phys. Rev. D",
    volume = "79",
    pages = "014015",
    year = "2009"
}

@article{Azizi:2016dhy,
    author = "Azizi, K. and Sarac, Y. and Sundu, H.",
    title = "{Analysis of $P_c^+(4380)$ and $P_c^+(4450)$ as pentaquark states in the molecular picture with QCD sum rules}",
    eprint = "1612.07479",
    archivePrefix = "arXiv",
    primaryClass = "hep-ph",
    doi = "10.1103/PhysRevD.95.094016",
    journal = "Phys. Rev. D",
    volume = "95",
    number = "9",
    pages = "094016",
    year = "2017"
}

@article{Azizi:2018bdv,
    author = "Azizi, K. and Sarac, Y. and Sundu, H.",
    title = "{Strong decay of $P_c(4380)$ pentaquark in a molecular picture}",
    eprint = "1802.01384",
    archivePrefix = "arXiv",
    primaryClass = "hep-ph",
    doi = "10.1016/j.physletb.2018.06.022",
    journal = "Phys. Lett. B",
    volume = "782",
    pages = "694--701",
    year = "2018"
}

@article{Azizi:2020ogm,
    author = "Azizi, K. and Sarac, Y. and Sundu, H.",
    title = "{Properties of the $P_c(4312)$ pentaquark and its bottom partner}",
    eprint = "2011.05828",
    archivePrefix = "arXiv",
    primaryClass = "hep-ph",
    doi = "10.1088/1674-1137/abe8ce",
    journal = "Chin. Phys. C",
    volume = "45",
    number = "5",
    pages = "053103",
    year = "2021"
}

@article{Azizi:2021utt,
    author = "Azizi, K. and Sarac, Y. and Sundu, H.",
    title = "{Investigation of $P_{cs}(4459)^0$ pentaquark via its strong decay to $\Lambda J/\Psi$}",
    eprint = "2101.07850",
    archivePrefix = "arXiv",
    primaryClass = "hep-ph",
    doi = "10.1103/PhysRevD.103.094033",
    journal = "Phys. Rev. D",
    volume = "103",
    number = "9",
    pages = "094033",
    year = "2021"
}

@article{Azizi:2018dva,
    author = "Azizi, K. and Sarac, Y. and Sundu, H.",
    title = "{Possible Molecular Pentaquark States with Different Spin and Quark Configurations}",
    eprint = "1805.06734",
    archivePrefix = "arXiv",
    primaryClass = "hep-ph",
    doi = "10.1103/PhysRevD.98.054002",
    journal = "Phys. Rev. D",
    volume = "98",
    number = "5",
    pages = "054002",
    year = "2018"
}

@article{Azizi:2017bgs,
    author = "Azizi, K. and Sarac, Y. and Sundu, H.",
    title = "{Hidden Bottom Pentaquark States with Spin 3/2 and 5/2}",
    eprint = "1707.01248",
    archivePrefix = "arXiv",
    primaryClass = "hep-ph",
    doi = "10.1103/PhysRevD.96.094030",
    journal = "Phys. Rev. D",
    volume = "96",
    number = "9",
    pages = "094030",
    year = "2017"
}

\end{document}